\documentclass[%
 reprint,
 superscriptaddress,
 amsmath,
 amssymb,
 aps,
 prc,
]{revtex4-2}

\usepackage{graphicx}
\usepackage{dcolumn}
\usepackage{bm}

\usepackage{multirow}

\newcommand{\st}{$\mathrm{^{\text{st}}}$}
\newcommand{\nd}{$\mathrm{^{\text{nd}}}$}
\newcommand{\rd}{$\mathrm{^{\text{rd}}}$}
\newcommand{\nth}{$\mathrm{^{\text{th}}}$}

\newcommand{\bnue}{$\mathrm{\bar{\nu}_e}$}

\newcommand{\Qb}{$\mathrm{Q_{\beta}}$}

\newcommand{\U}[1]{$\mathrm{^{#1}U}$}
\newcommand{\Pu}[1]{$\mathrm{^{#1}Pu}$}
\newcommand{\iso}[2]{$\mathrm{^{#1}{#2}}$}

\newcommand{\AME}{AME-2020}

\newcommand{\NUBASE}{NUBASE-2020}
\newcommand{\jeff}{JEFF-3.3}
\newcommand{\ENDF}{ENDF/B-VIII.0}

\newcommand{\JENDL}{JENDL-5}

\begin{document}

\title{A comprehensive revision of the summation method for the prediction of reactor \bnue{} fluxes and spectra}

\author{Lorenzo Périssé}
\altaffiliation[Now at ]{ILANCE, CNRS - University of Tokyo International Research Laboratory, Kashiwa, Chiba 277-8582, Japan}
\affiliation{IRFU, CEA, Universit\'{e} Paris-Saclay, F-91191 Gif-sur-Yvette, France}%
\author{Anthony Onillon}
\altaffiliation[Now at ]{Physik-department, Technische Universität München, D-85748 Garching, Germany}
\affiliation{IRFU, CEA, Universit\'{e} Paris-Saclay, F-91191 Gif-sur-Yvette, France}%
\author{Xavier Mougeot}
\affiliation{Universit{\'e} Paris-Saclay, CEA, List, Laboratoire National Henri Becquerel (LNE-LNHB), F-91120 Palaiseau, France}
\author{Matthieu Vivier}%
 \email{Corresponding author: matthieu.vivier@cea.fr}
\affiliation{IRFU, CEA, Universit\'{e} Paris-Saclay, F-91191 Gif-sur-Yvette, France}%
\author{Thierry Lasserre}
\affiliation{IRFU, CEA, Universit\'{e} Paris-Saclay, F-91191 Gif-sur-Yvette, France}%
\author{Alain Letourneau}
\affiliation{IRFU, CEA, Universit\'{e} Paris-Saclay, F-91191 Gif-sur-Yvette, France}%
\author{David Lhuillier}
\affiliation{IRFU, CEA, Universit\'{e} Paris-Saclay, F-91191 Gif-sur-Yvette, France}%
\author{Guillaume Mention}
\affiliation{IRFU, CEA, Universit\'{e} Paris-Saclay, F-91191 Gif-sur-Yvette, France}%


\date{\today}

\begin{abstract}
The summation method for the calculation of reactor \bnue{} fluxes and spectra is methodically revised and improved. For the first time, a complete uncertainty budget accounting for all known effects likely to impact these calculations is proposed. Uncertainties of a few percents at low energies and ranging up to 20\% at high energies are obtained on the calculation of a typical reactor \bnue{} spectrum. Although huge improvements have been achieved over the past decade, the quality and incompleteness of the present day evaluated nuclear decay data still limit the accuracy of the calculations and therefore dominate by far these uncertainties. Pushing the $\mathrm{\beta}$-decay modeling of the thousands of branches making a reactor \bnue{} spectrum to a high level of details comparatively brings modest changes. In particular, including nuclear structure calculations in the evaluation of the non-unique forbidden transitions gives a smaller impact than anticipated in past studies.~Finally, this new modeling is challenged against state-of-the-art predictions and measurements. While a good agreement is observed with the most recent Inverse Beta Decay measurements of reactor \bnue{} fluxes and spectra, it is unable to properly describe the reference aggregate $\mathrm{\beta}$ spectra measured at the Institut Laue-Langevin High-Flux reactor in the 80s. This result adds to recent suspicions about the reliability of these data and preferentially points toward a misprediction of the \U{235} \bnue{} spectrum.
\end{abstract}

\maketitle

\section{Introduction}
The determination of the \bnue{} flux and spectrum emitted by a nuclear reactor is a long standing problem, dating back from the first detection of the neutrino in 1956~\cite{reines1953detection,cowan1956detection}. Reactor \bnue{} mostly originate from the $\mathrm{\beta}$ decay of neutron-rich products following the fission of uranium and plutonium present in the nuclear fuel. They are also emitted in smaller amounts through the $\mathrm{\beta}$ decay of activation products following neutron irradiation of the fuel and/or structural materials present in the core. Past and present efforts focused on studying reactor \bnue{} both at commercial and research reactors. Commercial reactors, designed for the large scale production of electricity, mostly release energy through the fission of \U{235}, \Pu{239}, \Pu{241} and \U{238} isotopes while smaller research reactors run almost entirely through the fission of \U{235}.
The prediction of the \bnue{} flux and spectrum following the fission of each of these isotopes is a key ingredient to the study of the weak interaction and of the neutrino fundamental properties at such facilities, and still motivates huge experimental and theoretical efforts.~The first attempts to predict a reactor \bnue{} spectrum come from the conversion of measured aggregate $\mathrm{\beta}$ spectra resulting from the thermal fission of \U{235}~\cite{muehlhause1957antineutrino,carter1959free}. This pioneering work led to the so-called conversion method, which was since then refined both with new experimental data~\cite{Schreckenbach1982,Schreckenbach1985,Schreckenbach1989,Haag2014, Haag2014a} and an updated conversion procedure~\cite{vogel2007conversion}. Other calculations were also attempted by adding up the theoretical prediction of all known fission fragment $\mathrm{\beta}$/\bnue{} spectra using their corresponding yields and decay schemes~\cite{king1958,avignone1968,borovoi1977,davis1979,avignone1980,vogel1981}. As pointed out from the very beginning, this summation method however mostly suffers from the incompleteness of the evaluated nuclear databases, especially regarding fission fragments having a large total decay energy \Qb{}, hence giving a lower limit on the predicted \bnue{} flux. Until recently, the conversion method was therefore considered to provide the most accurate and robust predictions of \bnue{} fluxes and spectra at nuclear reactors.

In 2011, two independent revisions of the conversion procedure concluded to an increase in the expected reactor \bnue{} fluxes \cite{Mueller2011,Huber2011}, leading to a ($\mathrm{5.7 \pm 2.4}$)\% deficit in the detected \bnue{} rates with respect to predictions at short and middle baseline past reactor experiments~\cite{mention2011}. 
This reactor antineutrino anomaly (RAA) was confirmed by the last generation of Inverse Beta Decay (IBD) experiments (Daya Bay~\cite{Adey2019a}, Double Chooz~\cite{dekerret2020}, and RENO~\cite{yoon2021}) conducted at commercial reactors and aiming at measuring the $\mathrm{\theta_{13}}$ mixing angle. These experiments also highlighted an unexpected spectrum distortion in the 4.5 to 7.5 MeV energy regime with respect to predictions based on the Huber-Mueller (HM) revised calculations of the actinide fission \bnue{} spectrum of \U{235}, \U{238}, \Pu{239} and \Pu{241}~\cite{abe2014,An2016,choi2016}.
These anomalies motivated various experimental efforts at short baselines from research highly-enriched uranium (HEU) and commercial lowly-enriched uranium (LEU) reactors both to precisely reassess the \bnue{} absolute flux and spectrum as well as to search for a possible $\mathrm{\sim}$1 eV mass sterile neutrino able to explain the RAA (see e.g.~\cite{boser2020} and references therein). Although the sterile neutrino interpretation of the RAA is now strongly disfavored (see e.g.~\cite{dentler2017,dentler2018,giunti2019,berryman2021,berryman2022}), the discrepancies between predicted and observed \bnue{} fluxes and spectra have since been further confirmed, casting doubts on the reliability of the state-of-the-art conversion predictions.~In particular, the absolute normalization of the ILL aggregate $\mathrm{\beta}$ spectra they rely on has recently been seriously questioned~\cite{bergevin2019applied}. This suspicion was later confirmed by a measurement of the ratio of the \U{235} over the \Pu{239} aggregate $\mathrm{\beta}$ spectra, indicating for a (5.4 $\mathrm{\pm}$ 0.2)\% excess with respect to that same ratio constructed with the ILL reference data~\cite{Kopeikin2021}. At the same time, the RAA and the observed spectral anomalies have motivated many efforts in the nuclear physics community to improve the summation calculations, as this method is a powerful tool for dissecting and assessing systematic effects in reactor \bnue{} predictions~\cite{dwyer2015,sonzogni2015,sonzogni2017}. Particular focus have been paid on evaluating the role of the non-unique forbidden transitions~\cite{hayes2014,fang2015,hayen2019,hayen2019a} and of the quality and incompleteness of the evaluated nuclear databases~\cite{Fallot2012,Estienne2019,sonzogni2016,schmidt2021}. One notable finding is that the so-called Pandemonium effect, leading to a biased estimate of many of the fission fragment $\mathrm{\beta}$ decay schemes in modern nuclear evaluated data libraries, still contributes to discrepancies between summation predictions and state-of-the-art measurements of the \bnue{} flux at nuclear reactors~\cite{letourneau2022}.

Although significant progress has been made, especially with many improvements in the content and the quality of the nuclear evaluated data libraries, the summation method still suffers from approximations and lacks a comprehensive uncertainty budget to truly be a robust prediction tool. Together with the forthcoming generation of experiments aiming at studying Coherent Elastic Neutrino-Nucleus Scattering (see e.g.~\cite{CEvNSsnowmass2022}), the current and next generations of IBD experiments will reach increasing precision in the measurements of \bnue{} fluxes and spectra at reactors, making it necessary to extend and refine the summation calculations.~The following article therefore reports about a careful revision of the summation method, improving over current state-of-the-art predictions by using an advanced modeling of $\mathrm{\beta}$ decay and recent evaluated nuclear data. This revision work is based on the software Beta Energy Spectrum Tool for an Improved Optimal List of Elements (BESTIOLE), which was developed for the reevaluation of reactor antineutrino spectra in \cite{Mueller2011}. Further to these improvements, the largest systematic effects known to impact the calculations of reactor \bnue{} fluxes and spectra are here quantitatively studied to propose for the first time a detailed uncertainty model. As a cautionary word, it should already be stressed here that the construction of such an uncertainty model remains a difficult task which has to address both the limitations (quality and incompleteness) of the evaluated nuclear data libraries and the many approximations used in the modeling of the $\mathrm{\beta}$ branches.~Especially when the necessary input information are incomplete or missing, the uncertainty associated with each systematic effects is estimated following simple approaches using known and reliable data as a proxy, with the prime care of remaining conservative and as realistic as possible. The robustness and the limitations of this first uncertainty model for the summation method are in this regard discussed and criticized all along the article.

The article is structured as followed: section \ref{sec:SummationMethod} lays down the general principles of the summation method along with the uncertainty propagation formalism used throughout this work. Section \ref{sec:BetaDecayFormalism} details the improvements the present work makes over past modelings of $\mathrm{\beta}$ decay. In particular, the implementation of various electro-magnetic corrections to the Fermi theory as well as realistic calculations of the main forbidden non-unique transitions contributing to the detected flux at a reactor are presented. Section \ref{sec:evaluated_nuclear_data} describes the construction of a new base of nuclear input data for the modeling of each $\mathrm{\beta}$-branch contributing to a reactor \bnue{} spectrum, using recent Pandemonium-free evaluations available in the literature and in online databases.~Following these major revisions, section \ref{sec:Results} then presents the new summation calculations of the four major actinides \bnue{} fission spectra (\U{235}, \Pu{239}, \Pu{241}, \U{238}) along with a detailed breakdown of their uncertainties. In light of the RAA and the observed spectral anomalies, these new results are challenged against state-of-the-art predictions and measurements. Finally, section \ref{sec:Conclusions} summarizes the main results of this work and opens up possible improvements to the summation method, especially in view of addressing the RAA and providing a complete and robust prediction tool for the next generation of reactor \bnue{} experiments.
%
%
\section{The summation method}\label{sec:SummationMethod}
The summation calculation of a reactor $\mathrm{\beta}$ spectrum\footnote{Here and in the following, only the fission term is treated meaning that the activation term following from neutron irradiation of the fuel and/or structural materials is disregarded.} consists in adding up the $\mathrm{\beta}$ spectrum $\mathrm{S_p}$ of any fragment p populating the fission process of any of the four major actinides k = \U{235}, \Pu{239}, \Pu{241}, \U{238} present in the core:
\begin{equation}
    \mathrm{S_{tot}(E,t)} = \mathrm{ \sum_{k,p} \mathcal{A}^{k}_{p}(t) \, S_p (E),} \label{eq:reactor_spectrum}
\end{equation}
where $\mathrm{\mathcal{A}^{k}_{p}(t)}$ is the activity of the $\mathrm{p^{th}}$ fission fragment at irradiation time t and E the $\mathrm{\beta}$ particle kinetic energy. The fission fragment activities can be estimated using their cumulative yield $\mathrm{\mathcal{Y}^{k}_{p}(t)}$ after an irradiation time t, so that eq.~\ref{eq:reactor_spectrum} can also be written as:
\begin{eqnarray}
    \mathrm{S_{tot}(E,t)} &&= \mathrm{\sum_{k} f_k(t) \, \sum_{p} \mathcal{Y}^{k}_p (t) \, S_p (E)} \label{eq:reactor_spectrum_with_CFY}\\ 
    &&= \mathrm{\sum_{k} f_k(t) \, S_k(E,t),}\label{eq:simplified_reactor_spectrum}
\end{eqnarray}
where $\mathrm{f_k(t)}$ is the $\mathrm{k^{th}}$ actinide fission rate. The term $\mathrm{S_k(E,t)}$ is here the total $\mathrm{\beta}$ spectrum associated to one fission of the $\mathrm{k^{th}}$ actinide, and is usually denoted as actinide fission spectrum.
The $\mathrm{\beta}$ spectrum $\mathrm{S_p}$ of a fission fragment can generally be broken down to the superimposition of $\mathrm{N^{p}}$ transitions connecting either the ground state (GS) or an isomeric state (IS) to different excited levels b of its daughter nucleus:
\begin{equation}\label{eq:fission fragment spectrum}
\mathrm{S_p(E) = \sum_{b=1}^{N^{p}} \mathcal{B}^{p}_{b} \ S^{p}_{b}(E, E^{p}_{0,b}).}
\end{equation}
The probability $\mathrm{\mathcal{B}^{p}_{b}}$ to decay through a given transition is denoted here branching ratio. The transition branching ratios add up to unity if the fission fragment is a pure $\mathrm{\beta}$-emitter and to less than one otherwise. The endpoint energy $\mathrm{E^{p}_{0,b}}$ of a transition is usually expressed as:
\begin{equation}\label{eq:endpoint energy}
    \mathrm{E^{p}_{0,b} = Q^{p}_{\beta} + E^{p}_{IS} - E^{lvl}_{b},}
\end{equation}
where $\mathrm{Q^{p}_{\beta}}$ is the total $\mathrm{\beta}$ decay energy (corresponding to a ground state to ground state transition), $\mathrm{E^{p}_{IS}}$ is the energy of the parent nucleus isomeric state ($\mathrm{E^{p}_{IS}} = 0$ if the parent nucleus $\mathrm{\beta}$ decays from the ground state), and $\mathrm{E^{lvl}_{b}}$ is the energy of the daughter nucleus $\mathrm{b^{th}}$ excited state. 
Combining eq.~\ref{eq:reactor_spectrum}, \ref{eq:fission fragment spectrum} and \ref{eq:endpoint energy}, a reactor $\mathrm{\beta}$ spectrum can hence be considered as the sum of many individual $\mathrm{\beta}$-branch spectra, each requiring both an accurate $\mathrm{\beta}$-decay formalism and robust evaluated nuclear data to be modeled. The correspondence between a reactor $\mathrm{\beta}$ and \bnue{} spectrum is achieved at the $\mathrm{\beta}$-branch level using energy conservation and neglecting the daughter nucleus recoil energy by substituting in the above formula the $\mathrm{\beta}$ particle kinetic energy E with the \bnue{} energy $\mathrm{E_{\nu} = E^{p}_{0,b}-E}$. As opposed to the conversion method, the summation method then ensures a true and unique correspondence between reactor $\mathrm{\beta}$ and \bnue{} spectra while computing them with equal precision. 

In the following, many results will be expressed not only through raw \bnue{} flux and spectrum calculations, but also computing IBD yields in order to ease the comparison with state-of-the-art \bnue{} flux predictions and measurements (see e.g. section~\ref{sec:Results}). The IBD yield is here computed as:
\begin{equation}\label{eq:IBD_yield}
    \mathrm{\langle \sigma_{IBD} \rangle = \int_{E_{th}}^{E_{max}}\sigma_{IBD}(E_{\nu})\,S(E_{\nu})\,dE_{\nu},}
\end{equation}
where $\mathrm{\sigma_{IBD}(E_{\nu})}$ is the IBD cross-section taken from~\cite{Vogel1999} and $\mathrm{S(E_{\nu})}$ is a \bnue{} spectrum. Throughout this article, IBD yields could respectively be computed either at the level of $\mathrm{\beta}$ branch, at the level of a fission fragment or for a full actinide fission spectrum. The normalisation of the IBD cross-section was estimated using the Particle Data Group 2022 evaluation of the required fundamental constants~\cite{pdg2022}, also including radiative corrections as derived in~\cite{fayans1985}. The lower integration bound in eq.~\ref{eq:IBD_yield} was set to the IBD energy threshold $\mathrm{E_{th} = 1.806}$ MeV. The upper integration bound is especially relevant for the computation of the IBD yields associated to the fission of the \U{235}, \U{238}, \Pu{239} and \Pu{241} actinides, denoted hereafter as isotopic IBD yields. It was chosen as $\mathrm{E_{max}}$ = 10 MeV accordingly to~\cite{Giunti2022}. Extending this bound to energies higher than 10 MeV has been checked to give negligible (\textless 0.1\%) changes.

\begin{figure}[ht]
    \centering
    \includegraphics[width=0.47\textwidth]{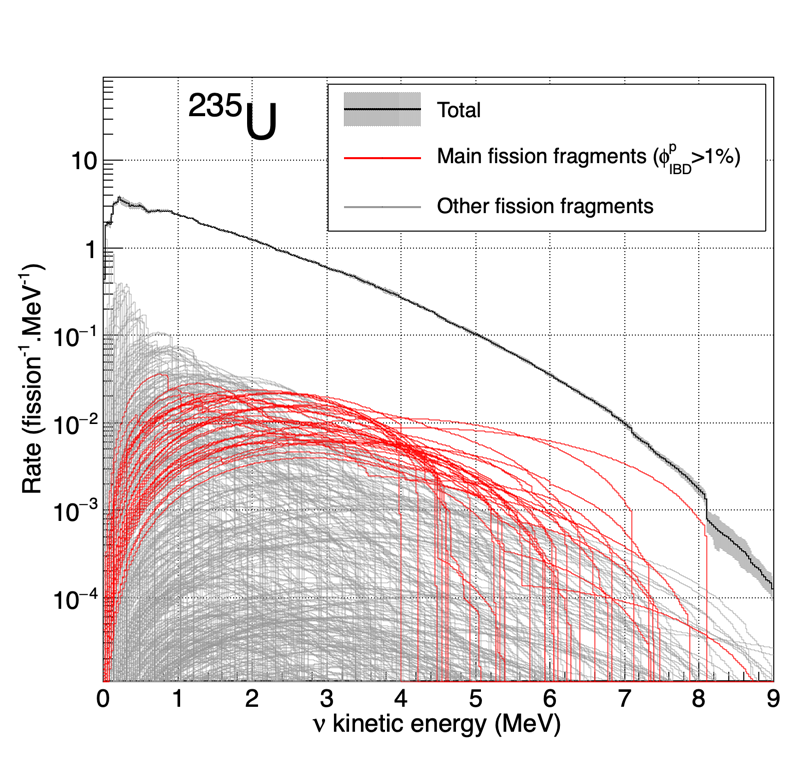}
    \caption{Summation calculation of the \bnue{} spectrum resulting from the thermal fission of \U{235}. The total spectrum (black solid line) is broken down to the contributions of all the fission fragments listed in the present day nuclear databases (grey solid lines). The dark grey area, especially visible at high energies, represents the associated total uncertainty. The red solid lines highlight the \bnue{} spectrum of 32 fission fragments, each contributing more than 1\% of the expected IBD yield (see eq.~\ref{eq:IBD_yield}). Summed all together, these fission fragments total 60\% of the \U{235} isotopic IBD yield.
    }
    \label{fig:fission_spectrum_235U}
\end{figure}

Figure~\ref{fig:fission_spectrum_235U} illustrates the summation calculation of a reactor \bnue{} spectrum originating from the thermal fission of \U{235}. A fission spectrum typically piles the individual $\mathrm{\beta}$/\bnue{} spectra of $\mathrm{\sim}$800 fragments, and totals more than 10,000 $\mathrm{\beta}$ branches. The exponential decrease of the spectrum as a function of energy results from the underlying \Qb{} distribution of the fission fragments, which makes fewer and fewer $\mathrm{\beta}$ branches to contribute toward high energies. Interestingly, only a handful of fission fragments dominates the associated IBD yield. For instance, about thirty fission fragments exhibit a \textgreater 1\% contribution to the \U{235} IBD yield. They are indicated by the red solid lines on Figure~\ref{fig:fission_spectrum_235U}.

The total uncertainty associated to the summation calculation of a reactor $\mathrm{\beta}$/\bnue{} spectrum combines uncertainties from many sources, which are propagated using the following covariance matrix formalism \cite{mathai1997jacobians}. 
The covariance matrix $\mathrm{V^{s}_p}$ associated to the binned spectrum $\mathrm{S_p(E)}$ of a fission fragment p for an uncertainty source s (see eq.~\ref{eq:fission fragment spectrum}) is given by:
\begin{equation}
\label{eq:cov_parameter_iso}
\mathrm{V_p^s = 
\sum_{i} V^{s}_{i} + 
\sum_{\substack{i,j \\ i \neq j}}  V^{s}_{ij},}
\end{equation}
where the first term sums the covariance matrix of the $\mathrm{i^{th}}$ branch spectrum for the parameter $\mathrm{s}$, and the second term sums the cross-term covariance matrices between the spectra of the $\mathrm{i^{th}}$ and $\mathrm{j^{th}}$ branches. Such cross-term covariance matrices encode correlations between branch spectra sourcing from correlations between parameters associated to one type of uncertainty source. The uncertainty sources considered in the present work are listed in Table \ref{tab:uncertainty budget}. They were defined and sometimes grouped such that they are independent, meaning that the total covariance matrix $\mathrm{V_p}$ of a fission fragment spectrum can be expressed as:
\begin{equation}
\label{eq:cov_iso_tot}
\mathrm{V_p = \sum_s V_p^s,}
\end{equation}
where the parameter s runs over the different uncertainty sources. The total uncertainty associated to each bin of a fission fragment spectrum is then given by the squared root of the $\mathrm{V_p}$ diagonal elements.~Applying the Jacobian matrix formalism to eq.~\ref{eq:reactor_spectrum_with_CFY} and eq.~\ref{eq:simplified_reactor_spectrum} for the propagation of uncertainties, the total covariance matrix $\mathrm{V_k}$ associated to an actinide fission spectrum $\mathrm{S_k(E)}$ can be approximated at first order as: 
\begin{equation}\label{eq:covariance_actinide_fission_spectrum}
\mathrm{V_k \simeq 
 \sum_{p, q}   \mathcal{Y}^{k}_{p}  \mathcal{Y}^{k}_{q} \  V_{pq}  +
\sum_{p, q}    V^{\mathcal{Y}^k_{p}   \mathcal{Y}^k_{q}} \ S_p(E) \ S_q(E).}
\end{equation}
where $\mathrm{V_{pq}}$ is the covariance of the binned spectra of the p\nth{} and q\nth{} fission fragments. As detailed later in section~\ref{sec:BetaDecayFormalism}, correlations between two fission fragment spectra $\mathrm{S_p}$ and $\mathrm{S_q}$ originate from common sources of uncertainty in their respective modeling. 
Finally, the term $\mathrm{V^{\mathcal{Y}^k_{p} \mathcal{Y}^k_{q}}}$ represents the covariance of the p\nth{} and q\nth{} fission fragment cumulative yields.
The possible sources of correlation among cumulative fission yields are especially discussed in section~\ref{subsec:fission yield data}. 
In eq.~\ref{eq:reactor_spectrum_with_CFY} and eq.~\ref{eq:simplified_reactor_spectrum}, fission rates $\mathrm{f_k}$ are usually estimated using reactor simulations. Modeling and nuclear data uncertainties can induce correlations between the estimated $\mathrm{f_k}$.
Furthermore, inter-actinide correlations also exist because a fission fragment can contribute to different actinide fission spectra. As such, the total covariance matrix of a reactor spectrum $\mathrm{S_{tot}(E)}$ is computed at first order as:
\begin{eqnarray}
\mathrm{V_{tot}\,}  &&\mathrm{\simeq 
\sum_{k,l}  V^{f_{k} f_{l}} \ S_k(E) \ S_l(E) + 
\sum_{k,l \neq k}  \sum_{p, q } f_{k} f_{l} \mathcal{Y}^{k}_p   \mathcal{Y}^{l}_q  V_{pq}}\nonumber \\
&&+ \mathrm{\sum_{k} f_{k}^{2} V_k,}
\end{eqnarray}
where $\mathrm{V^{f_{k} f_{l}}}$ is the covariance of the k\nth{} and l\nth{} actinide fission rates $\mathrm{f_k}$ and $\mathrm{f_l}$.
As in eq.~\ref{eq:covariance_actinide_fission_spectrum}, the covariance $\mathrm{V_{pq}}$ comes from a common source of uncertainty both present in the $\mathrm{p^{th}}$ and $\mathrm{q^{th}}$ fission fragment spectrum modeling. In the following work, the construction of the uncertainty budget distinguishes two main classes of uncertainty. The first one gathers uncertainty sources from the modeling of the $\mathrm{\beta}$ branches whereas the second one is related to evaluated nuclear data.
%
%
\section{Beta decay formalism}\label{sec:BetaDecayFormalism}
%
\subsection{Beta branch modeling}\label{subsec:beta_branch_modeling}
In the present work, the electron spectrum of a $\mathrm{\beta}$ branch is modeled according to the (V-A) theory of weak interaction using an advanced formalism developed by Behrens and B{\"u}hring \cite{Behrens1982}. The most general form of the $\beta$-decay Hamiltonian is usually expressed as the product of a lepton current encompassing all information related to the electron and neutrino wave functions and of a nuclear current encoding all the nuclear structure information of the parent and daughter nuclei. In this formalism, a multipole expansion of both these currents is performed to compute the transition matrix elements associated to a nuclear $\mathrm{\beta}$ decay, assuming a spherical symmetry of the system.

The transition matrix element is decomposed such that 
the lepton kinematic dependency is clearly separated from  
the pure nuclear structure term. This expansion introduces form factors that can be either of vector (V) or axial-vector (A) type. Their contribution in the transition depends on selection rules related to total angular momentum conservation. 
Each order of this multipole expansion is associated to a change $\emph{l}$~in angular momentum between the initial and final states. Usually, only terms with the lowest powers, i.e.~$\emph{l}$~and ($\emph{l+1}$), are kept in this expansion, the higher order terms being expected to be orders of magnitude smaller. The spin $\mathrm{\Delta J = |J_f-J_i|}$ and parity $\mathrm{\pi_f\pi_i}$ changes between the initial and final nuclear states then define which multipole expansion terms contribute to the $\mathrm{\beta}$-decay probability, leading to the well-known $\mathrm{\beta}$-decay selection rules \cite{behrens1969j}. In particular, transitions leaving out a single nuclear form factor in their multipole expansion are called unique transitions, whereas those having several nuclear form factors are called non-unique transitions.

The electron and neutrino wave functions in the lepton current are calculated taking into account the electromagnetic interaction of the outgoing $\mathrm{\beta}$ particle with the static Coulomb potential of the daughter nucleus. They are also expanded in terms of spherical harmonics. The combination of the previously described multipole decomposition of the transition matrix element and the available phase space then leads to the following usual expression for the electron spectrum:
\begin{eqnarray}\label{eq:branch_beta_spectrum}
\mathrm{S_b(W_e,W_0)\,} &&\mathrm{= K\, p_e W_e (W_{0,b} - W_e)^2 \times F(Z,W_e)}\nonumber \\
&&\mathrm{\times \, C(Z,W_e) \times (1 + \delta_R^{e^-} + \delta_{WM}),}
\end{eqnarray}
where $\mathrm{W_e = E_e/m_ec^2 + 1}$ and $\mathrm{p_e = \sqrt{W_e^2-1}}$ are respectively the electron total energy in units of its rest mass and associated momentum, Z is the atomic number of the daughter nucleus and K a factor normalizing the transition probability to unity. The term $\mathrm{F(Z, W_e)}$ denotes the Fermi function. It encodes the distortion of the electron wave function in the static Coulomb potential of the daughter nucleus. In the Behrens and B{\"u}hring formalism, the Fermi function is defined as:
\begin{equation}
    \mathrm{F(Z,W_e) = \frac{\alpha^2_{-1} + \alpha^2_{+1}}{2p_e^2},}
\end{equation}
where the $\mathrm{\alpha_k}$ quantities are called Coulomb amplitudes and are related to the normalization of the electron radial wave functions (ERWFs), the integer k being related to the angular momenta as defined in \cite{Behrens1982}. The calculation of the Coulomb amplitudes is detailed in section~\ref{subsec:coulomb-amplitudes}. The term $\mathrm{C(Z,W_e)}$ in eq.~\ref{eq:branch_beta_spectrum} is called the shape factor and includes the nuclear form factors originating from the multipole expansion of the nuclear current. At first order, the shape factor $\mathrm{C_L^U(Z,W_e)}$ of a unique-forbidden transition of degree L can be expressed in a rather simple way since the only contributing nuclear form factor can be factorized out into the normalisation constant K:
\begin{equation}\label{eq:unique-shape-factor}
    \mathrm{C_L^U(Z,W_e) = \sum_{k=1}^L \lambda_k \, \frac{p_e^{2(k-1)}p_{\nu}^{2(L-k)}}{(2k-1)!\,(2(L-k)+1)!}.}
\end{equation}
The $\mathrm{\lambda_k}$ parameters are called Coulomb functions and are related to the Coulomb amplitudes $\mathrm{\alpha_k}$ through the following relationship:
\begin{equation}
    \mathrm{\lambda_k = \frac{\alpha^2_{-k} + \alpha^2_{+k}}{\alpha^2_{-1} + \alpha^2_{+1}},}
\end{equation}
while $\mathrm{p_{\nu}}$ stands for the neutrino momentum. The computation of the shape factor for non-unique forbidden transitions is in the other hand more complicated. It usually requires information about the structure of the initial and final nuclear states. The treatment of the non-unique forbidden transitions is therefore discussed separately in section~\ref{subsec:non_unique_forbidden_transitions}. Finally, the terms $\mathrm{\delta^{e^-}_R}$ and $\mathrm{\delta_{WM}}$ respectively correspond to radiative and weak magnetism corrections. They are detailed in section~\ref{subsec:beta-branch-radiative-corrections} and in section~\ref{subsec:beta-branch-WM-corrections}.

It is worth mentioning at this stage that the present $\beta$-decay formalism includes many refinements with respect to past summation calculations of reactor \bnue{} spectra \cite{Mueller2011, Huber2011, Fallot2012}. First, the calculation of unique forbidden transitions, as described in eq.~\ref{eq:unique-shape-factor}, avoids the usual $\mathrm{\lambda_k=1}$ approximation. This approximation originates from the fact that the Coulomb functions $\mathrm{\lambda_k}$ are almost constant at high energy. However, it was shown to be systematically incorrect compared to unique forbidden shape factor calculations including the full energy dependence of the $\mathrm{\lambda_k}$ parameters \cite{Mougeot2015}. Second, the present formalism automatically includes the so-called nucleus finite-size and atomic screening effects through the calculation of the Coulomb amplitudes $\mathrm{\alpha_k}$ (see section~\ref{subsec:coulomb-amplitudes}). As opposed to past calculations, these electromagnetic corrections are here not only included in the calculation of the Fermi function, but also in the calculation of the shape factor. The impact of these two refinements is discussed in section~\ref{subsec:coulomb-amplitudes} for the calculation of unique forbidden shape factors. Last, the present summation calculations operate a new treatment of the non-unique forbidden transitions. As described in section~\ref{subsec:non_unique_forbidden_transitions}, detailed nuclear structure calculations are used to model the 23 most important non-unique branches contributing to the IBD yield detected at a nuclear reactor (see Table~\ref{tab:nu_transitions}), hence avoiding the systematic use of the so-called $\mathrm{\xi}$-approximation to compute their corresponding shape factor.
%
%
%
%
\subsection{Relativistic electron wave functions}\label{subsec:coulomb-amplitudes}
The Coulomb amplitudes $\mathrm{\alpha_k}$ are obtained by solving the Dirac equation for the electron in the static Coulomb potential of the daughter nucleus. The algorithm for the numerical solving of the Dirac equation follows the work of Behrens and B{\"u}hring \cite{Behrens1982}, and uses local power-series expansions of the ERWFs. The algorithm iteratively solves the Dirac equation on a spatial grid, starting from r = 0 up to a reconnection point $\mathrm{r = R_2}$ where the free electron wave functions associated to $\mathrm{r \rightarrow +\infty}$ can be safely recovered. The implementation of this algorithm has been validated against the published tables of Behrens and J{\"a}necke \cite{behrens1969j} up to the last possible decimal for an electrostatic potential generated by a uniformly charged sphere. Any reasonable change in the numerical algorithm parameters such as the choice of the cut-off in the power-series expansion of the ERWFs, the choice of the reconnection point $\mathrm{R_2}$ or the choice of the spatial grid size have been checked to give negligible changes in the value of the Coulomb amplitudes $\mathrm{\alpha_k}$ \cite{perisse:tel-03538198}. Numerical calculation errors coming from the solving of the Dirac equation are then safely neglected in the computation of a $\mathrm{\beta}$-branch spectrum.

In this work, the so-called nucleus finite-size and screening corrections to the calculation of a $\beta$-branch spectrum are taken into account by considering a nuclear potential V(r) generated by Z charges uniformly distributed within a spherical nucleus of radius R and screened by a cloud of ($\text{Z}-1$) atomic electrons. The potential V(r) takes different forms in the regions respectively delimited by the nuclear radius R and a point $\mathrm{R_1}$ beyond which the screened potential is smoothly transited up to an asymptotic point-like form until reaching the reconnection point $\mathrm{R_2}$. The complete expression of the screened potential used in this work can be found in \cite{Mougeot2014}. In the region $\mathrm{R \leq r \leq R_1}$, the screened potential is expressed as:
\begin{equation}
    \mathrm{V(r) = - \frac{\alpha Z}{r}\, \sum_{i=1}^N a_i e^{-\beta_i r},}
\end{equation}
where the parameters $\mathrm{a_i}$ and $\mathrm{\beta_i}$ are tabulated from fits of atomic screening functions calculated using the relativistic Dirac-Hartree-Fock-Slater formalism for atoms with Z=1-92 \cite{Salvat1987}. These screening functions improve over past evaluations, which often relied on the non-relativistic Thomas-Fermi statistical model of the atom. Although no uncertainties are reported in the literature, they have been shown to provide a very accurate modeling of screened potentials as compared to e.g. electron-atom scattering experimental data \cite{Salvat1987a}.
\begin{figure*}[!t]
    \centering
    \includegraphics[width=0.47\textwidth]{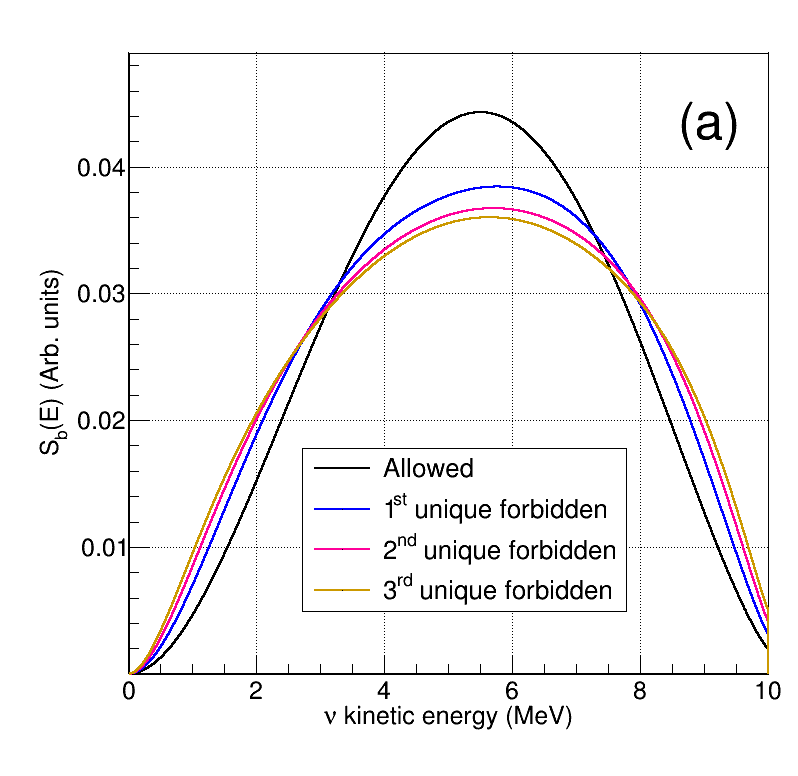}
    \includegraphics[width=0.47\textwidth]{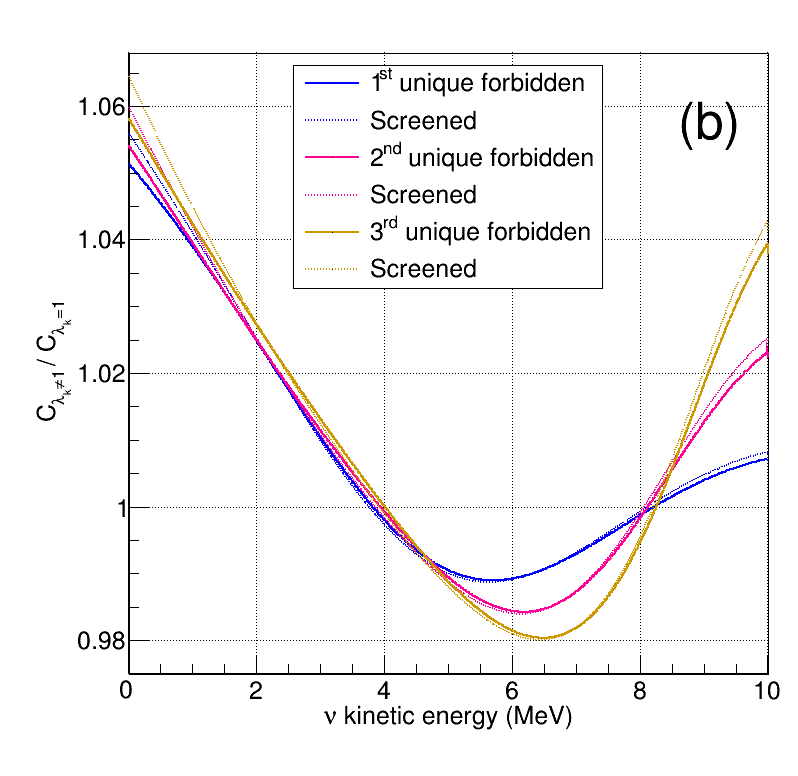}
    \caption{Comparison of allowed and forbidden unique transitions in the Behrens and Bühring $\mathrm{\beta}$ decay formalism. (a) \bnue{} spectra computed with accurate $\lambda_{k}$ Coulomb functions for a set of fictitious transitions of different forbiddenness degree with $\mathrm{Z} = 46$, $\mathrm{A} = 117$, and $\mathrm{E_0}$ = 10 MeV. (b) Impact of accurate $\lambda_{k}$ Coulomb functions compared to the $\lambda_{k}=1$ approximation on the calculation of shape factors for this same set of transitions. Solid lines include the nuclear finite-size effect into the computation of Coulomb functions, while dashed lines additionally include the atomic screening effect (see section~\ref{subsec:coulomb-amplitudes} for more details).
    }
    \label{fig:lambda_k_approximation_in_shape_factor}
\end{figure*}
As such, no uncertainty in the screened nucleus potential for the calculation of the Coulomb amplitudes $\mathrm{\alpha_k}$ is considered in this work. The nuclear radius R, also entering the definition of the nuclear Coulomb potential V(r), is either taken from a set of experimental root mean square charge radii evaluated in \cite{Angeli2013} or estimated using the prescription from \cite{Angeli1999} for nuclei off the stability line. This prescription slightly improves over the Elton formula commonly used in previous summation calculations of reactor \bnue{} spectra \cite{Fallot2012,Mueller2011} as it better reproduces experimental data for isotopes off the stability line. This new modeling of the nuclear radius has been found to give a negligible difference with respect to summation calculations using the Elton formula \cite{perisse:tel-03538198}. As such, the contribution of the evaluated nuclear radius uncertainties reported in \cite{Angeli2013}, which are smaller than this difference, are safely neglected.

For illustration purposes, a set of transitions with different forbidenness degrees is compared in Figure~\ref{fig:lambda_k_approximation_in_shape_factor} (a). Changes to unique forbidden shape factors when considering an appropriate calculation of the $\mathrm{\lambda_k}$ functions both including nuclear finite-size and atomic screening effects are depicted in Figure~\ref{fig:lambda_k_approximation_in_shape_factor} (b).
Compared to past calculations defaultly applying the $\mathrm{\lambda_k}$ = 1 approximation, the shape factor of a unique forbidden transition with a 10 MeV endpoint energy is typically corrected by a $\cal{O}$(5\%) factor, with even larger corrections for smaller endpoint energy transitions. For instance, unique transitions having an endpoint energy close to the IBD threshold can be corrected up to 60\% \cite{perisse:tel-03538198}.
Although these corrections are significant at the $\mathrm{\beta}$-branch level, they are expected to bring small changes to the computation of an actinide fission spectrum since unique forbidden transitions typically contribute to $\mathrm{\sim}$10\% of both the corresponding \bnue{} and IBD yield (see Table~\ref{tab:transition_contribution_breakdown} and Figure~\ref{fig:contribution_per_transition_type}).
%
\subsection{Treatment of non-unique forbidden transitions}\label{subsec:non_unique_forbidden_transitions}
\begin{figure*}[!t]
    \centering
    \includegraphics[width=0.47\textwidth]{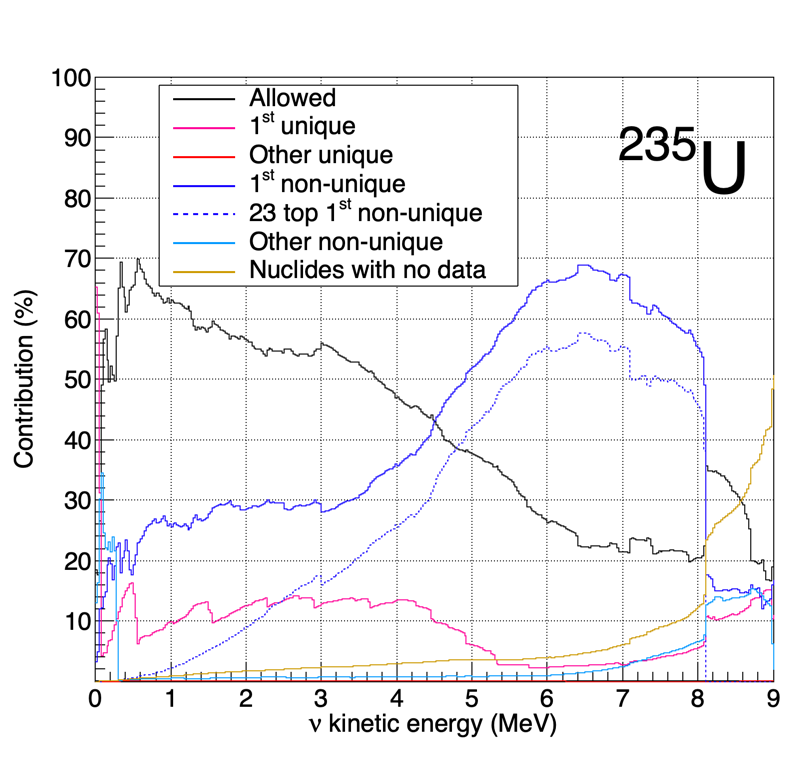}
    \includegraphics[width=0.47\textwidth]{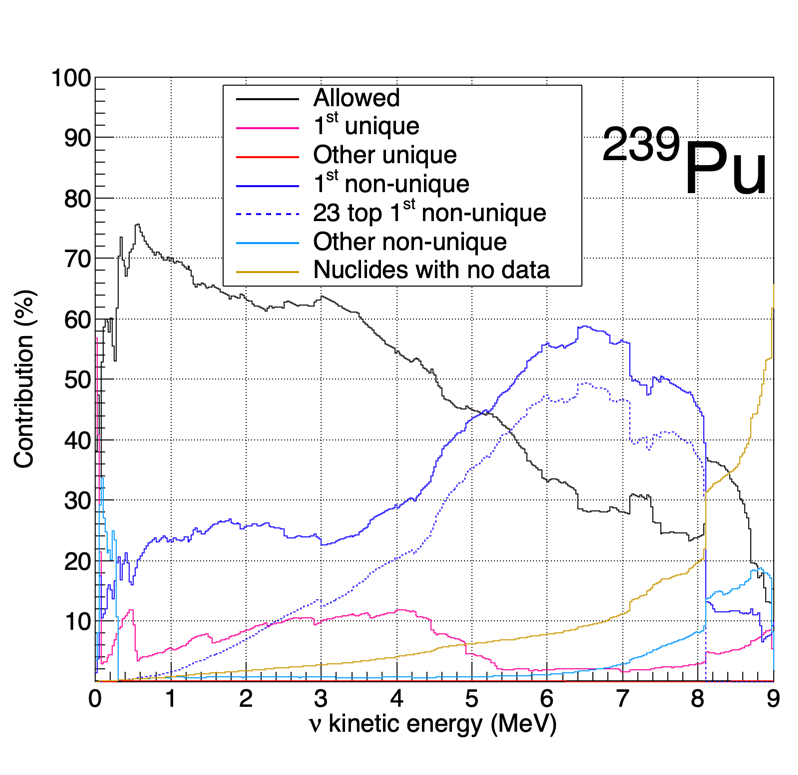}
    \caption{Contributions of the different types of $\mathrm{\beta^{-}}$ transition to the \U{235} and \Pu{239} fission \bnue{} spectra. Computations are done using cumulative fission yields from \jeff{}~\cite{JEFF33} and using nuclear decay data as described in section~\ref{subsec:decay data}. The solid line contributions add up to 100\%. The contribution from the 23 1\st~non-unique forbidden transitions modeled with nuclear structure calculations (dashed blue line) belongs to the 1\st~\ non-unique forbidden transition contribution (solid blue line). Contributions of other unique forbidden transitions (solid red lines) lie below 0.1\% and are not visible.}
    \label{fig:contribution_per_transition_type}
\end{figure*}

Non-unique forbidden transitions play an important role in the calculation of reactor \bnue{} spectra. For instance, Table~\ref{tab:transition_contribution_breakdown} details their contribution to the \U{235}, \U{238}, \Pu{239} and \Pu{241} \bnue{} fluxes and associated IBD yields, showing that they make a significant 30 to 40\% contribution to the latter. As shown by Figure~\ref{fig:contribution_per_transition_type}, these transitions also dominate the 4-8 MeV portion of a typical actinide fission spectrum. A precise calculation of their contribution is then relevant in light of the recently measured spectral distortions in that same energy range. As mentioned earlier in section~\ref{subsec:beta_branch_modeling}, non-unique forbidden transitions leave out several form factors in their associated $\mathrm{\beta}$ spectrum. Their modeling requires the use of a mathematical formalism based on nuclear structure calculations, which are complicated and numerically time-consuming to perform.
\begin{table*}[htp]
\centering
\begin{tabular}{lllll}
    \hline
    \hline
         & \U{235}  & \U{238}   & \Pu{239}  & \Pu{241}  \\
    \hline
        \multicolumn{1}{c}{\textbf{\bnue{} flux}  {\small [\%]} } & & & & \\ 
        Allowed                      &  57.9  &  62.9  &  64.8  &  66.9  \\
        1\st \ non-unique forbidden  &  26.1 (6.7)  &  22.8 (5.9)  &  22.3 (4.6)  &  21.1 (4.7)  \\
        1\st \ unique forbidden      &  11.4  &  7.8   &  8.0   &  6.4   \\
        Other non-unique forbidden   &  3.4   &  2.8   &  3.7   &  3.1   \\
        Other unique forbidden       &  0.1   &  0.1   &  0.1   &  0.1   \\
        Nuclides with no data        &  1.1   &  3.6   &  1.1   &  2.4   \\
        
    \hline
        \multicolumn{1}{c}{\textbf{IBD yield}  {\small [\%]} }  & & & & \\
        Allowed                      &  43.9  &  50.1  &  52.8  &  55.4  \\
        1\st \ non-unique forbidden  &  41.9 (30.6)  &  33.4 (20.9) &  33.2 (23.4)  &  30.0 (19.2)  \\
        1\st \ unique forbidden      &  10.0  &  5.7   &  8.4   &  5.4   \\
        Other non-unique forbidden   &  0.9   &  1.0   &  0.9   &  1.1   \\
        Other unique forbidden       &  0.1   &  0.1   &  0.1   &  0.1   \\
        Nuclides with no data        &  3.2   &  9.7   &  4.6   &  8.0   \\
        
    \hline
    \hline
\end{tabular}
\caption{Contributions of the different types of $\mathrm{\beta}$ decay transitions to the summation calculation of the \bnue{} fluxes (top) and IBD yields (bottom) from the fission of the \U{235}, \U{238}, \Pu{239} and \Pu{241} actinides. Computations are done using cumulative fission yields from \jeff{}~\cite{JEFF33} (see section~\ref{subsec:fission yield data}), and with nuclear structure and decay data as described in section~\ref{subsec:decay data}. The 1\st~non-unique forbidden line also displays in parenthesis the contribution of the 1\st~non-unique forbidden transitions computed with nuclear structure calculations (see section~\ref{subsubsec:main_forbidden_transitions} and Table~\ref{tab:nu_transitions}). The contribution of the nuclides having no decay information (see section~\ref{subsec: nuclei with no data}) is also displayed here for completeness, such that all contributions add up to 100\%.}
\label{tab:transition_contribution_breakdown}
\end{table*}
This section describes how the non-unique forbidden transitions are treated in the present work. Because nuclear structure calculations are computationally heavy, they were used to model the 23 most important non-unique branches participating to the typical IBD yield expected at a nuclear reactor (see Table~\ref{tab:nu_transitions}). This strategy ensured to compute a significant $\sim$70\% fraction of the total non-unique forbidden transition contribution to the IBD yield, hence providing a good compromise with respect to the overall accuracy of a reactor \bnue{} flux prediction. For perspective, the computation of about 70 (resp.~400) additional branches would be necessary to describe 90\% (resp.~99\%) of this contribution. In section ~\ref{subsubsec:xi_approximation}, the nuclear structure calculations of these 23 transitions are then used to construct an uncertainty model associated to the remaining non-unique transitions, which in the present work still miss realistic nuclear structure calculations and are computed using the $\mathrm{\xi}$-approximation.
\begin{table*}[t]
\centering
\begin{tabular}{>{\centering\arraybackslash}m{1cm}| 
>{\centering\arraybackslash}m{0.9cm} 
>{\centering\arraybackslash}m{0.9cm} 
>{\centering\arraybackslash}m{0.9cm} 
>{\centering\arraybackslash}m{1cm} 
>{\centering\arraybackslash}m{2cm} 
>{\centering\arraybackslash}m{1.4cm} 
>{\centering\arraybackslash}m{1.4cm} 
>{\centering\arraybackslash}m{1.4cm} 
}

\hline 
\hline
Nuclide & $\mathrm{Q_{\beta}}$ & $\mathrm{E_{IS}}$ & $\mathrm{E_{lvl}}$ & BR & $\mathrm{J_i^{\pi}\rightarrow J_f^{\pi}}$ & $\mathrm{\phi_{\bar{\nu}_e}}$ & $\mathrm{\phi_{IBD}}$ & $\mathrm{\Delta \phi_{IBD}}$ \\
 & \small [MeV] & \small [MeV] & \small [MeV] & \small [\%] & & \small [$\mathrm{10^{-1}}$ \%] & \small [\%] & \small [\%] \\
\hline \vspace{1mm}
$\mathrm{^{92}Rb^{\dagger}}$   & 8.095 & 0 & 0 & 87.50 & $\mathrm{0^-\rightarrow 0^+}$          & 4.98 & 5.44 & -0.08 \\
$\mathrm{^{96}Y^{\dagger}}$    & 7.109 & 0 & 0 & 96.60 & $\mathrm{0^-\rightarrow 0^+}$          & 6.62 & 5.29 &  1.80 \\
$\mathrm{^{142}Cs^{\dagger}}$  & 7.328 & 0 & 0 & 44.00 & $\mathrm{0^-\rightarrow 0^+}$          & 2.09 & 1.84 & -1.27 \\
$\mathrm{^{140}Cs^{\dagger}}$  & 6.219 & 0 & 0 & 35.50 & $\mathrm{1^-\rightarrow 0^+}$          & 3.13 & 1.65 & -9.31 \\
$\mathrm{^{137}I^{\dagger}}$   & 6.027 & 0 & 0 & 50.79 & $\mathrm{{7/2}^+\rightarrow {7/2}^-}$  & 2.87 & 1.38 & -8.79 \\
$\mathrm{^{139}Cs^{\dagger}}$  & 4.213 & 0 & 0 & 85.00 & $\mathrm{{7/2}^+\rightarrow {7/2}^-}$  & 8.33 & 1.28 & -14.70 \\
$\mathrm{^{95}Sr^{\dagger}}$   & 6.089 & 0 & 0 & 40.30 & $\mathrm{{1/2}^+\rightarrow {1/2}^-}$  & 3.04 & 1.28 & -23.04 \\
$\mathrm{^{135}Te}$            & 6.050 & 0 & 0 & 62.00 & $\mathrm{{7/2}^-\rightarrow {7/2}^+}$  & 3.31 & 1.17 & -32.41 \\
$\mathrm{^{90}Rb^{\dagger}}$   & 6.584 & 0 & 0 & 32.80 & $\mathrm{0^-\rightarrow 0^+}$          & 1.73 & 1.12 &  1.34 \\
$\mathrm{^{97}Y}$              & 6.821 & 0 & 0 & 40.00 & $\mathrm{1/2^-\rightarrow 1/2^+}$      & 1.32 & 1.09 &  14.53 \\
$\mathrm{^{93}Rb^{\dagger}}$   & 7.466 & 0 & 0 & 35.50 & $\mathrm{{5/2}^-\rightarrow {5/2}^+}$  & 1.64 & 1.02 & -43.69 \\
$\mathrm{^{136}I^m}$           & 6.883 & 0.201 & 1.891 & 71.00 & $\mathrm{6^-\rightarrow 6^+}$  & 2.25 & 0.61 & -10.76 \\
$\mathrm{^{94}Y^{\dagger}}$    & 4.918 & 0 & 0.919 & 34.18 & $\mathrm{2^-\rightarrow 2^+}$      & 3.13 & 0.56 &  22.08 \\
$\mathrm{^{91}Rb^{\dagger}}$   & 5.907 & 0 & 0.094 & 12.11 & $\mathrm{3/2^-\rightarrow 3/2^+}$  & 0.63 & 0.41 &  27.52 \\
$\mathrm{^{144}Pr}$            & 2.997 & 0 & 0 & 97.90 & $\mathrm{0^-\rightarrow 0^+}$          & 7.84 & 0.41 &  11.20 \\
$\mathrm{^{140}Cs^{\dagger}}$  & 6.219 & 0 & 0.602 & 10.55 & $\mathrm{1^-\rightarrow 2^+}$      & 0.93 & 0.38 & -5.71 \\
$\mathrm{^{91}Kr}$             & 6.771 & 0 & 0.108 & 18.00 & $\mathrm{5/2^+\rightarrow 5/2^-}$  & 0.72 & 0.36 & -33.09 \\
$\mathrm{^{141}Cs^{\dagger}}$  & 5.255 & 0 & 0.049 & 11.87 & $\mathrm{7/2^+\rightarrow 5/2^-}$  & 0.88 & 0.33 &  6.53 \\
$\mathrm{^{135}Te}$            & 6.050 & 0 & 0.604 & 19.00 & $\mathrm{7/2^-\rightarrow 5/2^+}$  & 1.02 & 0.31 & -14.69 \\
$\mathrm{^{91}Kr}$             & 6.771 & 0 & 0 & 9.00 & $\mathrm{5/2^+\rightarrow 3/2^-}$       & 0.36 & 0.31 &  20.27 \\
$\mathrm{^{91}Rb^{\dagger}}$   & 5.907 & 0 & 0 & 9.21 & $\mathrm{3/2^-\rightarrow 5/2^+}$       & 0.82 & 0.30 & -24.75 \\
$\mathrm{^{141}Cs^{\dagger}}$  & 5.255 & 0 & 0.055 & 11.87 & $\mathrm{7/2^+\rightarrow 7/2^-}$  & 0.88 & 0.29 & -4.65 \\
$\mathrm{^{90}Kr^{\dagger}}$   & 4.406 & 0 & 0 & 6.90 & $\mathrm{0^+\rightarrow 0^-}$           & 0.39 & 0.08 &  1.55 \\
\hline
\hline
\end{tabular}
\caption{Non-unique forbidden transitions computed with realistic nuclear structure calculations (see section~\ref{subsubsec:main_forbidden_transitions}) and listed by decreasing order of importance according to their relative contribution to the total IBD yield expected at a commercial nuclear reactor. The following fission fractions were assumed: 0.559 for \U{235}, 0.088 for \U{238}, 0.291 for \Pu{239} and 0.062 for \Pu{241}. Decay data are extracted from the ENSDF database \cite{ENSDF} and corrected with up-to-date TAGS data when indicated by ${}^{\dagger}$ (see section~\ref{subsubsec:TAGS_data}). The last three columns respectively indicate the contribution of these transitions to the total \bnue{} flux, the contribution of these transitions to the total IBD yield, and the difference between the transition IBD yield computed with nuclear structure calculations and under the $\mathrm{\xi}$-approximation, expressed relatively to the former. These transitions add up to 69.1\% of the total non-unique forbidden transition contribution. They total 26.9\% of the expected IBD yield.}
\label{tab:nu_transitions}
\end{table*}
\subsubsection{Calculations with nuclear structure} \label{subsubsec:main_forbidden_transitions}
In the Behrens and B{\"u}hring formalism \cite{Behrens1982}, the shape factor $\mathrm{C(Z,W_e)}$ in eq.~\ref{eq:branch_beta_spectrum} results from the previously mentioned multipole expansion of both the nuclear and lepton currents, assuming the so-called impulse approximation in which all nucleons are independent particles in a mean-field potential. The convolution of the nuclear structure and the lepton dynamics is embedded in the ad hoc terms $\mathrm{M_K(k_e,k_{\nu})}$ and $\mathrm{m_K(k_e,k_{\nu})}$ and is decomposed as a sum of the different multipole orders, giving the following general expression for the shape factor:
\begin{eqnarray}\label{eq:general_shape}
	\mathrm{C(Z,W_e)\,\,} &&\mathrm{= \sum\limits_{K,k_e,k_{\nu}} \lambda_{k_e} \biggl[ M_K^2(k_e,k_{\nu}) + m_K^2(k_e,k_{\nu})}\nonumber \\
 &&\mathrm{- \dfrac{2\mu_{k_e}\gamma_{k_e}}{k_e W_e} \, M_K(k_e,k_{\nu}) \, m_K(k_e,k_{\nu}) \biggr],}
\end{eqnarray}
where any quantities labeled by the lepton quantum numbers $\mathrm{k_e}$ and $\mathrm{k_{\nu}}$ depend on the electron and neutrino relativistic wave functions, respectively. The main multipole order K comes from the expansion of the nuclear current and is limited by the change in total angular momentum $\mathrm{\Delta J}$ between the initial and final nuclear states. It ranges from $\mathrm{K_{min} = \Delta J}$ to $\mathrm{K_{max} = J_i + J_f}$. In the present work, the prescription from \cite{Behrens1971} has been followed, considering only the dominant terms with respectively $\mathrm{K = K_{min}}$, $\mathrm{K_{min}+1}$ and $\mathrm{k_e+k_{\nu} = K+1}$, K+2 in eq.~\ref{eq:general_shape}. The next order terms for a set of high-energy transitions of interest were computed so as to check the validity of this prescription. Their contribution was found to be several orders of magnitude smaller, thus confirming they could be neglected.

In addition to the multipole expansion, the lepton wave functions are also expanded in the Behrens and B{\"u}hring formalism in powers of $\mathrm{(m_e R)}$, $\mathrm{(W_e R)}$ and $\mathrm{(\alpha Z)}$, with R the nuclear radius and $\mathrm{\alpha}$ the fine structure constant. This procedure avoids any overlap calculations between the nuclear and lepton wave functions. The nuclear matrix elements, also called form factor coefficients, become then independent of the lepton momenta, making the computation of the shape factor extremely fast, about a few milliseconds on a modern computer. Using this approach, the dominant term $\mathrm{M_K(k_e,k_{\nu})}$ in a non-unique forbidden transition reads:
\begin{widetext}
\begin{equation}\label{eq:MK_fnu}
\begin{array}{rl}
\mathrm{M_K(k_e,k_{\nu}) =} & \mathrm{\sqrt{\dfrac{1}{2}}\sqrt{\dfrac{(2K)!!}{(2K+1)!!}}\sqrt{\dfrac{1}{(2k_e-1)!(2k_{\nu}-1)!}} (p_e R)^{k_e-1} (p_{\nu} R)^{k_{\nu}-1}} \\
& \mathrm{\times \left\{ - \sqrt{\dfrac{2K+1}{K}} \, ^{\mathrm{V}}{F}{^{(0)}_{K,K-1,1}} - \dfrac{\alpha Z}{2k_e+1} \,^{\mathrm{V}}{F}{^{(0)}_{K,K,0}}(k_e,1,1,1) \right.} \\
& \mathrm{- \left( \dfrac{W_e R}{2k_e+1} + \dfrac{p_{\nu} R}{2k_{\nu}+1} \right) \, ^{\mathrm{V}}{F}{^{(0)}_{K,K,0}} - \dfrac{\alpha Z}{2k_e+1} \sqrt{\dfrac{K+1}{K}} \, ^{\mathrm{A}}{F}{^{(0)}_{K,K,1}}(k_e,1,1,1)} \\
& \mathrm{\left. - \left( \dfrac{W_e R}{2k_e+1} - \dfrac{p_{\nu} R}{2k_{\nu}+1} \right) \sqrt{\dfrac{K+1}{K}} \, ^{\mathrm{A}}{F}{^{(0)}_{K,K,1}} \right\}.}
\end{array}
\end{equation}
\end{widetext}
The form factor coefficients $\mathrm{^{V/A}{F}{^{(N)}_{K,L,s}}(k_e,m,n,\rho)}$ are either of vector or axial-vector type, as indicated by the left upper script. The L and s orders come with K in the development of the nuclear current and their contributions are summed in $\mathrm{M_K(k_e,k_{\nu})}$. The other dependencies $\mathrm{(N)}$ and $\mathrm{(k_e,m,n,\rho)}$ label the radial expansion of the lepton wave functions. When $\mathrm{\rho}=0$, the dependency in $\mathrm{(k_e,m,n,\rho)}$ disappears and is omitted in the notation, as shown in eq. \ref{eq:MK_fnu}. More details can be found in \cite{Behrens1971}. Such an expansion of the lepton wave functions is only possible for a simple Coulomb potential, e.g.~originating from a nucleus modeled as a uniformly charged sphere. In the present work, a different strategy is followed. The formulation of the $\mathrm{M_K(k_e,k_{\nu})}$ and $\mathrm{m_K(k_e,k_{\nu})}$ quantities is revisited in order to directly use the numerical ERWFs described in section~\ref{subsec:coulomb-amplitudes}. This approach ensures more accurate calculations of the shape factor because (i) both the finite-size nucleus and atomic screening effects are inherently taken into account in the calculation of the lepton wave functions and (ii) the lepton wave functions are not expanded anymore. The multipole expansion therefore only remains, the precision of which has been checked to be under control. The cost of such a full numerical treatment for the lepton current is the computational burden, each branch spectrum requiring several tens of minutes to be calculated. 

The form factor coefficients $\mathrm{^{V/A}{F}{^{(N)}_{K,L,s}}}$ can be calculated in a very simple way by reducing the $\beta$ transition to a single nucleon-nucleon transition. However, it is clear that such a description of the nuclear structure is neither realistic nor accurate. In the configuration mixing approach, the many-particle wave function of a nuclear state is described as a linear combination of the single-particle wave functions. The $\mathrm{\beta}$ decay transition amplitude is obtained by evaluating the corresponding one-body spherical tensor operator $\mathrm{T_{\lambda}}$ bracketed between the initial and final nuclear states, with $\mathrm{\lambda}$ the tensor rank. This total amplitude can be expressed as a weighted sum of all the single-particle transition amplitudes that play a role in the $\beta$ transition \cite{Suhonen2007}:
\begin{eqnarray}\label{eq:transition_amplitude}
	\mathrm{\langle \xi_f J_f | T_{\lambda} | \xi_i J_i \rangle \,} &&\mathrm{=  \dfrac{1}{\sqrt{2\lambda+1}}}\nonumber \\
    &&\mathrm{\times \sum_{a,b} \langle a | T_{\lambda} | b \rangle \, \langle \xi_f J_f | \left[c_{a}^{\dag} \tilde{c}_{b} \right]_{\lambda} | \xi_i J_i \rangle}.
\end{eqnarray}
The single-particle matrix element $\mathrm{\langle a | T_{\lambda} | b \rangle}$ describes a nucleon-nucleon transition, weighted by its one-body transition density (OBTD) $\mathrm{\langle \xi_f J_f | \left[c_{a}^{\dag} \tilde{c}_{b} \right]_{\lambda} |\xi_i J_i \rangle}$. Eventually, each nucleon-nucleon transition gives a specific form factor coefficient and all the contributions have to be summed up to determine $\mathrm{\mathrm{M_K(k_e,k_{\nu})}}$, as e.g. in eq.~\ref{eq:MK_fnu}.
In this work, the OBTD for each nucleon-nucleon transition is computed using the shell model numerical code NuShellX@MSU \cite{Brown2014}. This code can use different interaction Hamiltonians, each fitted in different mass regions to reproduce some experimental data. Nuclear structure wise, nucleons are distributed among a set of low-lying energy levels defining an inert core and a set of high-lying energy levels defining a valence space. Only nucleons present in the valence space can experience a transition in such a model. The proper definition of the inert core, the valence space and the interaction Hamiltonian depend on the nucleus mass number A. For nuclei with A \textless 100, the \emph{glepn} valence space above the doubly-magic $\mathrm{^{56}Ni}$ core as well as the recommended interaction with identical name is used \cite{Mach1990}. In this scheme, the valence nucleons fill the shells starting from $\mathrm{2p_{3/2}}$ up to $\mathrm{2d_{3/2}}$. The valence space is constrained such that the computation of the very large number of configurations remains tractable. It was thus restricted around the magic number 50: proton shells were free up to $\mathrm{1g_{9/2}}$ and higher shells were forced to be empty. Neutron shells were forced to be full up to $\mathrm{1g_{9/2}}$ and higher shells were free. For nuclei with A \textgreater 100, the \emph{jj56pn} valence space above the doubly-magic $\mathrm{^{132}Sn}$ core along with the recommended interaction \emph{khhe} adapted from \cite{Warburton1991} were selected. For nuclei with $A \leq 140$, no restriction was necessary on the valence space. For nuclei with A=141 and A=142, the valence space was constrained as follows, with (min,max) number of particles in a given shell: for protons, (2,4) in $\mathrm{1g_{7/2}}$ and (0,2) in the other shells; for neutrons, (1,4) in $\mathrm{1h_{9/2}}$, (0,4) in $\mathrm{1i_{13/2}}$, and (0,2) in the other shells. For nuclei with A=144, the proton shell were free except $\mathrm{1h_{11/2}}$ that was forced to be empty, and a maximum of three neutrons was allowed in all the orbitals except $\mathrm{1i_{13/2}}$ that was also forced to be empty. Once the nuclear levels determined, the OBTD have been calculated for the dominant K values in each $\mathrm{\beta}$ transition.

The single-particle matrix elements entering eq.~\ref{eq:transition_amplitude} are computed using nucleon wave functions extracted from harmonic oscillator potentials. For a realistic estimate of their corresponding frequencies, the method depicted in \cite{Towner1977} using the proton configurations provided by NuShellX was followed. When available, experimental root mean square charge radii have been taken from \cite{Angeli2013}. Otherwise, the fitted radius formula off the stability line from \cite{Angeli1999} was used. At this stage, it should be reminded that the present $\beta$-decay formalism is totally relativistic. The lepton and nucleon wave functions must thus exhibit large and small components resulting in non-relativistic transition matrix elements combining only large components, and in relativistic transition matrix elements involving small components. The expansion of $\mathrm{M_K(k_e,k_{\nu})}$ depicted in eq.~\ref{eq:MK_fnu} shows that the first form factor coefficient to arise is the relativistic vector matrix element $^{\mathrm{V}}{F}{_{K,K-1,1}}$, for which an accurate value is therefore of importance. Most of the nuclear structure models, including NuShellX, are however non-relativistic. To circumvent this issue, a simple solution consists in assuming that the large component of the nucleon wave function corresponds to the non-relativistic harmonic oscillator. An estimate of the small component is then deduced from the large one using the non-relativistic limit of the Dirac equation. Nevertheless, the inaccuracy of such an approach has been pointed out for decades (see e.g.~\cite{Sadler1993}). The best current approach consists in using the conserved vector current hypothesis (CVC), a property emerging from the gauge invariance of the weak interaction. The relativistic form factor coefficient $^{\mathrm{V}}{F}{_{K,K-1,1}}$ can then be related to the non-relativistic form factor coefficient $^{\mathrm{V}}{F}{_{K,K,0}}$ by~\cite{Behrens1982}:
\begin{eqnarray}
	\mathrm{^{\mathrm{V}}{F}{_{K,K-1,1}}} &&\mathrm{\simeq -\dfrac{R}{\sqrt{K(2K+1)}}\, ^{\mathrm{V}}{F}{_{K,K,0}}}\nonumber \\
    &&\mathrm{\times \biggl [W_0 - (m_n - m_p) + \Delta E_C \biggr ],}
\end{eqnarray}
where $\mathrm{m_n}$ and $\mathrm{m_p}$ are respectively the neutron and proton rest masses, $\mathrm{W_0}$ is the transition endpoint energy and $\mathrm{\Delta E_C}$ is the Coulomb displacement energy. The Coulomb displacement energy can be estimated by modeling the nucleus as a uniformly charged sphere:
\begin{equation}\label{eq:CoulEn}
	\mathrm{\Delta E_C = \dfrac{6}{5}\dfrac{\alpha Z}{R}.}
\end{equation}
This simple modeling is only an approximation and $\mathrm{\Delta E_C}$ was demonstrated to be sensitive to the mismatch between the initial and final nucleon wave functions~\cite{Damgaard1966}. The present calculations rather use the prescription from Behrens and B{\"u}hring, which assumes that the single-particle potential difference is determined by the average of the Coulomb potential \cite{Behrens1982}. The average of the Coulomb potential involves here the full numerical treatment of the lepton current and the nucleon wave functions as described previously. This Coulomb energy evaluation, which depends on the electron kinetic energy and is specific to each nucleon-nucleon transition, is expected to be more accurate than the usual formula given in eq.~\ref{eq:CoulEn}.

Finally, the axial-vector coupling constant $\mathrm{g_A}$, which appears in the definition of axial-vector form factor coefficients, may need to be adjusted to reproduce some experimental observables. Nucleon-nucleon transitions occur in nuclear matter, and a quenched value of $\mathrm{g_A}$ can partially correct for a mismodeling of the nuclear structure such as for instance the approximate treatment of the many-nucleon correlations. Varying $\mathrm{g_A}$ has been shown to modify the spectrum shape of some forbidden non-unique transitions, mostly for low-energy transitions \cite{Haa2016,Kostensalo2017,Kos17_2}. Without experimental spectrum to compare with, an effective $g_A$ value can be estimated from the quenching factor in infinite nuclear matter \cite{Suhonen2017}. However, the accuracy of this value is not guaranteed. Because the forbidden non-unique transitions of interest reported in Table~\ref{tab:nu_transitions} have large endpoint energies, a free-nucleon value of $\mathrm{g_A}$ = 1.2763 (15) resulting from the mean of two recent precise measurements \cite{Liu2010,Mund2013} has been chosen~\footnote{The latest recommended value of $\mathrm{g_A}$ = 1.2754 (13) recently released in \cite{pdg2022} is slightly different from the one adopted here. A few transitions have been recalculated with this value and no significant change has been noticed.}.

Calculation of the non-unique forbidden transitions as listed in Table~\ref{tab:nu_transitions} and using the previously described nuclear structure formalism is computationally heavy. The uncertainty associated to these transitions is therefore modeled in a rather simplistic but conservative way, by using the spectrum difference between the accurate modeling version and an allowed version of the transition obtained under the $\mathrm{\xi}$-approximation (see section~\ref{subsubsec:xi_approximation}). For each energy bin, the accurate non-unique forbidden and the $\xi$-approximated spectra are used to define the limits of a uniform distribution. The standard deviation of this distribution is then used to construct an associated covariance matrix. This covariance matrix is also constructed such that the total rate of the branch spectrum is conserved. Energy bins showing equal (resp. opposite) sign in the difference between the accurate modeling and the $\xi$-approximated version of the spectrum are treated as fully correlated (resp.~anti-correlated). With this uncertainty modeling, the IBD yield associated to a non-unique forbidden transition calculated with nuclear structure typically shows a $\mathrm{\mathcal{O}(10\%)}$ uncertainty. 
The uncertainties associated to these 23 non-unique forbidden transitions computed with nuclear structure induce an uncertainty of $\mathrm{\sim}$0.2\% on the isotopic IBD yields when they are assumed to be uncorrelated (see Table~\ref{tab:uncertainty budget}). The contribution of this source of uncertainty at the actinide fission spectrum level is also depicted on Figure~\ref{fig:fractional_uncertainty_modeling} for \U{235}.
The impact of fully correlating these 23 non-unique transitions has been checked, and resulted in a $\mathrm{\sim}$0.4\% uncertainty on the isotopic IBD yields. Although theoretical correlations may exist, no evidence allows to suggest that these branches are fully correlated. As such, the present work keeps these branches uncorrelated in the propagation of their corresponding uncertainties. The computation of the remaining non-unique forbidden transitions is done using the $\mathrm{\xi}$-approximation and is discussed in the next section.
\subsubsection{$\mathrm{\xi}$-approximation}\label{subsubsec:xi_approximation}
The accurate treatment of the 23 most important non-unique forbidden transitions as described in the previous section cannot be applied on a case-by-case basis to all the remaining hundreds of non-unique transitions present in a reactor \bnue{} spectrum. These remaining non-unique forbidden transitions were instead estimated using the $\mathrm{\xi}$-approximation.~The $\mathrm{\xi}$-approximation consists in treating a non-unique forbidden transition as its equivalent in term of total angular momentum variation but disregarding any parity change. This means that the shape factor of a $\mathrm{L^{th}}$ non-unique forbidden transition is approximated as the shape factor of a $\mathrm{(L-1)^{th}}$ unique forbidden transition. Because the vast majority of non-unique forbidden transitions present in a reactor \bnue{} spectrum are of the first kind, those were approximated as allowed transitions.\\
The origin and the validity of the $\mathrm{\xi}$-approximation remain poorly documented in the literature. They are discussed in the following. Going back to eq.~\ref{eq:general_shape} and keeping only the dominant terms in the multipole expansion, the shape factor of a first non-unique forbidden transition can be expressed as \cite{Behrens1982}:
\begin{equation}\label{eq:1stfnu_shape_factor}
    \mathrm{C(Z,W_e) = k_n (1 + aW_e + \mu_1 \gamma_1 b/W_e + cW_e^2),}
\end{equation}
where the parameters $\mathrm{k_n}$, a, b and c are linear combinations of form factor coefficients which are independent of the electron and neutrino momenta. These parameters are defined in Table~\ref{table:param_1stfnu}.
Historically, the notation $\xi = \alpha Z / 2R$ was first introduced in previous $\mathrm{\beta}$-decay formalisms \cite{Konopinski1941,Mahmoud1952,Kotani1959}. The $\mathrm{\xi}$-approximation assumes that the Coulomb energy of the $\mathrm{\beta}$ particle at the nuclear surface is large compared to the total decay energy, i.e. $\mathrm{2\xi \gg W_0}$. In this case, terms proportional to $\mathrm{(W_0 R)}$ and $\mathrm{(W_e R)}$ in the shape factor parameters of eq.~\ref{eq:1stfnu_shape_factor} can be neglected, and terms proportional to $\mathrm{(\alpha Z)}$ dominate. The quantities entering the calculation of $\mathrm{k_n}$, a, b and c (see Table~\ref{table:param_1stfnu}) then compare as:
\begin{eqnarray}\label{eq:xi_conditions}
	\mathrm{|A_0| \text{ or } |C_0| \gg |R B_0|\,} &&\mathrm{\simeq |W_0 R C_1| \simeq |R D_0|}\nonumber \\
    &&\mathrm{ \simeq |W_0 R E_0| \simeq |W_0 R F_0|}\nonumber \\
    &&\mathrm{\simeq |W_0 R G_0|,}
\end{eqnarray}
in which case the shape factor (eq.~\ref{eq:1stfnu_shape_factor}) simplifies to $\mathrm{C(Z,W_e) \simeq k_n (1 + \mu_1 \gamma_1 b/W_e)}$. This shape factor still does not resemble an allowed shape factor at this stage. By comparing this simplified shape factor to the general shape factor as expressed in eq.~\ref{eq:general_shape}, the second term in the former can be identified to the third term in the latter. This term is proportional to the $\mathrm{(M_K\, m_K})$ product, and also appears in allowed transitions with the same value $\mathrm{b = -2R (A_0B_0+C_0D_0)/(A_0^2+C_0^2)}$. As the $\mathrm{M_K}$ quantities are usually greater than $\mathrm{m_K}$ by an order of magnitude, terms connected with $\mathrm{m_K}$ can be neglected and the shape factor hence becomes independent of the electron energy $\mathrm{W_e}$. The $\mathrm{\xi}$-approximation is most likely to be correctly fulfilled for heavy nuclei. For instance, $\mathrm{2\xi}$ is already $\mathrm{\sim}$10~MeV for $\mathrm{^{70}}$Br and $\mathrm{\sim}$18~MeV for $\mathrm{^{241}}$Pu. Still, many transitions for which the $\mathrm{2\xi \gg W_0}$ criteria is met can exhibit a clear and significant deviation of their corresponding shape factor with respect to the allowed shape. For instance, a systematic comparison of experimentally measured shape factors with a theoretical prediction using the $\mathrm{\xi}$-approximation showed that the latter often failed to reproduce the former \cite{Mougeot2015}. In such cases, the nuclear structure of the initial and final states are most likely to be responsible for this failure.~This can be due to the so-called cancellation effect, where the form factor coefficients in $\mathrm{|A_0|}$ and $\mathrm{|C_0|}$ can compensate each other, making these quantities to be of the same order of magnitude than the others in eq.~\ref{eq:xi_conditions}. Another possible situation is a modification of the selection rules in strongly deformed nuclei, the $\mathrm{\beta}$ transition being then denoted as K-hindered \cite{Ejiri2017}, with K the projection of the total angular momentum on the symmetry axis.
\begin{table*}[!htp]
	\centering
	\begin{tabular}{rrcl}
		\hline \hline
		\rule[0.5cm]{0cm}{0cm} Shape factor & $\mathrm{C(Z,W_e)}$ & = & $\mathrm{k_n (1 + aW_e + \mu_1 \gamma_1 b/W_e + cW_e^2)}$ \\
		&&&\\ \hline
		\rule[0.5cm]{0cm}{0cm} with & $\mathrm{k_n}$ & = & $\mathrm{A_0^2 + C_0^2 - 2\mu_1 \gamma_1 R^2 C_1 D_0 + \frac{1}{9}(W_0R)^2(E_0^2 + G_0^2)}$ \\
		\rule[0.5cm]{0cm}{0cm} & a & = & $\mathrm{R \left\{ 2 C_0 C_1 - \frac{2}{9} (W_0 R)(E_0^2 + G_0^2) \right\}/k_n}$ \\ 
		\rule[0.5cm]{0cm}{0cm} & b & = & $\mathrm{-2R(A_0 B_0 + C_0 D_0)/k_n}$ \\ 
		\rule[0.5cm]{0cm}{0cm} & c  & = & $\mathrm{R^2 \left\{ C_1^2 + \frac{1}{9} (E_0^2 + G_0^2 + \lambda_2 F_0^2 + \lambda_2 G_0^2) \right\}/k_n}$ \\ 
		&&&\\ \hline
		\rule[0.5cm]{0cm}{0cm} where &$\mathrm{A_0}$ & = & $\mathrm{^{A}{F}{^{(0)}_{000}} - \frac{1}{3} \alpha Z \, ^{A}{F}{^{(0)}_{011}}(1,1,1,1) - \frac{1}{3} W_0 R \,^{A}{F}{^{(0)}_{011}}}$ \\
		\rule[0.5cm]{0cm}{0cm} &$\mathrm{B_0}$ & = & $\mathrm{- \frac{1}{3} \,^{A}{F}{^{(0)}_{011}}}$ \\
		\rule[0.5cm]{0cm}{0cm} & $\mathrm{C_0}$ & = & $\mathrm{-^{V}{F}{^{(0)}_{101}} - \frac{1}{3} \alpha Z \sqrt{\frac{1}{3}}\, ^{V}{F}{^{(0)}_{110}}(1,1,1,1) - \frac{1}{3} W_0 R \sqrt{\frac{1}{3}}\, ^{V}{F}{^{(0)}_{110}}}$ \\
		\rule[0.5cm]{0cm}{0cm} && & $- \frac{1}{3} \alpha Z \sqrt{\frac{2}{3}}\, ^{\mathrm{A}}{F}{^{(0)}_{111}}(1,1,1,1) + \frac{1}{3} W_0 R \sqrt{\frac{2}{3}}\, ^{\mathrm{A}}{F}{^{(0)}_{111}}$ \\
		\rule[0.5cm]{0cm}{0cm} &$\mathrm{C_1}$ & = & $\mathrm{-\frac{2}{3} \sqrt{\frac{2}{3}}\, ^{A}{F}{^{(0)}_{111}}}$ \\
		\rule[0.5cm]{0cm}{0cm} &$\mathrm{D_0}$ & = & $\mathrm{- \frac{1}{3} \left\{ \sqrt{\frac{1}{3}}\, ^{V}{F}{^{(0)}_{110}} + \sqrt{\frac{2}{3}}\, ^{A}{F}{^{(0)}_{111}} \right\}}$ \\
		\rule[0.5cm]{0cm}{0cm} &$\mathrm{E_0}$ & = & $\mathrm{\sqrt{\frac{2}{3}}\, ^{V}{F}{^{(0)}_{110}} + \sqrt{\frac{1}{3}}\, ^{A}{F}{^{(0)}_{111}}}$ \\
		\rule[0.5cm]{0cm}{0cm} &$\mathrm{F_0}$ & = & $\mathrm{\sqrt{\frac{2}{3}}\, ^{V}{F}{^{(0)}_{110}} - \sqrt{\frac{1}{3}}\, ^{A}{F}{^{(0)}_{111}}}$ \\
		\rule[0.5cm]{0cm}{0cm} &$\mathrm{G_0}$ & = & $\mathrm{- ^{A}{F}{^{(0)}_{211}}}$ \\ \\
		\hline \hline 	
	\end{tabular}
	\caption{Theoretical expression of the shape factor for a first forbidden non-unique transition, as re-written in \cite{Behrens1982}. Only dominant terms in the multipole expansion have been considered.}
	\label{table:param_1stfnu}
\end{table*}

\begin{figure}[!ht]
    \centering
    \includegraphics[width=0.47\textwidth]{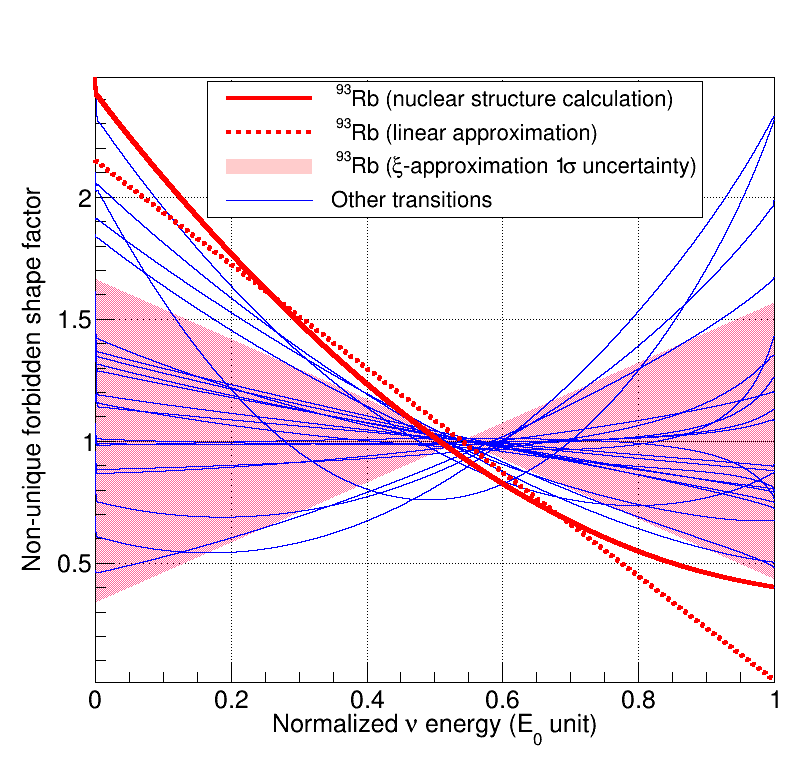}
    \caption{Ratio of \bnue{} spectra of the 23 non-unique forbidden transitions listed in Table~\ref{tab:nu_transitions} over their associated allowed shape obtained under the $\xi$-approximation. The x-axis is expressed in normalized $\mathrm{\bar{\nu}_e}$ kinetic energy $\mathrm{E_{\nu}/E_0}$, where $\mathrm{E_0}$ is the transition endpoint energy (see eq. \ref{eq:endpoint energy}). The solid red curve highlights the ratio obtained for the $\mathrm{{5/2}^-\rightarrow {5/2}^+}$ transition of \iso{93}{Rb}. The linear shape factor used to model the uncertainty resulting from the application of the $\xi$-approximation to this transition is represented by the dashed red curve (see text for more details). The application of the $\mathrm{\xi}$-approximation to this transition results in a $\mathrm{1\sigma}$ uncertainty band corresponding to the red area centered on unity.}
    \label{fig:non_unique_shape_factor}
\end{figure}

The same reasoning than the one explained here for first non-unique forbidden transitions can be applied to any higher order forbidden non-unique transitions. If the $\mathrm{\xi}$-approximation criterion is fulfilled, the contribution of the form factor coefficients becomes independent of the electron and neutrino energies and can be factored out of the shape factor. A $\mathrm{L^{th}}$ non-unique forbidden transition can then be treated as the corresponding $\mathrm{(L-1)^{th}}$ unique forbidden one. Following these theoretical considerations, a conservative criterion to ensure the $\mathrm{\xi}$-approximation to succeed for the calculation of non-unique forbidden transitions could be $\mathrm{2\xi/W_0 > 100}$, as prescribed in \cite{Mougeot2015}. Checking the decay data of the relevant transitions for the calculation of reactor \bnue{} spectra, this criterion is unfortunately met for transitions with endpoint energies smaller than 0.4 MeV only. Furthermore, relaxing this criterion to $\mathrm{2\xi/W_0 > 10}$ increases the maximum endpoint energy of these transitions to 3 MeV, showing that the $\mathrm{\xi}$-approximation will very likely fail for the vast majority of non-unique forbidden transitions. 
In order to account for these possible modeling errors, an uncertainty is applied to each of the $\mathrm{\xi}$-approximated transitions. In the same fashion that what is done in section~\ref{subsubsec:main_forbidden_transitions}, this uncertainty is constructed using the difference between a non-unique forbidden spectrum and its $\mathrm{\xi}$-approximated version. Because each of the non-unique forbidden transitions the $\mathrm{\xi}$-approximation is applied to are by definition not known, a simple strategy is to use a reference non-unique spectrum to compute this difference. This reference spectrum is modeled using the 23 non-unique forbidden transitions computed using nuclear structure calculations (see Table~\ref{tab:nu_transitions}). Figure~\ref{fig:non_unique_shape_factor} shows the deviation of their shape factors with respect to their corresponding $\mathrm{\xi}$-approximated version in the normalized \bnue{} energy representation $\mathrm{E_{\nu}}$/$\mathrm{E_0}$, where $\mathrm{E_0}$ is the endpoint energy. Because all of these transitions are first non-unique forbidden, these deviations are also deviations from an allowed shape. No general and systematic trend can be identified, making it difficult to describe them all with a unique and simple parametrisation. As such, all of these deviations are crudely approximated by a same linear function. This linear function is chosen such that it reasonably approximates the \iso{93}{Rb} shape factor, which exhibits the largest deviation among all of these transitions. As illustrated in Figure~\ref{fig:non_unique_shape_factor}, its amplitude reaches $150 \%$ at 0 MeV and vanishes at the endpoint energy. Constructed this way, this linear function bands the deviation of most of the calculated non-unique shape factors with their respective $\mathrm{\xi}$-approximated version. It should hence conservatively describe the modeling errors caused by the $\mathrm{\xi}$-approximation to most of the non-unique forbidden transitions still lacking nuclear structure calculations in the present computation of a reactor \bnue{} spectrum. The reference spectrum for each non-unique forbidden branch modeled with the $\xi$-approximation is then calculated multiplying this linear function by an allowed spectrum, and is used in the following way to construct the covariance matrix associated to a $\mathrm{\xi}$-approximated transition. For each energy bin, the previously described reference spectrum and the $\mathrm{\xi}$-approximated transition spectrum are used to define the limits of a uniform distribution. This uniform distribution is conservatively symmetrized to take into account that non-unique forbidden shape factors could either exhibit positively or negatively slopped deviations. The standard deviation of the resulting uniform distribution is then used to generate a covariance matrix. Furthermore, this covariance matrix is constructed such that energy bins showing equal (resp. opposite) sign in the difference between the reference and $\mathrm{\xi}$-approximated spectra are considered as fully correlated (resp. anti-correlated).

\begin{figure*}[!ht]
    \centering
    \includegraphics[width=0.47\textwidth]{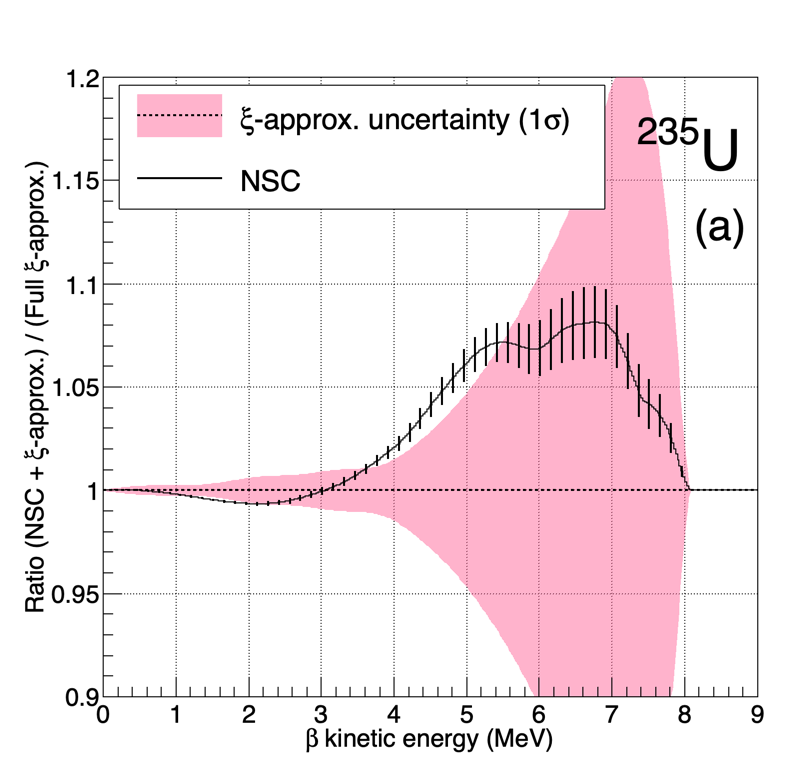}
    \includegraphics[width=0.47\textwidth]{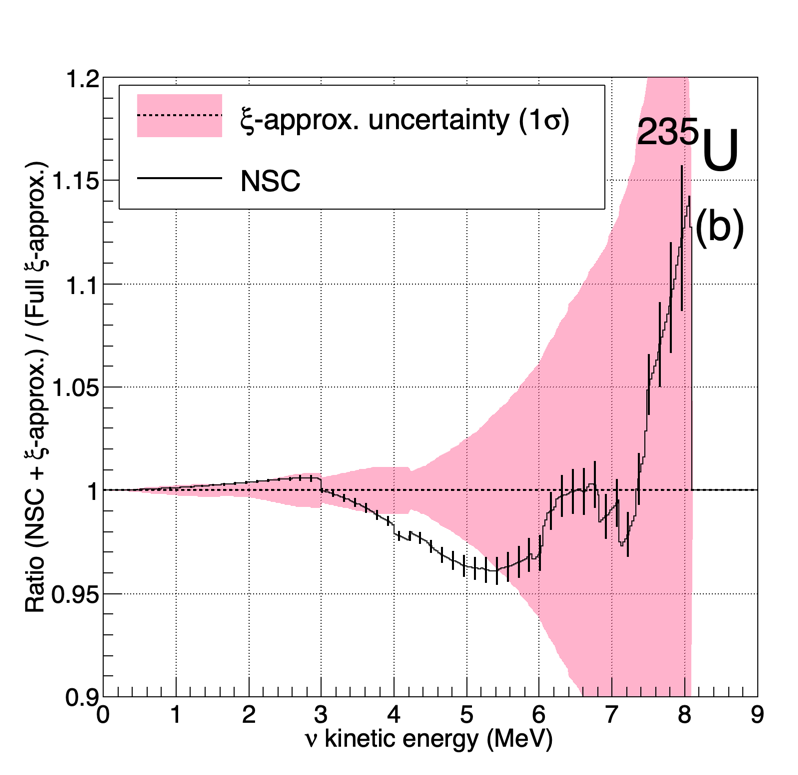}
    \caption{Impact of the $\mathrm{\xi}$-approximation to the calculation of the \U{235} fission \bnue{} spectrum. Panel (a) (resp.~panel (b)) plots the ratio of the $\beta$ (resp.~\bnue{}) spectrum including the 23 non-unique forbidden branches from Table~\ref{tab:nu_transitions} as obtained from the nuclear structure calculations (NSC) described in section~\ref{subsubsec:main_forbidden_transitions}, over the same spectrum where all non-unique transitions are treated under the $\mathrm{\xi}$-approximation. The red area represents the $\mathrm{1 \sigma}$ uncertainty band resulting from the application of the $\mathrm{\xi}$-approximation to these 23 non-unique forbidden branches. Black error bars represent the uncertainty derived from the nuclear structure calculations if applied to these same 23 transitions instead.
    }
    \label{fig:non_unique_impact_on_U235}
\end{figure*}

At the level of a branch, the $\mathrm{\xi}$-approximation leads to a typical 30\% uncertainty on the corresponding IBD yield. The contribution of $\mathrm{\xi}$-approximated branches to isotopic IBD yields is at the level of 10-15\% (see Table~\ref{tab:transition_contribution_breakdown}) and their corresponding uncertainty shrinks to $\mathrm{\sim}$0.4\% (see Table~\ref{tab:uncertainty budget}). The associated fractional uncertainty obtained for the \U{235} fission spectrum is displayed in Figure~\ref{fig:fractional_uncertainty_modeling}, and shows that the uncertainty related to the $\mathrm{\xi}$-approximation mostly dominates the total modeling uncertainty budget at low energies. No correlation among the $\mathrm{\xi}$-approximated transitions has been considered in the previously quoted uncertainties. The impact of fully correlating these transitions has nevertheless been checked. It was found to be significant, increasing the uncertainty at the isotopic IBD yield level to $\mathrm{\sim}$3\%. Yet, and for the same reason than the one discussed in the previous section, the present work assumes no transition-to-transition correlations. Finally, the robustness of this uncertainty modeling has been checked by computing the \U{235} fission spectrum with (i) all non-unique forbidden transitions as treated by the $\mathrm{\xi}$-approximation and (ii) including the 23 most contributing non-unique forbidden transitions as computed in section~\ref{subsubsec:main_forbidden_transitions}, the rest of the non-unique forbidden transitions remaining $\mathrm{\xi}$-approximated. Figure~\ref{fig:non_unique_impact_on_U235} depicts the ratios of these two spectra both in the $\beta$ and the \bnue{} energy representations. The application of the $\mathrm{\xi}$-approximation to these 23 non-unique forbidden branches results in a $\mathrm{1\sigma}$ uncertainty band (represented in red) which mostly covers the observed differences in both cases. Moreover, the 1.4\% decrease in the \U{235} IBD yield when going from case (i) to case (ii) is well covered by the 1.8\% uncertainty resulting from the application of the $\mathrm{\xi}$-approximation to all non-unique transitions. Similar decreases have also been obtained for the \U{238}, \Pu{239} and \Pu{241} isotopic IBD yields, which equally amount to 1.1\%. They are also well covered by the previously described uncertainty resulting from the $\mathrm{\xi}$-approximation.

%
\subsection{Radiative corrections}\label{subsec:beta-branch-radiative-corrections}
Radiative corrections come from non-static Coulomb processes arising at higher perturbation levels such as virtual photon exchange and inner brehmsstrahlung, in which one or multiple photons can be emitted in the final state. These effects can be important in the calculation of a $\mathrm{\beta}$ spectrum, especially for transitions with endpoint energy much larger than the electron mass. Radiative corrections are usually separated between an inner and outer part. The latter being the only one to include energy-dependent terms, no inner radiative correction is then considered in this work. Only outer radiative corrections of order $\mathrm{\alpha}$ are taken into account. The higher order terms in $\mathrm{\alpha^m Z^n}$ ($\mathrm{m > n}$) are much smaller and they can be neglected to a first approximation \cite{Hayen2018}. 
The ${\cal{O}}$($\mathrm{\alpha}$) outer radiative correction for the electron can be found in \cite{Sirlin1967}.
Although the neutrino is insensitive to the nucleus Coulomb field, virtual photon exchange and energy conservation in inner brehmsstrahlung processes can still indirectly change its kinetic energy. The total ${\cal{O}}$($\mathrm{\alpha}$) radiative correction for the neutrino has been explicitly calculated in \cite{Sirlin2011}, and somewhat differs from the electron case because of a different inner brehmsstrahlung contribution. In the calculation of a \bnue{} spectrum at the $\mathrm{\beta}$-branch level, the electron correction must therefore be replaced by the neutrino correction.
Radiative corrections only depend on the transition endpoint energy. Electron radiative corrections typically range from less than 1\% for transitions with endpoint energies close to the IBD threshold up to $\mathrm{\sim}$10\% for transitions with endpoint energies reaching 10 MeV. On the other hand, neutrino radiative corrections barely reach 1\% for any transition with endpoint energy below 10 MeV. Once propagated to the calculation of a full reactor \bnue{} spectrum (see eq.~\ref{eq:reactor_spectrum}), they total a $\mathrm{\lesssim 0.5\%}$ shape correction and a $\mathrm{\lesssim 0.2\%}$ correction on the corresponding IBD yield since most of the $\beta$-decay transitions exhibit low endpoint energies. Additionally, transitions with endpoint energies above the IBD threshold display a large range of endpoint energies up to 10 MeV and thus transition spectra average out in a reactor \bnue{} spectrum.

The radiative corrections to the calculation of an electron and \bnue{} spectrum at the $\mathrm{\beta}$-branch level have been derived for allowed transitions only \cite{Sirlin1967, Sirlin2011}. The procedure followed in this work and also in other previous summation models \cite{Mueller2011,Huber2011,Fallot2012} is to apply these same corrections to any other type of $\beta$-decay transition. The validity of this procedure is not well-known. 
As such, an uncertainty associated to the outer radiative correction modeling is conservatively built similarly to what is done for the accurate modeling of non-unique forbidden transitions in section~\ref{subsec:non_unique_forbidden_transitions}. The associated covariance matrix is here constructed using the difference between an actinide fission spectrum (see eq.~\ref{eq:simplified_reactor_spectrum}) calculated with and without the application of outer radiative corrections.
When propagated to the computation of a full reactor \bnue{} spectrum, this source of uncertainty is treated as fully correlated between fission spectra since the corrections always behave in a similar fashion.
Using this approach, the radiative correction uncertainty is $\mathrm{\sim}$0.1\% when propagated to the IBD yield of each actinide (see Table~\ref{tab:uncertainty budget}). The associated fractional uncertainty obtained in the case of the \U{235} fission spectrum is displayed in Figure~\ref{fig:fractional_uncertainty_modeling}, showing that this uncertainty source is negligible with respect to the other $\mathrm{\beta}$-decay modeling uncertainties.
\subsection{Weak magnetism correction}\label{subsec:beta-branch-WM-corrections}
\begin{figure}[!ht]
    \centering
    \includegraphics[width=0.47\textwidth]{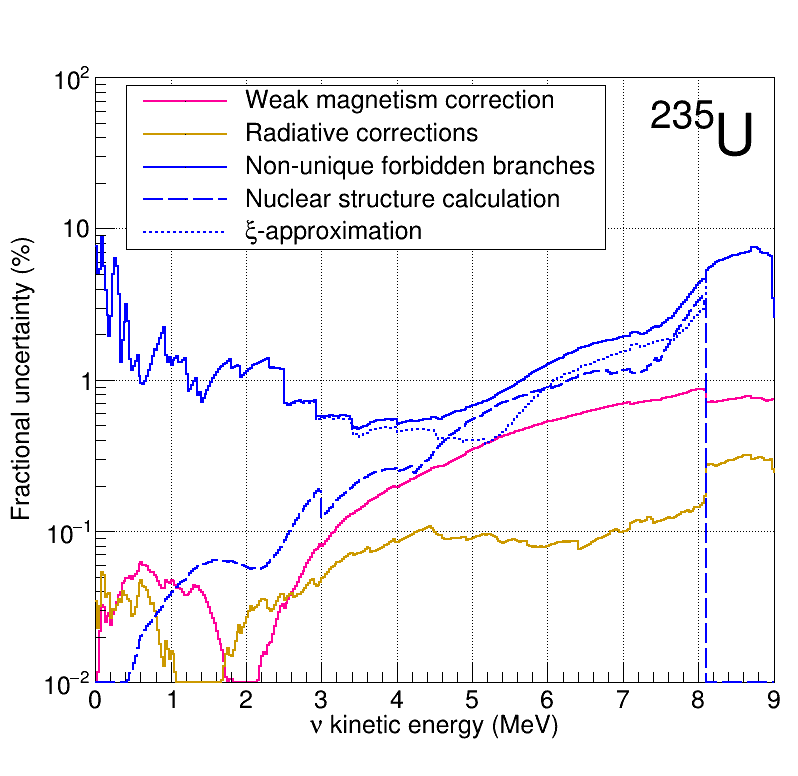}
    \caption{Breakdown of fractional uncertainties associated to the modeling of $\mathrm{\beta}$ branches in the summation calculation of the \U{235} fission \bnue{} spectrum. Uncertainties associated to the treatment of non-unique forbidden transitions are further broken down between the 23 branches computed with nuclear structure and described in Table~\ref{tab:nu_transitions} (blue dashed line) and the rest being $\mathrm{\xi}$-approximated (blue dotted line).}
    \label{fig:fractional_uncertainty_modeling}
\end{figure}

Weak magnetism (WM) refers to the dominant contribution of a class of additional induced-nuclear currents appearing in the vector part of the $\beta$-decay Hamiltonian when taking into account the finite-size and the internal structure of the nucleons. In the Behrens and B{\"u}hring formalism \cite{Behrens1982}, a weak magnetism correction to the calculation of a $\mathrm{\beta}$/\bnue{} spectrum at the branch level in principle only applies to Gamow-Teller allowed and non-unique forbidden transitions, as forbidden transitions of the unique type solely depend on axial nuclear form factors.  Equivalent WM corrections for allowed Gamow-Teller transitions were consecutively derived in \cite{Vogel1984, fayans1985}, and were extensively applied to any type of transitions in the past summation calculations of reactor \bnue{} spectra \cite{Mueller2011,Huber2011,Fallot2012}. The present work considers the weak magnetism correction derived in \cite{hayes2014} for allowed Gamow-Teller transitions.
Because no clear prescription about weak magnetism in non-unique transitions yet exists in the literature, the correction for allowed branches is indistinctively applied to the corresponding spectra.
The WM correction exhibits a similar dependency to the transition endpoint energy than the radiative corrections (see section~\ref{subsec:beta-branch-radiative-corrections}). The magnitude of the correction is ${\cal{O}}$(1\%) (resp.~$\mathrm{\sim}$4\%) for transitions with $\mathrm{E_{0}}$ close to the IBD threshold (resp. 10 MeV). Once propagated to the calculation of an actinide fission \bnue{} spectrum, weak magnetism typically introduces up to a +0.1\% correction below 2 MeV and a linearly decreasing correction reaching -1.5\% at 8 MeV \cite{perisse:tel-03538198}. In the same fashion than for radiative corrections, an uncertainty associated to the WM correction is built up by comparing actinide fission \bnue{} spectra calculated with or without this correction, and constructing the associated covariance matrix such that the total rate is conserved. This source of uncertainty is fully correlated between each actinide fission spectra since their respective WM correction are similar. This prescription results in a relative uncertainty of about $\mathrm{\sim}$0.3\% for each isotopic IBD yield (see Table~\ref{tab:uncertainty budget}). The associated fractional uncertainty obtained in the case of the \U{235} fission spectrum is displayed in Figure~\ref{fig:fractional_uncertainty_modeling}. Similarly to the radiative corrections, the weak magnetism correction plays a minor role in the combined uncertainty of a reactor \bnue{} spectrum.
%
%
%
\section{Evaluated nuclear data}\label{sec:evaluated_nuclear_data}
The summation calculation of a reactor \bnue{} spectrum, as depicted in eq.~\ref{eq:reactor_spectrum}, eq.~\ref{eq:fission fragment spectrum} and eq.~\ref{eq:endpoint energy} requires a large set of various nuclear data. In the present work, a new database parsing recent evaluated nuclear databases available online has been built. The construction of this database starts from the fission yield data (section~\ref{subsec:fission yield data}), which defines the list and probability of occurence of all possible products for a fissioning system irradiated by neutrons, such as one of the four major actinides \U{235}, \Pu{239}, \Pu{241} and \U{238} present in a reactor core. After the identification of all the $\mathrm{\beta}$ emitters, a database of evaluated decay information is constructed using the Evaluated Nuclear Structure Data Files (ENSDF) \cite{ENSDF} as described in section~\ref{subsec:decay data}. All modern evaluated nuclear decay databases are known to suffer from the so-called Pandemonium effect, which can significantly impact the calculation of reactor \bnue{} fluxes and spectra. Section~\ref{subsec: correction for the pandemonium effect} describes the correction of several important fission products decay information using recent Total Absorption Gamma-ray Spectrometry (TAGS) data available first, and then integral $\mathrm{\beta}$ spectrum measurements. The incompleteness of the evaluated nuclear decay databases leaves many fission products with no decay information. A new modeling of their contribution to the calculation of an actinide fission spectrum is proposed in section~\ref{subsec: nuclei with no data}. All along these sections, a particular attention is paid to precisely describe the usage and the uncertainty treatment of all the evaluated nuclear data so as to detail and understand how they contribute to the combined uncertainty budget of a fission actinide \bnue{} spectrum and flux.
%
%
\subsection{Fission yield information}\label{subsec:fission yield data}
In eq.~\ref{eq:reactor_spectrum}, the activity of each fission product is necessary to properly weight their corresponding contribution into the calculation of a reactor $\mathrm{\beta}$/\bnue{} spectrum. The total activity of each fission product can be computed over a reactor cycle using independent fission yields together with a reactor evolution code able to estimate a core inventory at irradiation time t. An independent fission yield $\mathrm{\mathcal{I}^{k}_{p}}$ refers to the probability that a particular nuclide p will be produced directly from the fission of the k\nth{} actinide. The activity $\mathrm{\mathcal{A}^{k}_{p}(t)}$ of such a fission product can be estimated using a cumulative fission yield $\mathrm{\mathcal{Y}^{k}_{p}(t)}$ such that:
\begin{equation}\label{eq:FP_activity}
    \mathrm{\mathcal{A}^{k}_{p}(t) = f_k(t) \times \mathcal{Y}^{k}_{p}(t),}
\end{equation}
where $\mathrm{f_k(t)}$ is the fission rate of the k\nth{} actinide at irradiation time t. The cumulative yield $\mathrm{\mathcal{Y}^{k}_{p}(t)}$ can here be understood as the total probability that the nuclide p is present after a time t, meaning it either is due to direct production from a fission event
or comes from the decay of a parent fission product. 

The fission yield (FY) information required to the computation of a reactor \bnue{} spectrum can be extracted from several nuclear data libraries, the most popular being the Joint Evaluated Fission and Fusion Files (JEFF, EU) \cite{JEFF_web}, the Evaluated Nuclear Data Files (ENDF, USA) \cite{ENDF_web} and the Japanese Evaluated Nuclear Data Library (JENDL, Japan) \cite{JENDL_web}. For a given fissionning system, both independent and cumulative FYs each estimated for a set of three neutron energies (thermal, epithermal and fast) are provided. The cumulative FYs are those estimated at infinite irradiation time. For a fissioning system k, they are recursively computed as:
\begin{eqnarray}
    \mathrm{\mathcal{Y}^{k,\infty}_p} &&\mathrm{=\lim_{t\to\infty} \mathcal{Y}^{k}_{p}(t)}\nonumber \\
    &&\mathrm{=\mathcal{I}^{k}_{p} + \sum_m b_{mp}\,\mathcal{Y}^k_m,}\label{eq:CFY}
\end{eqnarray}
where $\mathrm{b_{mp}}$ is the probability that the parent nuclide m decays to the daughter nuclide p. The FY evaluation process in these libraries pretty much follows the same methodology, where basically a set of selected experimental data are combined with semi-empirical fission models to coherently assess all the independent and cumulative products yields for a given fissioning system. Moreover, the FY evaluated data can nearly be identical from one library to another. A good example is JENDL, which until very recently used to include the FY data from the ENDF evaluation \cite{Tsubakihara2021}.

In the present work, and unless otherwise indicated, thermal (resp. fast) cumulative fission yields at infinite irradiation time are used to compute the \U{235}, \Pu{239} and \Pu{241} (resp. \U{238}) actinide fission spectra. In other words, neither off-equilibrium corrections nor the impact of different fission neutron energies are taken into account. Assessing the impact of both these effects and including them into the present calculations is beyond the scope of this article. However, off-equilibrium effects have already been studied in the past \cite{Mueller2011, Kopeikin2003}. They typically yield a small $\mathrm{\lesssim 2}$\% negative correction to the 1.8-3 MeV portion of a \bnue{} spectrum emitted by commercial pressurized water reactors burning fuel over a typical 12- to 18-month cycle. The variation of the fission product yields with neutron energies has recently been studied by simulating and comparing low-enriched and highly-enriched reactor core designs \cite{Littlejohn2018}. The corresponding actinide fission \bnue{} spectra showed small $\mathrm{\mathcal{O}(1\%)}$ changes below $\mathrm{\sim}$5 MeV and up to $\mathrm{\sim}$10\% beyond. Their corresponding IBD yields were found to change by less than 1\%. Both these corrections then fall well within the final uncertainty budget of an actinide fission spectrum calculation (see e.g.~figure~\ref{fig:fractional_uncertainty_data_sources}). They also give an IBD yield correction negligible with respect to its typical uncertainty (see Table~\ref{tab:uncertainty budget}).

The latest FY evaluations from the JEFF, ENDF and JENDL librairies are here considered. The FY data are extracted respectively from the \jeff{} \cite{JEFF33}, the \ENDF{} \cite{ENDF8} and the \JENDL{} \cite{JENDL5} releases. Following the prescription of \cite{ENDF8_correction}, some of the FY data of the \ENDF{} release have been corrected from an erroneous evaluation, mostly leading to anomalously large uncertainties. As opposed to past releases, the \JENDL{} release includes now for the first time FY information which do not rely anymore on the ENDF evaluation \cite{Tsubakihara2021}. Although sharing common experimental data and similarities in the underlying models to describe the fission process, the FY evaluation of these three database releases were conducted independently, making then worth a comparison for the calculation of reactor \bnue{} spectra. Any of these libraries ever included information about the correlations between the fission product yield information, whether these correlations come from the experimental data or the evaluation method. The complexity of determining FY correlations is a worldwide recognized problem and has been investigated over the last decade by many groups \cite{Fiorito2016, Rochman2016, Terranova2017}. A set of matrices estimating the covariances of the independent and cumulative FY for the fission of the \U{235}, \U{238}, \Pu{239} and \Pu{241} actinides and exclusively sourcing from the evaluation process have been recently estimated in \cite{Matthews2021} for the \jeff{} and \ENDF{} releases. Covariance matrices originating from the FY evaluation process are also included in the \JENDL{} library \cite{Tsubakihara2021}. They are used here as an attempt to assess the impact of FY correlations among fission products.

\begin{figure*}[!ht]
    \centering
    \includegraphics[width=0.47\textwidth]{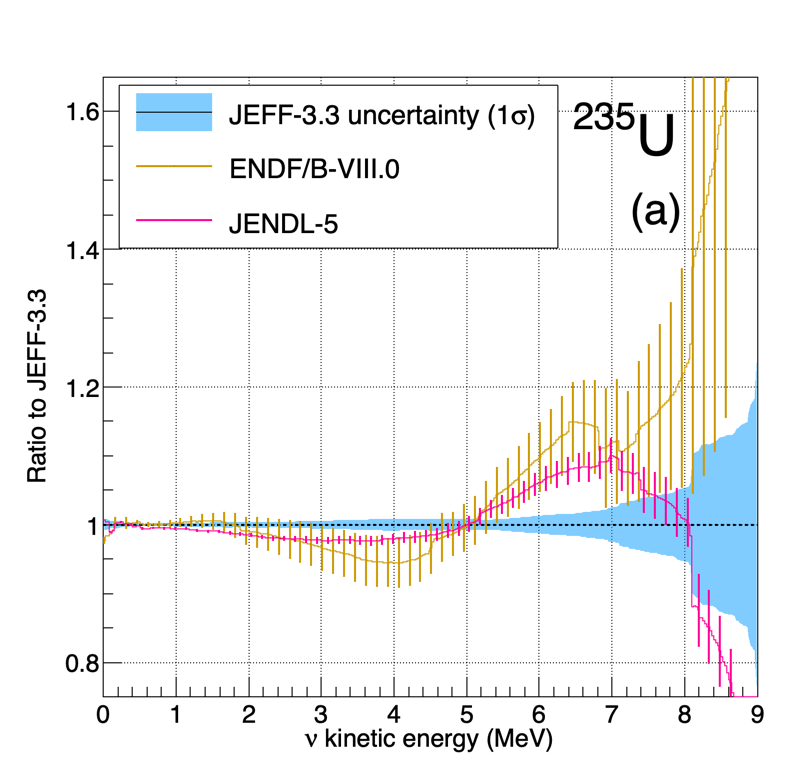}
    \includegraphics[width=0.47\textwidth]{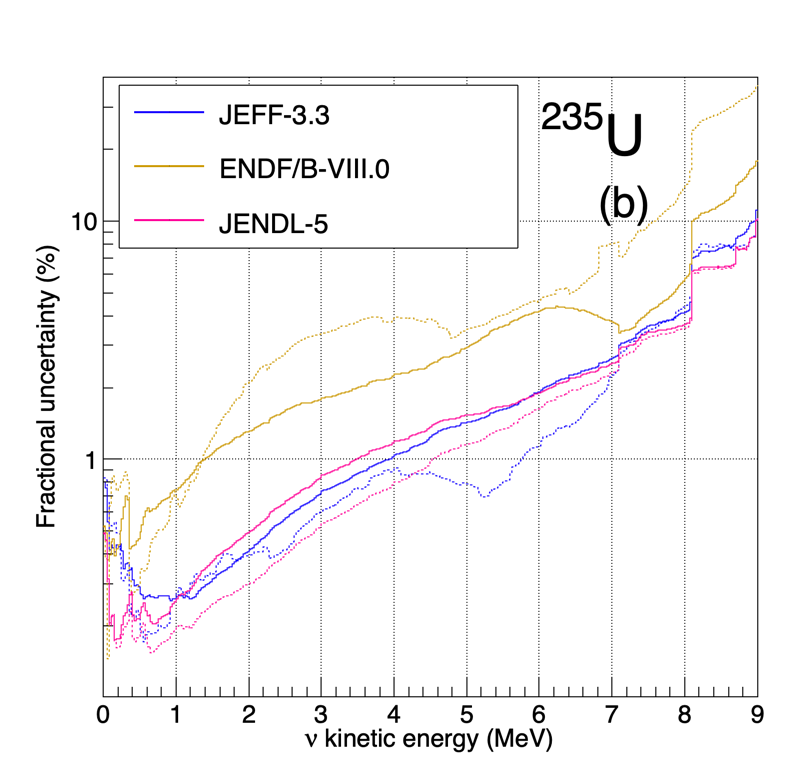}
    \caption{Impact of the evaluated fission yield libraries on the computation of the \bnue{} spectrum from the thermal fission of \U{235}. (a) Ratio of the \bnue{} spectra using the \ENDF{} and \JENDL{} libraries over the one using the \jeff{} library. Uncertainties are propagated using correlations originating from the fission yield evaluation process (see text for more details). (b) Systematic uncertainty on the \U{235} \bnue{} spectrum induced by the FY uncertainties for these three libraries. The dashed lines correspond to an uncertainty propagation procedure using correlations originating from the FY evaluation process, as opposed to the solid lines which assume the FY information to be uncorrelated.
    }
    \label{fig:TBS_ratio_FY_libraries}
\end{figure*}

Figure~\ref{fig:TBS_ratio_FY_libraries} (a) compares the \bnue{} spectra from the thermal fission of \U{235} using cumulative FY from these 3 libraries. The most notable differences are driven by a limited number of fission fragments having important yields which can differ up to an order of magnitude. For instance, the decrease in the ENDF/JEFF ratio around 4 MeV is mostly due to \iso{102}{Tc}. A list of the most relevant isotopes having notable and significant FY differences among these libraries is shown in Table~\ref{tab:FY library isotope differences}. These differences, combined with a slightly different list of FPs in these three libraries, induce isotopic IBD yield variations at the level of $\mathrm{\sim}$0.5\% for \U{235}, $\mathrm{\sim}$2\% for \Pu{239} and \Pu{241}, and $\mathrm{\sim}$3\% for \U{238}. These IBD yield variations indicate a small tension between these three librairies when considering the uncertainty budget associated to FY as shown in Table~\ref{tab:uncertainty budget}. Moreover, the different fission \bnue{} spectra are not always consistent with each other within their 1$\mathrm{\sigma}$ uncertainty bars, as seen in Figure~\ref{fig:TBS_ratio_FY_libraries} (a).
The relative uncertainty on the \U{235} \bnue{} spectrum induced by the cumulative FY uncertainty is illustrated in Figure~\ref{fig:TBS_ratio_FY_libraries} (b) for the three libraries. For each library, the FY uncertainties are propagated either by assuming the FY uncorrelated or by using the covariance matrices as described above. FY-to-FY correlations, induced by the evaluation process in the \jeff{} and \JENDL{} libraries, decrease the \U{235} fractional uncertainty in the 2-7 MeV region. A $\mathrm{1.1\rightarrow0.8}$\% decrease in the corresponding IBD yield uncertainty is observed. Similar effects are also found for the \U{238}, \Pu{239} and \Pu{241} \bnue{} spectra and IBD yields. In the opposite way, the FY uncertainties and covariances evaluated in the ENDF/B.VIII.0 library increase both the fractional uncertainty budget and the IBD yield ($\mathrm{2.2\rightarrow4.0}$\%) of the \U{235} spectrum. This behavior has not been further investigated at the present stage. It should however be noted that the ENDF/B.VIII.0 FY evaluation dates back from the 2000s, and then may well be outdated with respect to the more recent \jeff{} and \JENDL{} evaluations.

To ease any future comparison, the FY evaluation from the \jeff{} library is here chosen, as it is the the most commonly used among the past and current summation predictions~\cite{Mueller2011,Fallot2012,Estienne2019}. Furthermore, the following summation calculations conservatively keep uncorrelated FY information among the fission fragments because (i) the previously mentioned set of covariance matrices are incomplete as they do not include any correlations sourcing from the experimental data the FY evaluation are based on and (ii) the use of these matrices results in a decrease of the uncertainties both on the isotopic IBD yields and the actinide fission \bnue{} spectra.~Beyond the existing correlations between individual fission product yields, correlations between different fissioning isotopes such as \U{235}, \U{238}, \Pu{239} and \Pu{241} in the present study, are also expected~\cite{Rochman2016}. No evaluation could however be found neither among the existing libraries nor in the literature. They are then disregarded in the present work.
\begin{table*}[htp]
\centering
\begin{tabular}{lcccc}
    \hline
    \hline
        Nuclide &  \jeff{}  &  \ENDF{}  & \JENDL{}  &  \Qb{} [MeV] \\
    \hline
        \iso{102}{Tc}     & $\mathrm{(4.29 \pm 0.06) \times 10^{-2}}$  & $\mathrm{(2.24 \pm 0.36) \times 10^{-3}}$    & $\mathrm{(4.29 \pm 0.04) \times 10^{-2}}$   &  4.53   \\
        \iso{102}{Tc^{*}} & $\mathrm{(4.18 \pm 1.51) \times 10^{-5}}$  & $\mathrm{(4.29 \pm 0.47) \times 10^{-2}}$    & $\mathrm{(2.16 \pm 0.78) \times 10^{-5}}$   &  4.58    \\
        \iso{97}{Y}       & $\mathrm{(2.16 \pm 0.25) \times 10^{-2}}$  & $\mathrm{(4.89 \pm 1.12) \times 10^{-2}}$    & $\mathrm{(1.90 \pm 0.26) \times 10^{-2}}$   &  6.82    \\
        \iso{97}{Y^{*}}   & $\mathrm{(2.52 \pm 0.20) \times 10^{-2}}$  & 0                                            & $\mathrm{(3.06 \pm 0.27) \times 10^{-2}}$   &  7.49    \\
        \iso{100}{Nb}     & $\mathrm{(5.54 \pm 0.10) \times 10^{-2}}$  & $\mathrm{(3.11 \pm 1.00) \times 10^{-2}}$    & $\mathrm{(5.71 \pm 0.13) \times 10^{-2}}$   &  6.40   \\
        \iso{100}{Nb^{*}} & $\mathrm{(6.18 \pm 1.95) \times 10^{-3}}$  & $\mathrm{(3.11 \pm 1.00) \times 10^{-2}}$    & $\mathrm{(4.87 \pm 1.25) \times 10^{-3}}$   &  6.71   \\
        \iso{96}{Y}       & $\mathrm{(4.66 \pm 0.15) \times 10^{-2}}$  & $\mathrm{(6.00 \pm 0.96) \times 10^{-2}}$    & $\mathrm{(5.73 \pm 0.14) \times 10^{-2}}$   &  7.11   \\
        \iso{86}{Ge}      & $\mathrm{(8.67 \pm 3.01) \times 10^{-6}}$  & $\mathrm{(6.29 \pm 1.01) \times 10^{-3}}$    & $\mathrm{(3.45 \pm 1.20) \times 10^{-7}}$   &  9.56   \\
        \iso{88}{As}      & $\mathrm{(1.57 \pm 0.61) \times 10^{-5}}$  & $\mathrm{(1.24 \pm 0.56) \times 10^{-3}}$    & $\mathrm{(2.56 \pm 0.99) \times 10^{-5}}$   &  13.43   \\
        \iso{92}{Rb}      & $\mathrm{(4.37 \pm 0.19) \times 10^{-2}}$  & $\mathrm{(4.82 \pm 0.07) \times 10^{-2}}$    & $\mathrm{(5.01 \pm 0.22) \times 10^{-2}}$   &  8.09   \\
    \hline
    \hline
\end{tabular}
\caption{List of fragments having relevant cumulative fission yield differences for \U{235} among the \jeff, the \ENDF{} and \JENDL{} libraries. These fission fragments are mainly responsible for the differences seen in the associated ratios in Figure~\ref{fig:TBS_ratio_FY_libraries} (a). 
}
\label{tab:FY library isotope differences}
\end{table*}
%
%
\subsection{Nuclear structure and decay data}\label{subsec:decay data}
As shown by eq.~\ref{eq:fission fragment spectrum}, \ref{eq:endpoint energy} and \ref{eq:branch_beta_spectrum}, the nuclear structure and decay information of all the fission products making up a reactor \bnue{} spectrum are necessary. In the present work, they are mostly extracted from the 2020, June $\mathrm{29^{th}}$ release of ENSDF \cite{ENSDF}, which includes nuclear structure and decay data for over 3000 nuclides. An ENSDF evaluation can be found for approximately 70\% of the fission products entering the composition of a reactor spectrum, which corresponds to more than 600 isotopes and several thousands of transitions. The nuclear level properties and $\mathrm{\beta}$ transition information are extracted using the ENSDF++ program \cite{ENSDFpp}. The following sections present the different types of evaluated nuclear and decay data necessary for the computation of a fission product $\mathrm{\beta}$/\bnue{} spectrum, and how their associated uncertainties are treated. Most importantly, many fission products have missing or incomplete decay information. When necessary, the procedure used to circumvent this lack of data is also described.
\subsubsection{Branching ratio and $\mathrm{\beta^{-}}$ intensity}
Branching ratios (BR) and $\mathrm{\beta^{-}}$ intensities along with their respective uncertainties are read from ENSDF. In the computation of a fission fragment spectrum, the sum of BRs among all listed transitions is normalized to its $\mathrm{\beta^{-}}$ intensity. Corresponding uncertainties are propagated using a Monte Carlo method. All transitions having complete BR information are treated as Gaussian distributed random variables. The same goes for transitions missing a BR uncertainty, which are assigned a relative 10\% uncertainty. This choice corresponds to the median value of the relative BR uncertainty distribution among all known transitions in ENSDF. Finally, transitions having only an upper (resp. a lower limit) as information about their corresponding BR are treated as random variables following an upper bounded (resp. lower bounded) uniform distribution. The total uncertainty of an isotope $\mathrm{\beta}$/\bnue{} flux originating from the BR information must equal the reported $\mathrm{\beta^{-}}$ intensity uncertainty. Because any information on BR correlations are available in ENSDF, artificial correlations are introduced to meet this constraint. During this step, the phase space of possible correlations is randomly probed to pick the set of correlation parameters maximizing the corresponding isotope IBD yield uncertainty. Further details about the BR and $\mathrm{\beta^{-}}$ uncertainty propagation method are given in \cite{perisse:tel-03538198}. Figure~\ref{fig:fractional_uncertainty_data_sources} illustrates the contribution of this uncertainty source to the fractional uncertainty of the \U{235} \bnue{} fission spectrum, showing that it is among the least important with a $\mathrm{\mathcal{O}(1\%)}$ contribution below 7 MeV. These uncertainties combine to a $\mathrm{\sim}$0.4\% uncertainty once propagated to the calculation of the corresponding IBD yield.
\subsubsection{Endpoint energy, nuclear level spin and parity information}
The endpoint energy of a transition is estimated using eq.~\ref{eq:endpoint energy}, and needs the total $\mathrm{\beta}$ decay energy \Qb{}, the parent isotope isomeric state level $\mathrm{E_{IS}}$ and the daughter isotope level $\mathrm{E^{lvl}}$ information. Fission product $\mathrm{Q_{\beta}}$ information are extracted from the 2020 release of the atomic mass evaluation (AME-2020) database \cite{Wang_2021}, as the proposed evaluation is based on a least square analysis using multiple experimental data, and therefore seems more robust than a single measurement. As a result, the corresponding uncertainty is slightly reduced compared to the analytical \Qb{} calculation based on the mass difference. The metastable parent nucleus energy level $\mathrm{E_{IS}}$ and the corresponding uncertainty are taken from ENSDF. If missing, this information is then retrieved from the 2020 release of the NUBASE database \cite{Kondev_2021}. Finally, the different daughter $\mathrm{\beta}$-feeding state energies along with their uncertainty are extracted from ENSDF. The transition endpoint energy uncertainty is then estimated quadratically summing the $\mathrm{Q_{\beta}}$, $\mathrm{E_{IS}}$ and $\mathrm{E^{lvl}}$ uncertainties.
Spin and parity of the parent and daughter nuclear levels are necessary input information to determine the type of a transition. These information are also extracted from the ENSDF database. Should they may be missing or not fully determined, the following choices are made for the computation of a transition spectrum. When a transition exhibits multiple spin/parity combinations, the corresponding $\mathrm{\beta}$/\bnue{} spectrum is computed as the spectrum average of all associated forbiddeness degree (FD) unique branches. If the spin of either the parent or daughter nucleus is missing, the transition spectrum is defaulty computed as an allowed one.
The uncertainties sourcing from endpoint energy, nuclear level spin and parity information are simultaneously propagated using a Monte Carlo method in the computation of a fission fragment spectrum. Therefore, they are combined all together into a single covariance matrix. Endpoint energy is treated as a Gaussian distributed random variable, and is constrained to yield only positive values. Because the endpoint energy uncertainty of a transition is usually dominated by the fission fragment \Qb{} uncertainty, and also because the \Qb{} value is used to compute the endpoint energies of all fission fragment transitions, endpoint energy information is considered to be fully correlated among those transitions. After an endpoint energy is sampled, incompleteness in the spin/parity information of a transition is then considered.~When many spin/parity combinations are possible, all resulting unique FD are evenly sampled. If a transition misses a spin information, the transition type is randomly sampled between allowed, 1\st{}, 2\nd{} and 3\rd{} unique forbidden transitions.
Similarly to the BR and $\mathrm{\beta^{-}}$ intensity information, uncertainties associated to endpoint energy, nuclear level spin and parity information are found to negligibly contribute to the total uncertainty budget of an actinide fission $\mathrm{\beta}/$\bnue{} spectrum (see Figure~\ref{fig:fractional_uncertainty_data_sources}). They typically induce a $\mathrm{\sim}$0.1\% uncertainty on the \U{235}, \U{238}, \Pu{239} and \Pu{241} isotopic IBD yields (see Table~\ref{tab:uncertainty budget}).
%
%
%
\subsection{Correction for the Pandemonium effect}\label{subsec: correction for the pandemonium effect}
The decay scheme of a parent radionuclide is usually inferred by measuring the intensity and energy of the $\mathrm{\gamma}$-rays emitted in the deexcitation cascade of the daughter nucleus. In the past decades, experimental apparatuses widely used High Purity Germanium detectors (HPGe) for the detection and measurement of these $\mathrm{\gamma}$ rays in coincidence with the $\mathrm{\beta}$ particles, because of their excellent energy reconstruction performances. However, these devices mostly suffered both from limitations and from an incorrect characterization of their detection efficiency at high energies, often leading to an underestimate of the daughter nucleus level density at high excitation energies and hence biasing the parent nuclide decay scheme. This effect, first pointed out in \cite{Hardy1977}, is called the Pandemonium effect and is known to be widely present in modern evaluated nuclear databases. At the level of a $\mathrm{\beta}$ decaying isotope, the Pandemonium effect underestimates (resp. overestimates) the $\mathrm{\beta}$ and \bnue{} spectra at low energies (resp. high energies). Beyond the summation prediction of reactor \bnue{} spectra, having reliable decay data is also important for nuclear reactor operation and safety considerations~\cite{rykaczewski2010}. Therefore, several experimental efforts are conducted to correct the nuclear structure information and the decay scheme of the most important nuclides known to be Pandemonium-affected. The following sections discuss the different sources of Pandemonium-corrected decay information, the extraction of the relevant information and how they are applied to correct the library of nuclear structure and decay information used in the present actinide fission \bnue{} spectrum calculations.
\subsubsection{Total Absorption Gamma-ray Spectrometry data}\label{subsubsec:TAGS_data}
The most reliable Pandemonium-free sources of decay data come from Total Absorption Gamma ray Spectroscopy (TAGS) measurements. The TAGS experimental technique generally uses an arrangement of high-efficiency $\mathrm{\gamma}$-ray detectors (typically NaI or $\mathrm{BaF_2}$ scintillating crystals) with a nearly 4$\mathrm{\pi}$ coverage able to fully reconstruct the $\mathrm{\gamma}$ cascade following a nuclear $\mathrm{\beta}$ decay. As opposed to HPGe devices, such $\mathrm{\gamma}$ detectors have a modest energy resolution, thus requiring the use of deconvolution techniques to properly assess the decay scheme of the parent nucleus. In this work, and whenever it is possible, TAGS data are prioritized over the ENSDF data of any known Pandemonium-affected nuclide.~A first campaign of TAGS measurements was conducted in 1997 at the INEL ISOL facility using NaI(Tl) scintillation detectors by the group of Greenwood et al. \cite{Greenwood1997}. Using these data, the decay schemes of 49 short-lived fission products extracted from ENSDF were corrected. 

More recently, the emerging and increasingly pressing needs of reliable decay information for the predictions of \bnue{} fluxes and spectra emitted at nuclear reactors further accelerated the experimental efforts to measure (or remeasure) a selection of Pandemonium-affected radionuclides with the TAGS technique \cite{TAGSmeeting2009, TAGSmeeting2014}. In this work, the decay schemes of 45 radionuclides have been then retrieved and corrected when necessary using these recent TAGS data. They are listed in Table~\ref{tab:list TAGS}. Among these, 6 radionuclides (\iso{89}{Rb}, \iso{90}{Rb}, \iso{90}{Rb}$\mathrm{^m}$, \iso{91}{Rb}, \iso{94}{Sr}, \iso{140}{Cs}) were already measured by the group of Greenwood et al., which the more recent TAGS data have then been prioritized over. 
Moreover, the \iso{137}{Xe} ENSDF data have been validated by the TAGS measurement reported in \cite{Rasco2017}.
Finally, the ENSDF decay data of \iso{99}{Zr} and \iso{87}{Kr} remained uncorrected in the present work since their corresponding TAGS data were either incomplete or not available in a usable format. 
For information, \iso{99}{Zr} (\iso{87}{Kr}) represents 1.0\%, 0.7\%, 1.3\% and 0.9\% (0.3\%, 0.1\%, 0.2\% and 0.1\%) of respectively the \U{235}, \U{238}, \Pu{239} and \Pu{241} expected IBD yields.

\begin{table}[htp]
\centering
\begin{tabular}{c|c}
    \hline
    \hline
        Isotope & Reference \\
    \hline
        \iso{76}{Ga} & \cite{Dombos2016} \\
        \iso{84}{Br}, \iso{85}{Br}  & \cite{Goetz2017} \\
        \iso{86}{Br}, \iso{91}{Rb} & \cite{Rice2017}  \\
        \iso{87}{Br}, \iso{88}{Br}, \iso{94}{Rb}  & \cite{Valencia2017}\\
        \iso{89}{Kr}, \iso{89}{Rb}, \iso{90}{Kr}, \iso{90}{Rb}, \iso{90}{Rb}$\mathrm{^m}$, \iso{92}{Rb} & \cite{Fijalkowska2017}\\
        \iso{93}{Rb}, \iso{139}{Xe}  & \cite{Fijalkowska2017}\\
        \iso{94}{Kr} & \cite{Miernik2016} \\
        \iso{94}{Sr}  & \cite{Rice2014} \\
        \iso{95}{Rb}, \iso{137}{I}  & \cite{Guadilla2019a} \\
        \iso{96}{Y}, \iso{96}{Y}$\mathrm{^m}$  & \cite{Guadilla2022} \\
        \iso{98}{Nb}  & \cite{Rasco2022} \\
        \iso{142}{Cs} & \cite{Rasco2016} \\
        \iso{100}{Nb}, \iso{100}{Nb}$\mathrm{^m}$, \iso{102}{Nb}, \iso{102}{Nb}$\mathrm{^m}$  & \cite{Guadilla2019} \\
        \iso{101}{Nb}, \iso{105}{Mo}, \iso{106}{Tc}, \iso{107}{Tc}  &  \cite{Algora2010} \\
        \iso{100}{Tc} & \cite{Guadilla2017b} \\
        \iso{102}{Tc}, \iso{104}{Tc}, \iso{105}{Tc} & \cite{Jordan2013} \\
        \iso{101}{Zr}, \iso{102}{Zr}, \iso{109}{Tc} &  \cite{Dombos2021} \\
        \iso{103}{Mo}, \iso{103}{Tc}, \iso{140}{Cs} & \cite{GuadillaGomez2017} \\
        \iso{103}{Nb}, \iso{104}{Nb}$\mathrm{^m}$   &  \cite{Gombas2021} \\
        \iso{137}{Xe} & \cite{Rasco2017} \\
    \hline
    \hline
\end{tabular}
\caption{List of fission fragments whose decay information are corrected following recent TAGS measurements.}
\label{tab:list TAGS}
\end{table}
All selected TAGS data sets have been here updated using the \AME \ database \cite{Wang_2021} for the estimate of the \Qb \ energies, and with the \NUBASE \ database \cite{Kondev_2021} for the determination of the metastable level energies and $\mathrm{\beta^{-}}$ intensities. In addition, TAGS data alone often miss spin and parity information. Therefore, correspondences between the reported nuclear level energies in the TAGS data with those documented in ENSDF have been systematically searched for to preserve these information when using the decay scheme of a TAGS-measured isotope. If no such correspondence could be identified, the corresponding transition was arbitrarily treated as allowed.
\begin{table*}[!htp]
\centering
\begin{tabular}{lllll}
    \hline
    \hline
        & \U{235}  & \U{238}   & \Pu{239}  & \Pu{241}  \\
    \hline
        \multicolumn{1}{l}{\textbf{Number of fission products}  } & & & & \\ 
        TAGS                            &  84   &  83   &  84   &  84   \\
        Tengblad                        &  44   &  44   &  44   &  44    \\
        Nuclides with no data           &  217  &  232  &  216  &  247    \\
        ENSDF                           &  448 (29)  &  419 (29)  &  507 (29)  &  485 (29)   \\
    \hline
        \multicolumn{1}{l}{\textbf{\bnue{} flux contribution}  {\small [\%]} } & & & & \\ 
        TAGS                            &  36.8  &  34.7  &  34.9  &  33.5   \\
        Tengblad                        &  8.0   &  11.0  &  6.2   &  7.6    \\
        Nuclides with no data           &  1.1   &  3.6   &  1.2   &  2.5    \\
        ENSDF                           &  54.1 (14.9) &  50.7 (14.0) &  57.7 (12.0) &  56.4 (11.9)   \\
    \hline
        \multicolumn{1}{l}{\textbf{IBD yield contribution}  {\small [\%]} }  & & & &  \\
        TAGS                            &  57.8  &  42.1  &  60.0  &  47.6  \\
        Tengblad                        &  13.3  &  17.5  &  10.5  &  12.6   \\
        Nuclides with no data           &  3.2   &  9.7   &  4.7   &  8.0    \\
        ENSDF                           &  25.7 (12.3)  &  30.7 (10.0)  &  24.8 (10.6)  &  31.8 (10.0)   \\
    \hline
    \hline
\end{tabular}
\caption{Importance of the different sources of evaluated decay information to the summation calculation of the \U{235}, \U{238}, \Pu{239} and \Pu{241} fission spectra, sorted either according to the number of fission products (top) or the contribution to the total \bnue{} flux (middle) or the contribution to the IBD yield (bottom). The \bnue{} flux and IBD yield contributions are estimated using cumulative fission yields from \jeff~\cite{JEFF33} (see section~\ref{subsec:fission yield data}). The contribution of fission fragments potentially having a remaining Pandemonium effect in their respective decay data (see section~\ref{subsubsec: uncorrected pandemonium effect} for more details) is displayed in parenthesis in the ENSDF line.}
\label{tab:contribution breakdown}
\end{table*}
\begin{figure*}[ht!]
    \centering
    \includegraphics[width=0.47\textwidth]{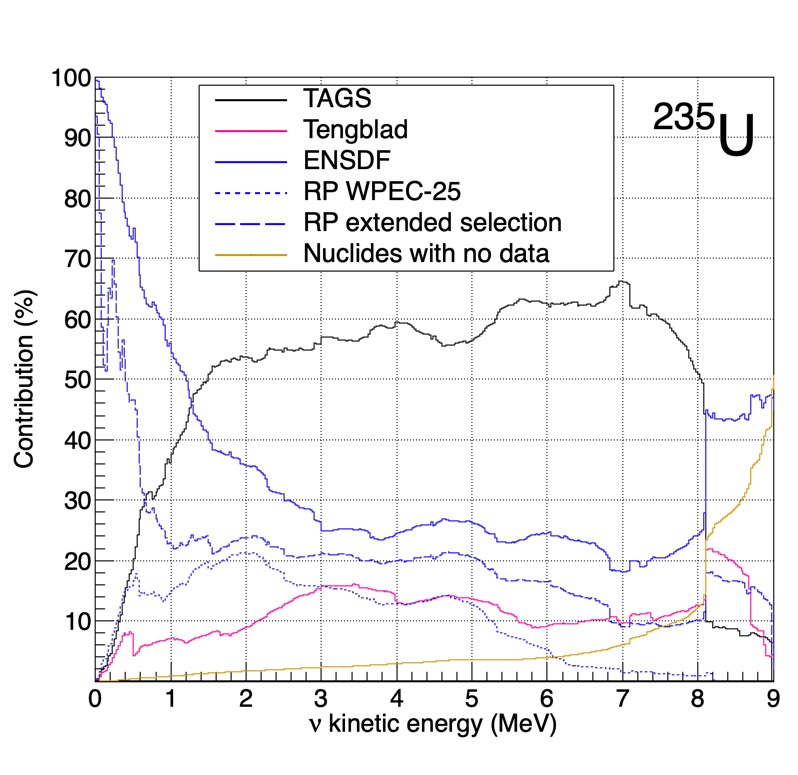}
    \includegraphics[width=0.47\textwidth]{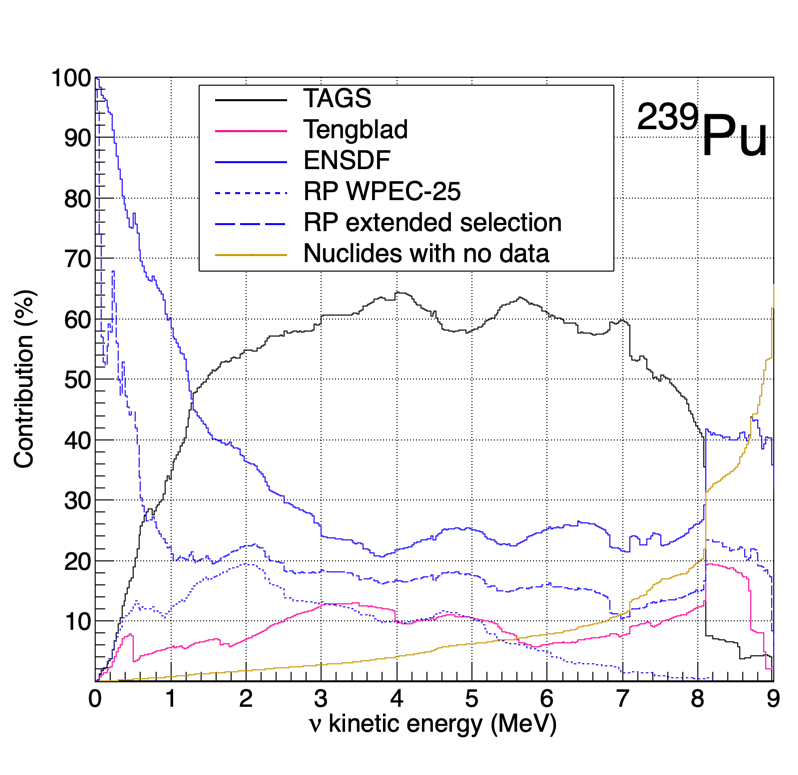}
\caption{Contributions of the different sources of decay information used to compute the \U{235} and \Pu{239} fission \bnue{} spectra. Contributions associated to solid lines add up to 100\%. Contributions from isotopes possibly affected by a residual Pandemonium effect (dotted or dashed blue lines) belongs to the ENSDF data contribution (solid blue line). The RP WPEC-25 case corresponds to the contribution of 29 radionuclides identified by the WPEC-25 group potentially having a Pandemonium effect while the RP extended case additionally includes hundreds of isotopes based on a broader selection (see section~\ref{subsubsec: uncorrected pandemonium effect} for further details).}
\label{fig:contribution spectrum}
\end{figure*}

\begin{figure*}[ht!]
    \centering
    \includegraphics[width=0.47\textwidth]{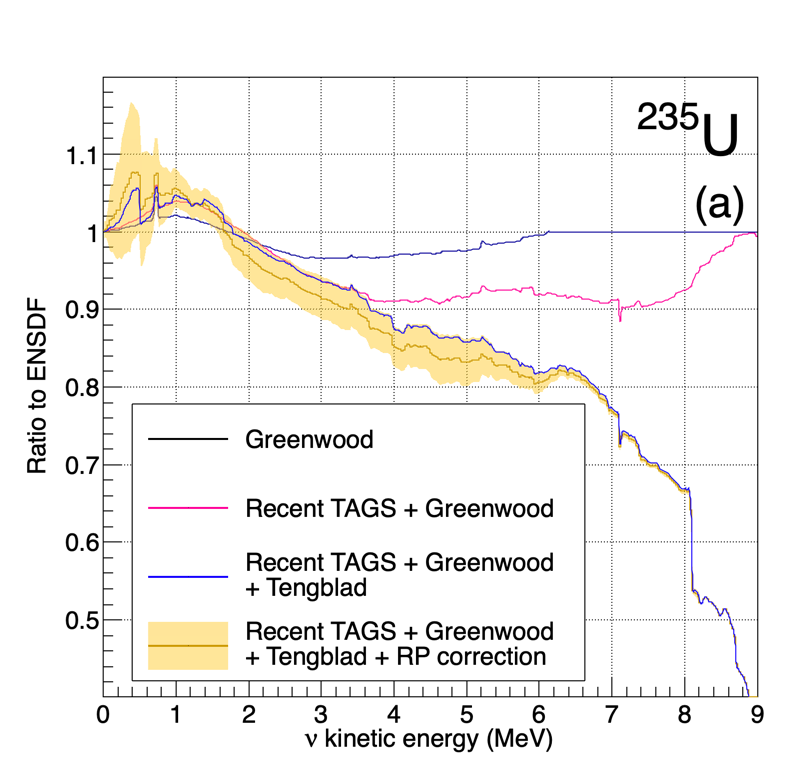}
    \includegraphics[width=0.47\textwidth]{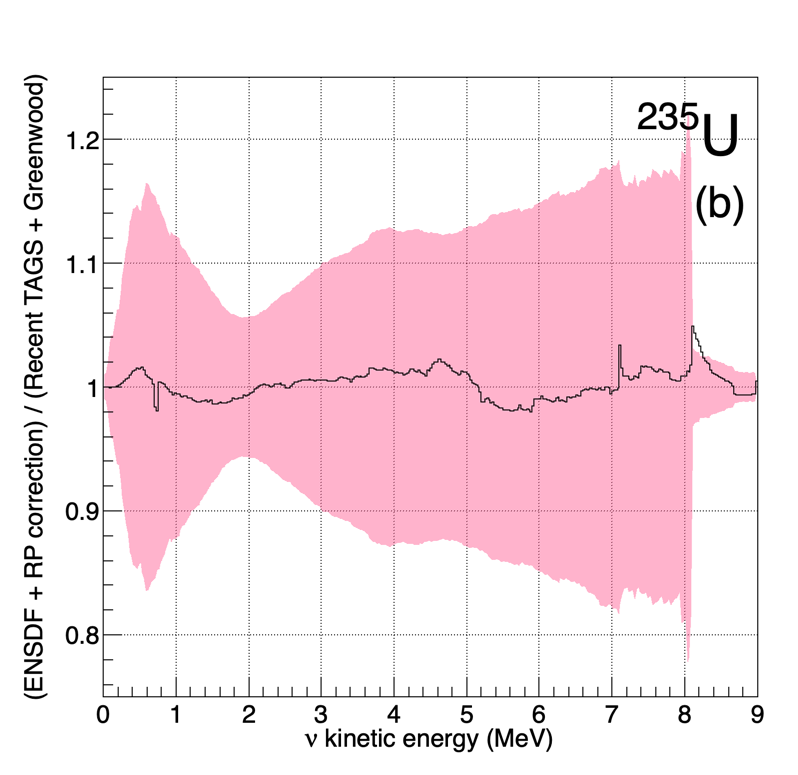}
\caption{Correction of the Pandemonium effect in the calculation of the \U{235} \bnue{} fission spectrum. (a) Impact of different sources of Pandemonium-free decay data. The ratios of spectra computed with different nuclear decay data sources over a spectrum calculated using ENSDF extracted decay data only are displayed. The Greenwood, Recent TAGS, Tengblad and RP correction labels respectively correspond to TAGS data coming from \cite{Greenwood1997}, to TAGS-corrected radionuclides as listed in Table~\ref{tab:list TAGS}, to integral $\mathrm{\beta}$ spectrum measurements from \cite{Tengblad1989} and to the 29 isotopes which the residual Pandemonium correction has been applied to. The yellow band shows the $\mathrm{1 \sigma}$ uncertainty resulting from applying the residual Pandemonium correction. (b) Coherence test of the residual Pandemonium correction, as derived in section~\ref{subsubsec: uncorrected pandemonium effect}. The black curve compares a \U{235} spectrum computed with TAGS-corrected data (Greenwood and Recent TAGS) to a spectrum computed using ENSDF extracted decay data only but applying the residual Pandemonium correction to these same isotopes instead of using their TAGS data. The red band is the $1 \sigma$ uncertainty resulting from the residual Pandemonium correction.}
\label{fig:Pandemonium_correction}
\end{figure*}

Table~\ref{tab:contribution breakdown} breaks down the calculation of each actinide fission \bnue{} spectrum and flux by source of decay information, especially showing that $\mathrm{\sim}$60\% (resp. $\mathrm{\sim}$45\%) of the \U{235} and \Pu{239} (resp. \U{238} and \Pu{241}) IBD yield is calculated using TAGS-corrected nuclear decay information. In a complementary way, Figure~\ref{fig:contribution spectrum} shows the contribution of the TAGS source of decay information to the calculation of the \U{235} and \Pu{239} fission \bnue{} spectra, showing that they amount to $\mathrm{\sim}$60\% of the spectrum in the 2-8 MeV energy range. The impact of correcting the ENSDF decay data from the Pandemonium effect is illustrated on Figure~\ref{fig:Pandemonium_correction} (a) for the calculation of the \U{235} \bnue{} spectrum. A very similar behavior is observed for the corresponding $\mathrm{\beta}$ spectrum. The use of the data from Greenwood et al. and from the recent TAGS measurements decreases (resp. increases) the \bnue{} spectrum above (resp. below) $\mathrm{\sim}$2 MeV by a few percents. This trend directly stems from the fact that Pandemonium-affected fission fragments have their corresponding $\mathrm{\beta}$/\bnue{} spectra underestimated (resp. overestimated) at low (resp. high) energies. As a result, the corresponding \U{235} IBD yield decreases by 2.4\% after including the TAGS data from Greenwood et al., and by another 5.4\% after adding the most recent TAGS data as listed in Table~\ref{tab:list TAGS}. The Pandemonium effect impacts the other actinide fission $\mathrm{\beta}$/\bnue{} spectrum and flux in a very similar way. Including and using the TAGS data as described previously result in a $\mathrm{\sim}$6-8\% decrease of their respective IBD yield.
%
%
\subsubsection{Integral $\mathrm{\beta}$ spectrum measurements}
\label{subsubsec: direct beta measurements}
Besides TAGS data, another valuable set of Pandemonium-free data comes from integral $\mathrm{\beta}$ spectrum measurements. Following the work of \cite{Mueller2011}, measurements of the continuous $\mathrm{\beta}$ and $\mathrm{\gamma}$-ray spectra emitted after the decay of 111 fission products by the group of Tengblad et al. \cite{Tengblad1989,Rudstam1990} were here considered to further correct nuclear decay data extracted from ENSDF. These 111 fission products are mostly short-lived radionuclides with large \Qb{} energies, and make up for $\mathrm{\sim}$90\% of a typical nuclear reactor \bnue{} flux above 6 MeV \cite{Tengblad1989}. Among these 111 measured $\mathrm{\beta}$ spectra, 44 were found to be consistent with calculations using ENSDF decay data. The corresponding integral $\mathrm{\beta}$ spectrum data were hence disregarded. Among the remaining 67 isotopes, 23 isotopes have also recently been measured with the TAGS method as described in the previous section. As opposed to the TAGS technique, integral $\mathrm{\beta}$ measurements only give access to the full $\mathrm{\beta}$ spectrum of a radionuclide without any information about the underlying $\mathrm{\beta}$ decay scheme. Therefore, the integral $\mathrm{\beta}$ spectrum data set associated to these 23 radionuclides has also been disregarded. The integral $\mathrm{\beta}$ spectrum measurements of 44 isotopes finally remain after this selection. They are listed in Table~\ref{tab:list Tengblad}. They have then been used to model their \bnue{} contribution instead of directly using the corresponding ENSDF extracted decay information.
\begin{table}[!htp]
\centering
\begin{tabular}{c} 
    \hline
    \hline
        Isotope  \\
    \hline
        \iso{80}{Ga},
        \iso{81}{Ga},
        \iso{82}{Ga},
        \iso{83}{Ge},
        \iso{79}{As},
        \iso{81}{As},
        \iso{82}{As},
        \iso{83}{As},\\
        \iso{85}{As},
        \iso{86}{As},
        \iso{83}{Se},
        \iso{83}{Se}*,
        \iso{89}{Br},
        \iso{90}{Br},
        \iso{87}{Kr},
        \iso{96}{Rb},\\
        \iso{97}{Sr},
        \iso{97}{Y}*,
        \iso{98}{Y},   
        \iso{99}{Y},
        \iso{99}{Nb},
        \iso{130}{Sn},
        \iso{130}{Sn}*,
        \iso{131}{Sn},\\
        \iso{133}{Sn},
        \iso{131}{Sb},
        \iso{133}{Sb},
        \iso{134}{Sb},
        \iso{135}{Sb},  
        \iso{136}{Sb},
        \iso{137}{Sb},
        \iso{136}{Te},\\
        \iso{137}{Te},
        \iso{135}{I},
        \iso{136}{I},
        \iso{138}{I},
        \iso{139}{I},
        \iso{140}{I},
        \iso{137}{Xe},  
        \iso{143}{Cs}, \\
        \iso{144}{Cs},  
        \iso{146}{Cs},
        \iso{146}{Ba},
        \iso{146}{La} \\
    \hline
    \hline 
\end{tabular}
\caption{List of fission fragments whose integral $\beta$ spectra are taken from \cite{Tengblad1989}.}
\label{tab:list Tengblad}
\end{table}

\begin{figure}[ht]
    \centering
    \includegraphics[width=0.5\textwidth]{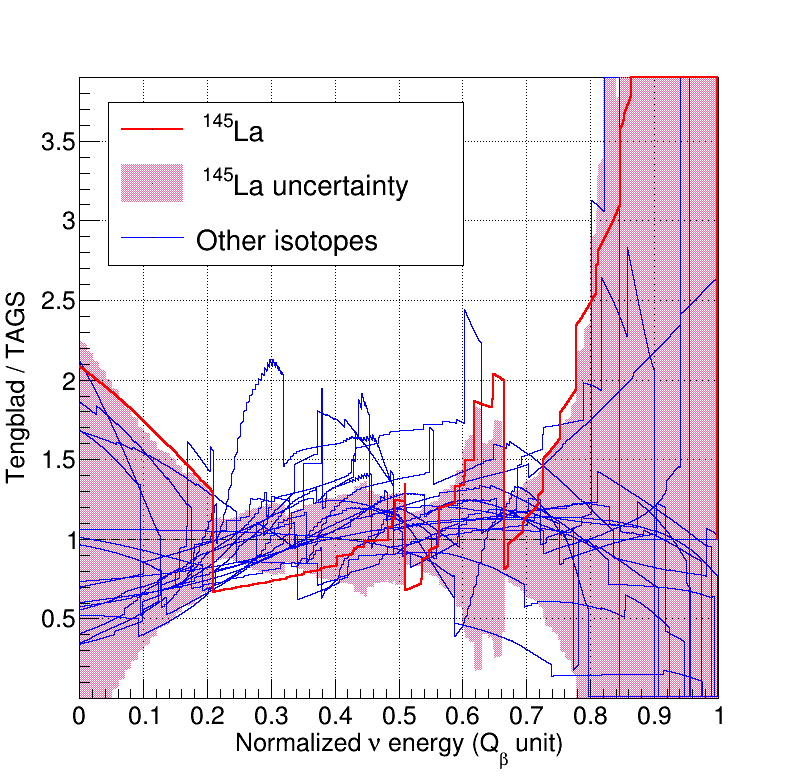}
    \caption{Miscalculation of \bnue{} spectra resulting from the conversion procedure applied to integral $\mathrm{\beta}$ spectrum measurements of a set of 24 fission fragments also having TAGS data (blue lines). The x-axis is expressed in normalized $\mathrm{\bar{\nu}_e}$ kinetic energy $\mathrm{E_{\nu}}$/\Qb{}. 
    The red line is the ratio associated to \iso{145}{La}. The red band represents the 1$\mathrm{\sigma}$ uncertainty modeled to cover calculation errors in the case of \iso{145}{La} (see text for more details).
    }
    \label{fig:ratio Tengblad}
\end{figure}

The \bnue{} counterpart of each of these 44 fission products has been modeled by applying the so-called conversion method, in which the associated $\mathrm{\beta}$ spectrum is adjusted by a set of 2 to 6 virtual allowed branches, each with a branching ratio and an endpoint energy as free parameters~\cite{Mueller2011}. The corresponding \bnue{} spectrum is then obtained by applying energy conservation to the adjusted virtual $\mathrm{\beta}$ branches, i.e. by substituting the $\mathrm{\beta}$ particle kinetic energy E with the \bnue{} energy $\mathrm{E_{\nu} = E_{0}-E}$, where $\mathrm{E_{0}}$ is the adjusted endpoint energy. Applying this procedure, these 44 fission fragments are found to make a 10-17\% contribution to each isotopic IBD yield (see Table~\ref{tab:contribution breakdown}) and a $\mathrm{\sim}$10\% contribution to an \bnue{} actinide fission spectrum (see Figure~\ref{fig:contribution spectrum}). Although the conversion procedure reproduces each isotope experimental $\mathrm{\beta}$ spectrum to less than a percent over the whole energy range, nothing ensures that the associated \bnue{} spectrum is accurately described by the set of adjusted virtual branches. As an example, Figure~\ref{fig:ratio Tengblad} depicts the ratio of \bnue{} spectra computed using the conversion procedure on integral $\mathrm{\beta}$ spectrum measurements over the same spectra using TAGS data for the previously mentioned 23 fission products both sharing these sources of data. Local excesses exceeding $\mathrm{\sim}$50\% are clearly visible. An uncertainty accounting for a potential bias originating from the conversion procedure must therefore be derived and applied to the 44 fission product \bnue{} spectra converted from the integral $\mathrm{\beta}$ measurements. The construction of the associated covariance matrix uses the 23 fission products both sharing integral $\mathrm{\beta}$ spectrum and TAGS measurements, and proceeds as follow: (i) each of these 23 fission products is assigned an individual covariance matrix computed using the relative difference between spectra expressed in normalized \bnue{} energy $\mathrm{E_{\nu}/Q_{\beta}}$ from converted integral $\mathrm{\beta}$ measurements and from using TAGS data (see Figure~\ref{fig:ratio Tengblad}) (ii) a single covariance matrix is computed by averaging out these 23 individual covariance matrices. The obtained covariance matrix encompasses then the average relative difference between a \bnue{} spectrum derived from integral $\mathrm{\beta}$ measurement conversion and from using TAGS data. It is then applied to model the uncertainty of each of the 44 \bnue{} spectra converted from integral $\mathrm{\beta}$ measurements. As an example, the red band in Figure~\ref{fig:ratio Tengblad} shows the 1$\mathrm{\sigma}$ uncertainty derived for \iso{145}{La}. Because the distribution of the relative difference between the \bnue{} spectrum converted from integral $\mathrm{\beta}$ spectrum measurements and the \bnue{} spectrum calculated using TAGS data does not show any particular pattern, the uncertainty applied to the converted integral $\mathrm{\beta}$ spectra is treated as uncorrelated among the 44 associated isotopes. Using this approach, the uncertainty derived on an individual isotope IBD yield is typically $\mathrm{\sim}$20\%. At the isotopic IBD yield level, the associated uncertainty totals $\mathrm{\sim}$1.5\% of the uncertainty budget, making it the second most important source of uncertainty (see Table~\ref{tab:uncertainty budget}). Figure~\ref{fig:fractional_uncertainty_data_sources} displays the associated fractional uncertainty for the calculation of the \U{235} fission \bnue{} spectrum. It shows that the conversion of the integral $\mathrm{\beta}$ spectrum measurements is the dominant source of uncertainties above 5 MeV.

Finally, the impact of incorporating the data from the integral $\mathrm{\beta}$ measurements of Tengblad et al.~\cite{Tengblad1989} in the calculation of the \U{235} actinide fission \bnue{} fission spectrum is illustarted by the blue curve in Figure~\ref{fig:Pandemonium_correction} (a). These data brings a significant \textgreater 10\% decrease to the $\mathrm{\gtrsim}$4~MeV portion of the spectrum, especially because as stated previously, they correspond to radionuclides with large \Qb{} energies. They bring a slightly larger correction to the calculation of the uranium (4.9\% and 6.2\% decrease for \U{235} and \U{238}, respectively) than the plutonium (2.9\% and 3.7\% decrease for \Pu{239} and \Pu{241}, respectively) isotopic IBD yields.
\subsubsection{Uncorrected decay information}
\label{subsubsec: uncorrected pandemonium effect}
Although the decay information of the most important fission products entering the computation of a reactor \bnue{} spectrum have been reassessed and corrected when necessary, a substantial amount of the remaining fission products may still be potentially affected and thus be uncorrected from the Pandemonium effect. 
Such radionuclides are usually identified by comparing the energy $\mathrm{E^{lvl}_{max}}$ of the highest recorded level of the daughter nucleus to the total energy \Qb{} available for the $\mathrm{\beta}$ decay.
If a significant difference between $\mathrm{E^{lvl}_{max}}$ and \Qb{} is observed, a radionuclide is then highly suspected to be Pandemonium-affected.
Using a similar criterion, 29 isotopes present in ENSDF have been identified and priorily selected by the Working Party on International Evaluation Co-operation of the Nuclear Energy Agency (WPEC-25) for a new measurement of their corresponding $\mathrm{\beta}$ decay scheme with the TAGS technique~\cite{TAGSmeeting2009, TAGSmeeting2014}.
As shown by Table~\ref{tab:contribution breakdown}, these isotopes represent 10-12\% of an isotopic IBD yield. For curiosity, this selection was extended using a very loose criterion,~i.e. searching the full ENSDF database for fission products with complete decay schemes and having at least a 20\% difference between $\mathrm{E^{lvl}_{max}}$ and $\mathrm{Q_{\beta}}$. This criterion selects 172, 160, 202 and 190 additional isotopes for \U{235}, \U{238}, \Pu{239} and \Pu{241} respectively, meaning that in a worst case scenario, 17 to 20\% of an isotopic IBD yield is impacted by a residual Pandemonium effect. The contribution of these isotopes to the calculation of the \U{235} and \Pu{239} \bnue{} fission spectra are also depicted on Figure~\ref{fig:contribution spectrum}. The blue short-dashed (resp. long-dashed) lines shows the WPEC-25 (resp. extended) selection, showing that the residual Pandemonium effect would mostly impact the $\mathrm{\lesssim}$ 5 MeV portion of a \bnue{} spectrum.

Because no other source than ENSDF is yet available, directly correcting these isotopes for the Pandemonium effect is impossible. The followed strategy is instead to apply an average correction, which is constructed by considering the average impact of the Pandemonium effect among a set of 81 radionuclides showing different decay information between their ENSDF and TAGS records. The average ratio of spectra computed with TAGS data over spectra computed with ENSDF data is taken as the correction. A covariance matrix is derived  by considering the dispersion of the 81 ratios with respect to the average ratio.
Using such approach, the IBD yield uncertainty for a fission fragment is $\mathrm{\sim}$20\% and safely covers the amplitude of the associated residual Pandemonium correction, which on average amounts to 18\% for the 81 radionuclides discussed previously.
At the actinide fission spectrum calculation stage, the uncertainty derived for the residual Pandemonium effect is treated as fully correlated among the relevant fission products. This choice is motivated by the fact that the Pandemonium effect systematically impacts the concerned fission fragment $\mathrm{\beta}$/\bnue{} spectra in the same way.
The robustness of the residual Pandemonium correction construction has been checked by comparing the computation of a \U{235} fission spectrum (i) correcting the 81 fission fragments with their TAGS data and (ii) correcting these same fission fragments with the residual Pandemonium correction instead. Results are displayed in Figure~\ref{fig:Pandemonium_correction} (b) and demonstrate that the residual Pandemonium correction reproduces fairly well the (true) correction of the spectrum using TAGS data to better than 3\% below 8 MeV. Furthermore, the difference is largely covered by its associated uncertainty, as depicted by the red band on that same figure.
Because the rather loose selection discussed above might misidentify a fair amount of fission fragments having a wrong decay scheme, the residual Pandemonium correction and associated covariance matrix is here only applied to the 29 fission fragments selected by the WPEC-25 group.
Applied this way, it decreases the \U{235}, \U{238}, \Pu{239} and \Pu{241} IBD yields by ($2.3 \pm 2.6$)\%, ($1.9 \pm 2.0$)\%, ($2.0 \pm 2.3$)\% and ($1.8 \pm 2.1$)\%, respectively. The amplitude of the residual Pandemonium correction to these isotopic IBD yields is then largely covered by the associated uncertainty. This is not surprising given the way the correction was constructed, but reassuring though given that this correction may not be perfectly suited to these 29 fission fragments. The residual Pandemonium correction uncertainty amounts to $\mathrm{\sim}$2.5\% when propagated to the calculation of the \U{235}, \U{238}, \Pu{239} and \Pu{241} IBD yields, making it by far dominant among all considered sources of uncertainty (see Table~\ref{tab:uncertainty budget}). 
The corresponding fractional uncertainty obtained in the computation of the \U{235} fission \bnue{} spectrum is also illustrated in Figure~\ref{fig:fractional_uncertainty_data_sources}, showing that it prevails below 5~MeV.
%
%
%
\subsection{Nuclides with no data}\label{subsec: nuclei with no data}
The crossing of the FP list extracted from the \jeff{} evaluated FY database \cite{JEFF33} with ENSDF \cite{ENSDF} leaves hundreds of emitters with neither nuclear structure information nor decay data. Table~\ref{tab:contribution breakdown} shows the estimated contribution of these nuclides with no data (NND) to the calculation of the \U{235}, \U{238}, \Pu{239} and \Pu{241} \bnue{} flux. The NND are usually short-lived and exhibit high \Qb{} energies far above the IBD threshold. They can therefore play a non-negligible role in the calculation of a reactor \bnue{} spectrum, with a typical $\mathrm{\sim}$5-10 \% contribution to the IBD flux. The modeling of the NND component in past summation calculations was usually done either using the Gross Theory of $\mathrm{\beta}$-decay~\cite{Takahashi1969,Nakata1997} or the so-called \Qb{} effective modeling~\cite{davis1979,Mueller2011} or a mix of both approaches~\cite{Fallot2012,Estienne2019}. None of these effective modelings is expected to accurately predict this contribution. Most importantly, no uncertainty treatment relative to this contribution has ever been proposed in the literature.

\begin{figure}[ht]
    \centering
    \includegraphics[width=0.47\textwidth]{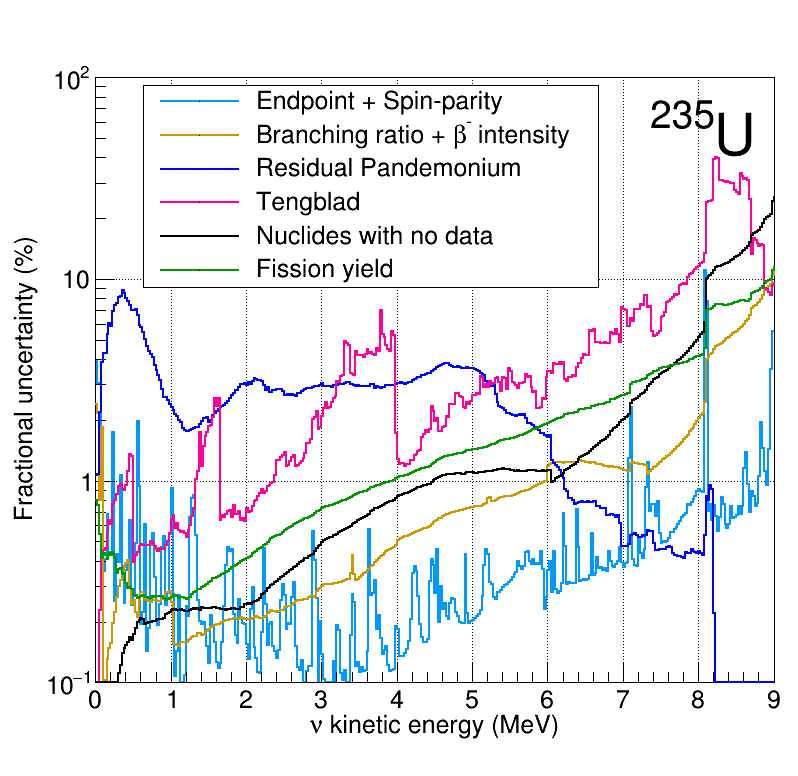}
    \caption{Uncertainties in the summation calculation of the \U{235} fission \bnue{} spectrum, broken down by source and type of evaluated nuclear data.}
    \label{fig:fractional_uncertainty_data_sources}
\end{figure}

In the present work, another approach to compute the contribution of the NND has been attempted. In a nutshell, this approach computes and averages the summation spectra associated to several pools of fission fragments to estimate the NND contribution, and uses the dispersion among these spectra to build an associated covariance matrix. In a first step, each NND is attributed a pool of several nuclides having a complete decay scheme and chosen such that their total $\mathrm{\beta}$-decay energy \Qb{} ranges within $\mathrm{\pm}$10\% of the NND \Qb{}. The resulting pool size typically varies from a few tens of nuclides up to a hundred. Then, each pool has a reference $\mathrm{\beta}$/\bnue{} spectrum estimated by averaging out the individual nuclide spectra. A covariance matrix is also built by using the dispersion of the individual nuclide spectra around the pool average spectrum. Summing each of these NND emulated spectra with their corresponding cumulative FY and $\mathrm{\beta^{-}}$ intensities then gives an estimate of their contribution to the calculation of an actinide fission \bnue{} spectrum. The uncertainty derived with this pool modeling on each of the NND IBD yield is typically at the level of $\mathrm{\sim}$60\%. No correlations among the calculated NND spectra are considered when propagating these uncertainties. At the fission actinide level, the NND contribution uncertainty to the IBD yield typically amounts to $\mathrm{\sim}$1\%, except for \U{235}. As shown by Table~\ref{tab:contribution breakdown}, NND have a smaller contribution of only 3.2\% to the \U{235} IBD yield, thus leading to a total uncertainty of 0.6\% (see Table~\ref{tab:uncertainty budget}). The associated fractional uncertainty obtained in the case of the \U{235} actinide fission \bnue{} spectrum is displayed in Figure~\ref{fig:fractional_uncertainty_data_sources}, with a rising trend following the contribution of the NND to the \bnue{} flux (see left panel of Figure~\ref{fig:contribution spectrum}). The NND source of uncertainty starts to significantly contribute in the high-energy portion ($\mathrm{\gtrsim}$ 6 MeV) of the spectrum.
\begin{figure*}[t!]
    \centering
    \includegraphics[width=0.47\textwidth]{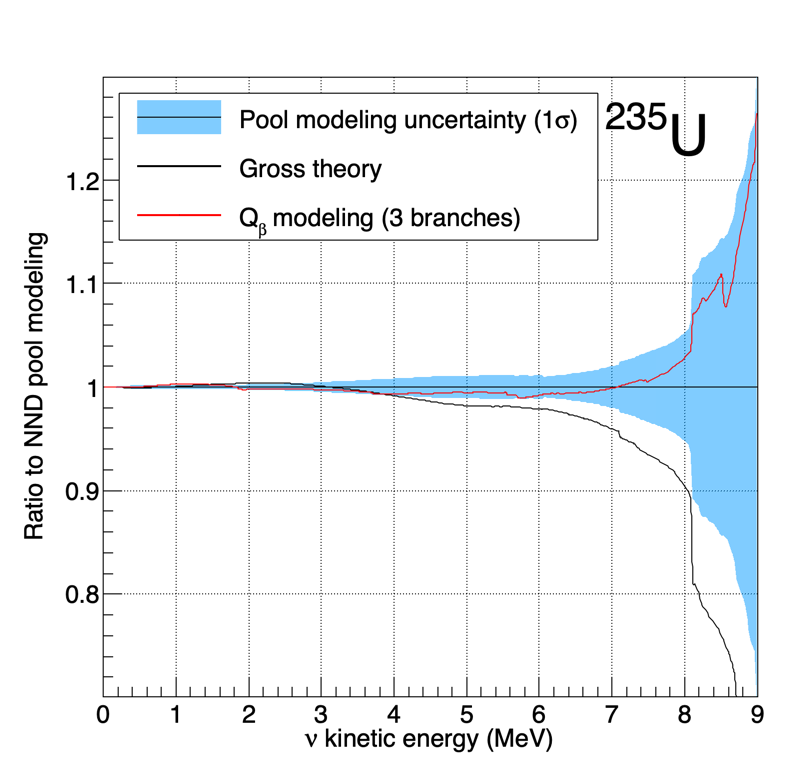}
    \includegraphics[width=0.47\textwidth]{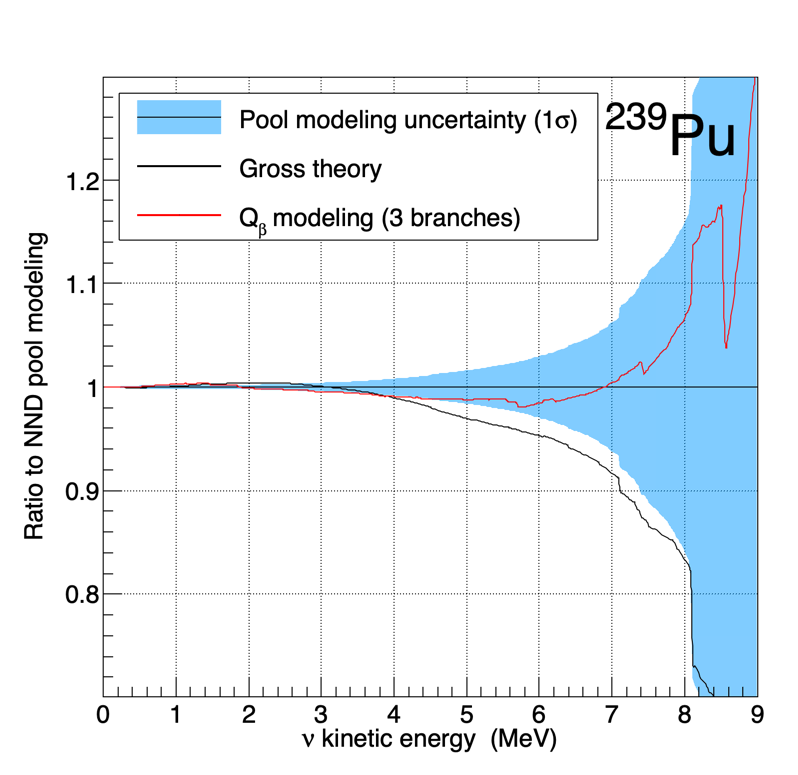}
\caption{Impact of different approaches to compute the contribution of the fission fragments with no decay information to the summation calculation of the \U{235} and \Pu{239} fission \bnue{} spectra. The comparison of the Gross Theory and \Qb{} effective modeling is made against the pool modeling described in section~\ref{subsec: nuclei with no data}. The blue area centered on unity represents the 1$\mathrm{\sigma}$ uncertainty band derived for the pool modeling.}
\label{fig:ratio_impact_NND}
\end{figure*}

The impact of the pool method for modeling the NND component in the summation calculation of the \U{235} and \Pu{239} fission \bnue{} spectra is investigated by comparing a modeling using the Gross Theory of allowed $\beta$-decay and a \Qb{} effective modeling using 3 evenly distributed transitions having the same branching ratios. As illustrated on Figure~\ref{fig:ratio_impact_NND}, notable changes are visible in the high energy part of the spectra above 6-7 MeV. In this energy regime, the Gross Theory (resp. \Qb{} effective modeling) predicts smaller (resp. larger) fluxes than the previously described modeling of the NND component. The corresponding $\mathrm{1 \sigma}$ uncertainty, pictured by the blue band on Figure~\ref{fig:ratio_impact_NND}, pretty much (resp. hardly) covers the reported differences with respect to the \Qb{} effective modeling (resp. the Gross Theory calculation). Both the Gross Theory and the \Qb{} effective modeling bring a negative 1-3\% change to the \U{238}, \Pu{239} and \Pu{241} IBD yields with respect to the present calculation of the NND component. The corresponding changes to the \U{235} IBD yield are however smaller than 1\%. This can be traced to a smaller contribution of the NND for this actinide (see e.g. Table~\ref{tab:contribution breakdown} or Figure~\ref{fig:contribution spectrum}). As shown by Table~\ref{tab:uncertainty budget}, the present pool modeling of the NND component induces an uncertainty of about 1\% on each isotopic IBD yield.
%
%
\section{Results and discussion}\label{sec:Results}
All along the previous sections, the summation calculation of the \bnue{} spectrum and flux following the fission of the \U{235}, \U{238}, \Pu{239} and \Pu{241} actinides has been methodically dissected and revised, keeping as one of the main objective the construction of a detailed and realistic uncertainty budget. The main results following this work are manyfold. They are discussed here after and challenged against state-of-the-art predictions and measurements.

Figure~\ref{fig:fractional_uncertainty_modeling} and Figure~\ref{fig:fractional_uncertainty_data_sources} summarize how the uncertainty budget of the \U{235} actinide fission \bnue{} spectrum respectively breaks down according to $\mathrm{\beta}$-decay modeling and evaluated nuclear data sources of uncertainty. The main lesson learned is that the quality and the incompleteness of the evaluated nuclear databases dominate by far the uncertainties. This finding equally applies to the other actinides. Although this fact was already known for decades, it has here been rigorously assessed and quantified. In more details, the $\mathrm{\beta}$ decay scheme correction of the known Pandemonium-affected fission fragments significantly changes the spectrum, with an approximate -4\%/MeV linear correction between 2 and 7 MeV (see Figure~\ref{fig:Pandemonium_correction} (a)). This correction combines different sets of nuclear data, and exhibits an uncertainty budget which is largely dominated by the conversion of the integral $\mathrm{\beta}$ spectrum measurements of Tengblad et al.~\cite{Tengblad1989}. This uncertainty especially prevails the total uncertainty budget of the \U{235} \bnue{} spectrum above $\mathrm{\sim}$5 MeV. The very likely possibility that many other fission fragments are left with an incorrect decay scheme led to the ad hoc construction of the residual Pandemonium correction, as described in section~\ref{subsubsec: uncorrected pandemonium effect}. Although the magnitude of this correction is far smaller than the correction arising from the combined sets of TAGS and integral $\mathrm{\beta}$ spectrum data (see Figure~\ref{fig:Pandemonium_correction} (a)), its associated uncertainty has been constructed in a conservative way, and therefore dominates the final uncertainty budget of the present calculations below 5 MeV. Finally, the incompleteness of the fission fragment decay information, mostly represented here by the contribution of the nuclides without any decay data (NND, see section~\ref{subsec: nuclei with no data}), only impacts the present calculations above 6-8 MeV depending on the considered fission actinide (see e.g. Figure~\ref{fig:contribution spectrum}). Consequently, the associated uncertainty prevails the combined uncertainty budget in that same energy regime. Nuclear decay data wise, a sizeable 20-25\% portion of an actinide fission spectrum is then still weakly modeled and points out the necessity of priorily measuring both the Tengblad et al. (see Table~\ref{tab:list Tengblad}) and WPEC-25 (see section~\ref{subsubsec: uncorrected pandemonium effect}) selections of fission fragments with the TAGS technique in order to improve the summation method. In the other hand, the NND component of the actinide fission spectra, which emerges in the very high energy portion of the spectrum, could well be better understood and tackled taking advantage of high energy measurements of reactor \bnue{}, such as the one recently released by the Daya Bay experiment~\cite{An2022b}.
Summation calculations are also sensitive to the details of the fission yield evaluation. Small tensions between the most recent evaluations have been observed (see e.g. Figure~\ref{fig:TBS_ratio_FY_libraries}), and therefore calls for a careful review of the used nuclear input data. The lack of a complete covariance matrix estimate for any of the available evaluations especially makes the uncertainty budget associated to FY information incomplete. Furthermore, the yields populating either the ground state or the isomeric state of a fission product are also weakly evaluated information, when available~\cite{Sears2021}. They can significantly differ from one library to another, and can therefore considerably impact the calculations, especially in the high energy portion of the spectrum.
Beta decay wise, pushing the branch modeling to a high level of refinement has been found to give a smaller impact (see section~\ref{sec:BetaDecayFormalism}). In particular, using advanced nuclear structure calculation to realistically model the 23 main non-unique forbidden transitions, totaling all together $\mathrm{\sim}$25\% of the expected IBD yield, gives a modest $\mathrm{\lesssim}$ 5\% change to the \bnue{} fission spectra when compared to the widely used $\mathrm{\xi}$-approximation treatment (see Figure~\ref{fig:non_unique_impact_on_U235}). This change is especially visible in the $\mathrm{\gtrsim}$ 4 MeV portion of the spectrum, where the contribution of these non-unique forbidden transitions takes over. Overall, the improved treatment of the non-unique transitions presented in this work gives a 0.5-5\% uncertainty in the computation of the \bnue{} fission spectra above 2 MeV, which is about 2-3 times less than the uncertainties arising from the quality and incompleteness of the evaluated nuclear decay data discussed above.

\begin{figure*}[ht!]
    \centering
     \includegraphics[width=0.85\textwidth]{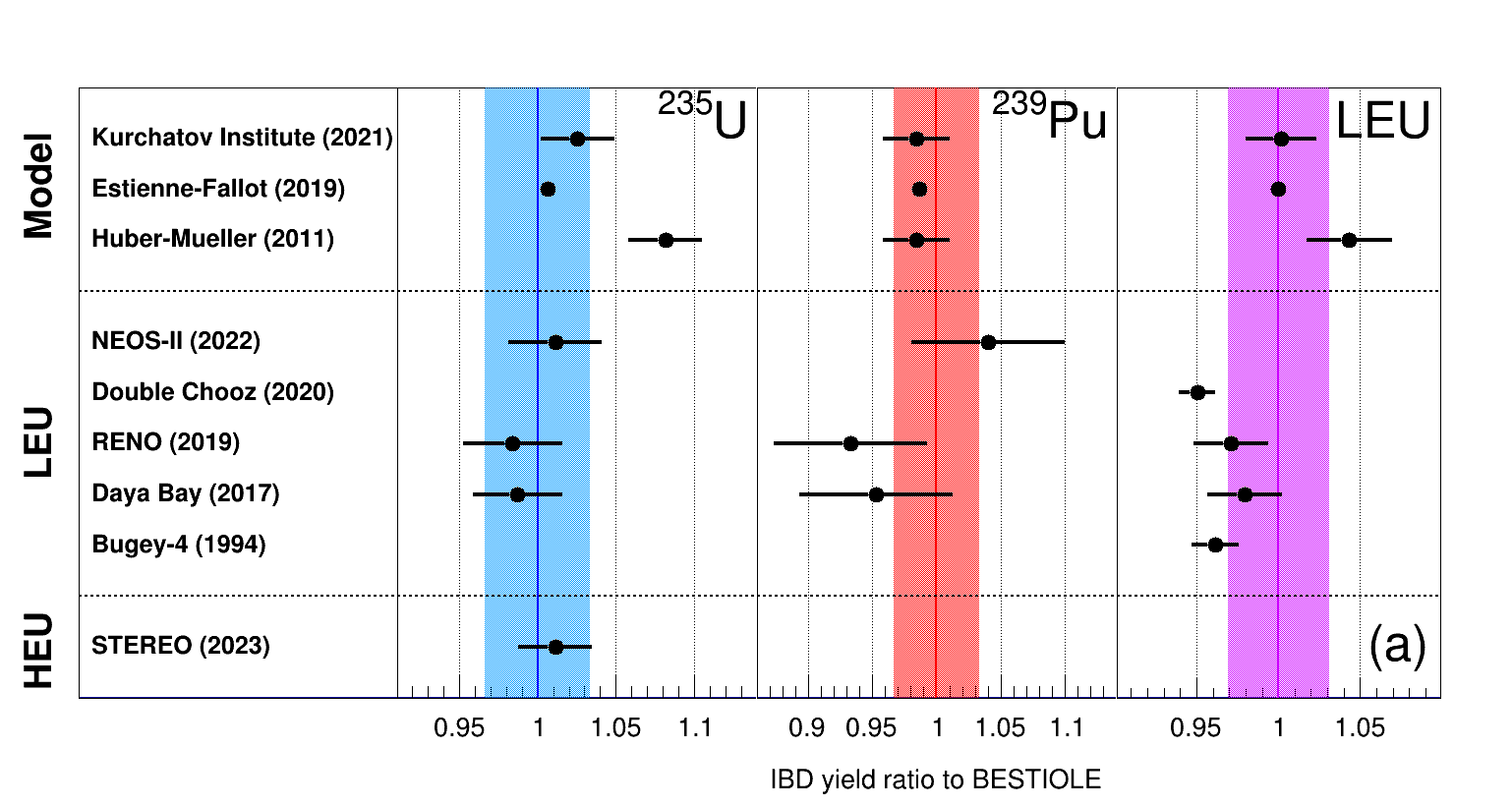}
     \includegraphics[width=0.6\textwidth]{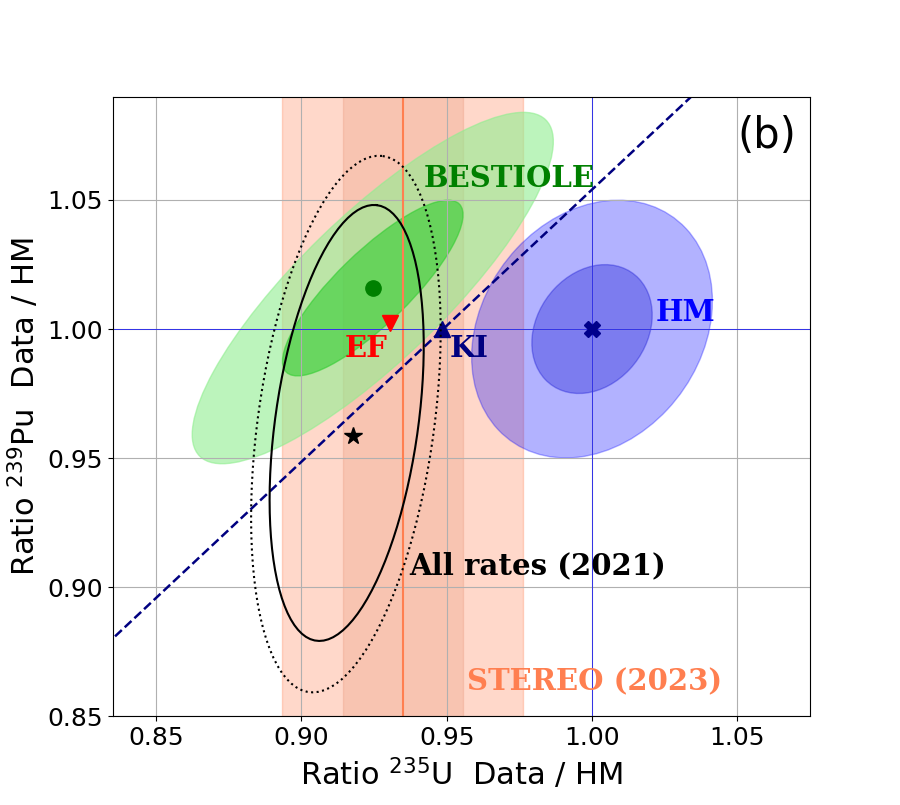} 
    \caption{Comparison of the IBD yields as obtained with BESTIOLE in the present work to a selection of state-of-the-art predictions and measurements. (a) Comparison of the isotopic IBD yields for \U{235}, \Pu{239} and the combination of \U{235}, \U{238}, \Pu{239} and \Pu{241} as measured at LEU commercial reactors. The shaded areas correspond to the 1$\mathrm{\sigma}$ uncertainty band estimated from the present summation calculations. The EF IBD yield predictions miss an uncertainty, because those were not evaluated in~\cite{Fallot2012,Estienne2019}. (b) Comparison of IBD yields expressed relatively to the HM prediction in the (\U{235},\Pu{239}) plane. The HM conversion prediction is pictured by the blue cross. The green dot and red inverted pyramid respectively correspond to the present summation calculations from BESTIOLE and from the EF prediction. The dark (light) shades are the 68\% CL (95\% CL) contours for the BESTIOLE summation calculations (green) and the HM prediction (blue). The latest STEREO measurement of the \U{235} IBD yield~\cite{Almazan2023} is pictured by the orange vertical line. The light and dark shaded bands are respectively the 68\% CL and 95\% CL associated uncertainty. The solid line (dotted line) ellipses correspond to 95\% CL (99\% CL) contours from a global analysis using fuel evolution and absolute rate measurements at LEU and HEU reactors~\cite{berryman2021}. The dashed blue line corresponds to the $\mathrm{(S^5/S^9)}$ aggregate beta spectrum ratio measured at the Kurchatov Institute. The blue triangle lying on this line corresponds to the KI prediction. The corresponding 68\% and 95\% CL contours are not displayed not to overload the figure. They are exactly the same than those of the HM prediction.}
    \label{fig:IBD_yield_comparison}
\end{figure*}

These conclusions hold for the three other actinides, and can be further appreciated while computing their corresponding IBD yields. 
Table~\ref{tab:uncertainty budget} presents the four isotopic IBD yields as obtained in the present work, along with a detailed breakdown of their respective uncertainty budget. They are found to have a total $\mathrm{\sim}$3\% relative uncertainty, which is slightly larger than the uncertainty budget corresponding to the latest conversion predictions~\cite{Mueller2011,Huber2011}. 
\begin{table*}[!htp]
\centering
\begin{tabular}{clllll}
    \hline
    \hline
        &       & \U{235}  & \U{238}   & \Pu{239}  & \Pu{241}  \\
    \hline
    \multicolumn{2}{l}{IBD yield {\small [$\mathrm{10^{-43} \ {cm^2/fission}}$]}}  & \multicolumn{1}{c}{6.25} & \multicolumn{1}{c}{10.01} & \multicolumn{1}{c}{4.48} & \multicolumn{1}{c}{6.58} \\
    \hline
    \multicolumn{2}{l}{Uncertainty {\small [\%]}}     &     &     &     &    \\
    \hline
\multirow{6}{*}{Data}     & 
          Endpoint + Spin-parity          &  0.1  &  0.1  &  0.1  &  0.1  \\
& Branching ratio + $\beta^{-}$ intensity &  0.4  &  0.3  &  0.4  &  0.3  \\
        & Residual Pandemonium            &  2.6  &  2.1  &  2.3  &  2.1  \\
        & Tengblad                        &  1.4  &  1.7  &  1.6  &  1.4  \\
        & Nuclides with no data           &  0.6  &  1.2  &  0.9  &  1.3  \\
        & Fission yield                   &  1.1  &  1.2  &  1.3  &  1.3  \\
    \hline
    \multicolumn{1}{l}{\multirow{6}{*}{Modeling}} & 
          Weak magnetism                  &  0.3  &  0.3  &  0.2  &  0.2  \\
        & Radiative correction            &  0.1  &  0.1  &  0.1  &  0.1  \\
        & Nuclear structure calculation   &  0.3  &  0.2  &  0.2  &  0.2  \\
        & $\mathrm{\xi}$-approximation    &  0.3  &  0.5  &  0.3  &  0.4  \\
        & IBD cross-section               &  0.1  &  0.1  &  0.1  &  0.1  \\
    \hline
        & Total                           &  3.3  &  3.2  &  3.3  &  3.2  \\
    \hline
    \hline
\end{tabular}
\caption{Isotopic IBD yields and their corresponding uncertainties, broken down according to uncertainty sources from evaluated nuclear data and modeling of $\mathrm{\beta}$ branches.}
\label{tab:uncertainty budget}
\end{table*}
%
%
Figure~\ref{fig:IBD_yield_comparison} (a) compares the \U{235} and \Pu{239} IBD yields against the Estienne-Fallot (EF) summation prediction~\cite{Fallot2012,Estienne2019}, the Huber-Mueller (HM)~\cite{Mueller2011,Huber2011} and the Kurchatov Institute (KI)~\cite{Kopeikin2021} conversion predictions, as well as a selection of measurements achieved by the Daya Bay~\cite{An2017}, NEOS-II~\cite{NEOSII}, RENO~\cite{Bak2019}, Double Chooz~\cite{dekerret2020}, Bugey-4~\cite{Giunti2022,Declais1994} and STEREO~\cite{Almazan2023} IBD experiments. 
The EF and HM IBD yields are here evaluated by considering fission spectra estimated at 450 days of irradiation time. They are expected to negligibly differ with respect to a spectrum computed under full equilibrium conditions. The EF fission spectra are directly taken from \cite{Estienne2019}, while off-equilibrium corrections at 450 days estimated in~\cite{Mueller2011} are applied to the HM prediction. Moreover, HM fission spectra are completed below 1.825 MeV and above 8.125 MeV using the present summation calculations, which gives a $\mathrm{\sim}$0.5\% correction to the corresponding isotopic IBD yields.
The comparison of the present IBD yield calculations to the EF prediction does not show any significant discrepancies within the reported uncertainties. Although these two summation predictions differ in e.g.~the $\mathrm{\beta}$ decay modeling details or the treatment of the NND component, this result is not surprising because they use very similar sets of Pandemonium corrected nuclear decay data. As stated earlier, the correction of the evaluated nuclear decay data from the Pandemonium effect is the most impactful to the summation calculations of reactor \bnue{} spectra.
A (7.5 $\mathrm{\pm}$ 3.9)\% difference in the \U{235} IBD yield is observed with respect to the HM prediction, while \Pu{239} IBD yields show a very good agreement. This difference is not highly significant, but is in line with the recent \U{235} IBD yield measurements conducted by the LEU and HEU experiments cited above.
Figure~\ref{fig:IBD_yield_comparison} (b) further illustrates this observation. Together with the latest STEREO measurement of the \U{235} IBD yield~\cite{Almazan2023}, it expresses in the (\U{235},\Pu{239}) plane the result of a global rate analysis using IBD data from HEU and LEU experiments relatively to the HM prediction~\cite{berryman2021}, and shows that both are in tension with the latter. The IBD yields obtained through the summation method with BESTIOLE in the present work and in the EF prediction are also shown for comparison. They come in a very good agreement with these experimental results, and all together suggest that the RAA is mosty caused by an overestimate of the \U{235} \bnue{} flux. The dashed blue line, which indicates how the \Pu{239}/\U{235} IBD yield ratio would scale according to the recent KI measurement of the $\mathrm{(S^5/S^9)}$ aggregate beta spectrum ratio, further supports this interpretation. In particular, the KI prediction assumes that the (5.4 $\mathrm{\pm}$ 0.2)\% offset measured in this ratio is entirely caused by a wrong normalisation of the ILL original \U{235} aggregate beta spectrum measurement. As shown by Figure~\ref{fig:IBD_yield_comparison} (b), it exhibits a much better agreement with the IBD yield experimental data than the HM prediction does. 
Finally, the experimental uncertainties reported in the recent measurements of the \U{235} and \Pu{239} IBD yields are of the same order of magnitude than those obtained in the present summation prediction. As can be seen in Figure~\ref{fig:IBD_yield_comparison}, they cannot yet fully conclude about the origin of the RAA, demonstrating that further improvements both in future reactor \bnue{} flux predictions and measurements are necessary.

%
%

\begin{figure*}[t!]
    \centering
    \includegraphics[width=0.47\textwidth]{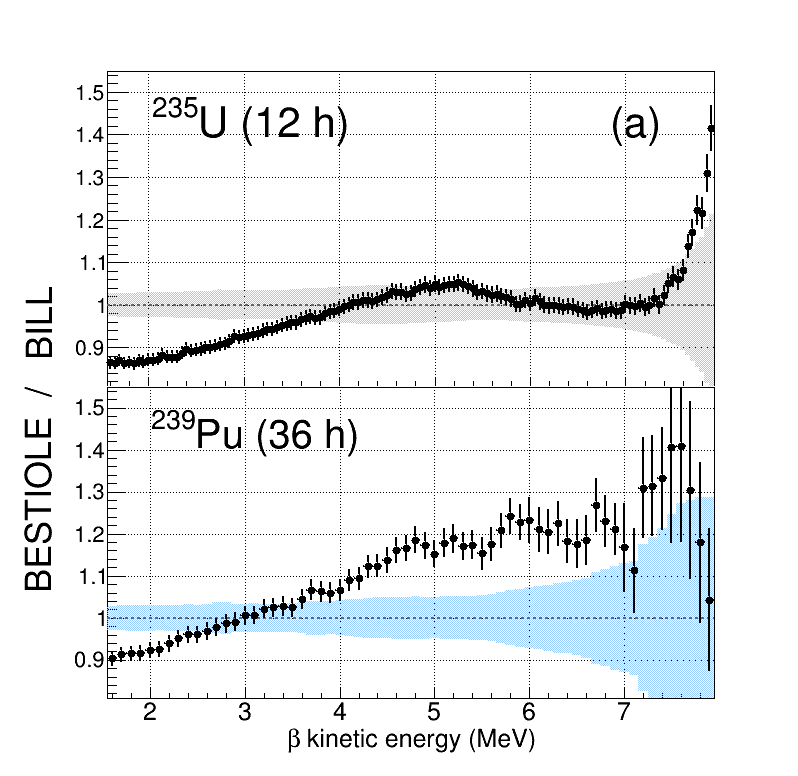}
    \includegraphics[width=0.47\textwidth]{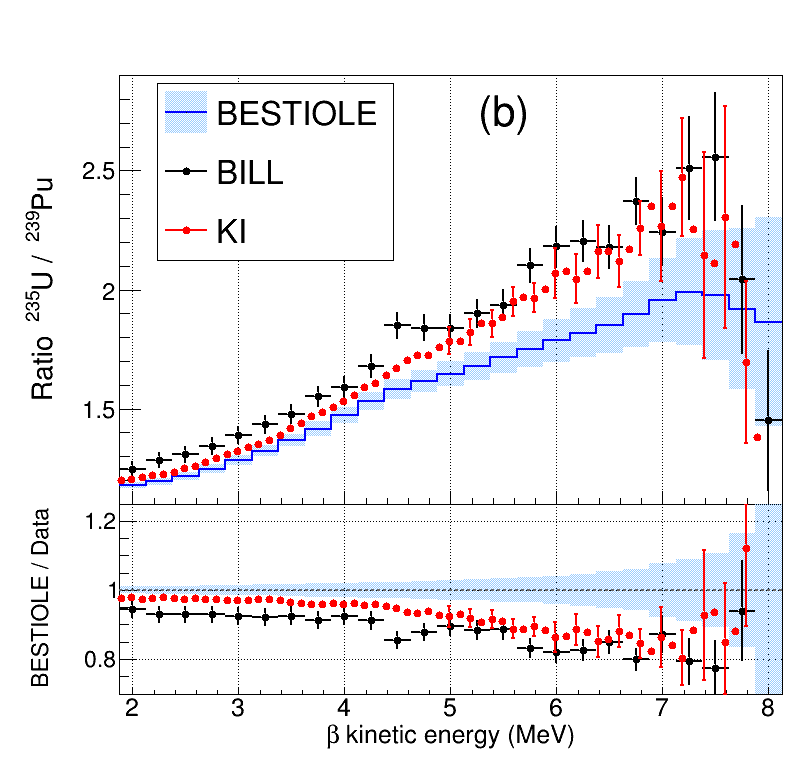}
\caption{Comparison of the BESTIOLE summation prediction to the ILL aggregate $\mathrm{\beta}$ spectrum data~\cite{Schreckenbach1982,Schreckenbach1985}. Prediction uncertainties are represented by the blue band. Black error bars correspond to experimental uncertainties. (a) Ratios to the \U{235} (top) and \Pu{239} (bottom) ILL $\mathrm{\beta}$ spectra. (b) The top panel shows the $\mathrm{(S^5/S^9)}$ $\mathrm{\beta}$ spectrum ratios both coming from the summation prediction (BESTIOLE), from the ILL data (BILL), and from the recent $\mathrm{(S^5/S^9)}$ ratio measurement performed at the Kurchatov Institute (KI)~\cite{Kopeikin2021}. The corresponding $\mathrm{(S^5/S^9)^{BESTIOLE}}$/$\mathrm{(S^5/S^9)^{BILL,KI}}$ double ratios are displayed in the bottom panel.}
\label{fig:Beta_spectrum_comparison}
\end{figure*}

To investigate a step further the IBD yield differences with respect to the HM model, Figure~\ref{fig:Beta_spectrum_comparison} (a) compares the present summation calculations to the aggregate $\mathrm{\beta}$ spectra measured in the 1980s at the Institut Laue-Langevin (ILL) with the BILL magnetic spectrometer, for the thermal fission of \U{235} and \Pu{239} respectively. Off-equilibrium effects at low energies caused by the short neutron irradiation time of the ILL measurements are taken into account by using the FISPACT-II numerical code~\cite{Sublet2017} together with IFYs from the \jeff{} database~\cite{JEFF33}. The FP activities were calculated after a 12-h and a 36-h irradiation time for \U{235} and \Pu{239}, respectively. As a first approximation, their associated uncertainties were taken as those coming from the CFY evaluation. Significant discrepancies, linearly ranging from -15\% up to +5\% for \U{235} and -10\% up to +20\% for \Pu{239} can be observed between 1 and 5 MeV. Figure~\ref{fig:Beta_spectrum_comparison} (b) plots the $\mathrm{S^5/S^9}$ aggregate $\mathrm{\beta}$ spectrum ratios constructed from the ILL data and from the present summation calculations. Although a (7.0 $\mathrm{\pm}$ 3.0)\% mean offset is still present, these $\mathrm{S^5/S^9}$ ratios are in a closer agreement. The linear deviations observed in Figure~\ref{fig:Beta_spectrum_comparison} (a) then hint at a possible systematic effect both present in the \U{235} and \Pu{239} $\mathrm{\beta}$ spectra, and partially compensating when constructing the $\mathrm{S^5/S^9}$ ratio. At the present stage, nothing indicates whether this systematic effect comes from the ILL data or from the present calculations. For further comparison, data points from the KI $\mathrm{(S^5/S^9)}$ measurement are also superimposed on Figure~\ref{fig:Beta_spectrum_comparison} (b). They exhibit a better agreement than the ILL data when compared to the present summation calculations. Still, significant deviations can be seen, especially at high energies. 

\begin{figure}[!ht]
    \centering
    \includegraphics[width=0.47\textwidth]{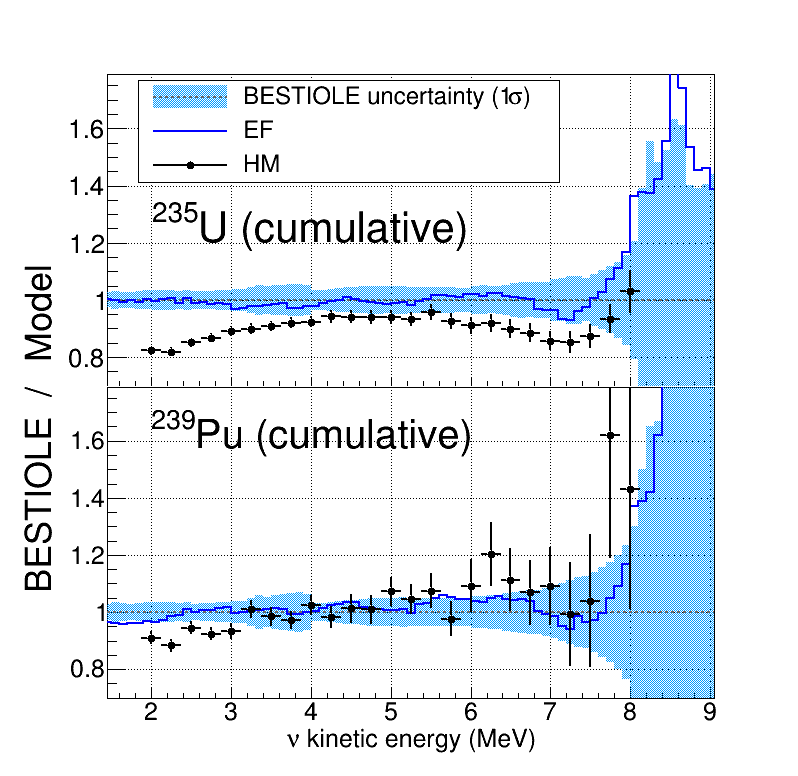}
    \caption{Comparison of the BESTIOLE summation calculations to the EF and HM \U{235} and \Pu{239} \bnue{}  fission spectra. Prediction uncertainties from BESTIOLE are represented by the blue band. Black uncertainty bars correspond to HM uncertainties. No uncertainties are available for the EF prediction.}
    \label{fig:EF_and_HM_nu_spectrum_comparison}
\end{figure}

The IBD yield comparison discussed above is further examined in Figure~\ref{fig:EF_and_HM_nu_spectrum_comparison}, which displays the ratio of both the \U{235} and \Pu{239} \bnue{} spectra as predicted with BESTIOLE in the present work to either the HM or the EF model. Unsurprisingly, similar deviations than those observed in the direct comparison to the ILL data (see Figure~\ref{fig:Beta_spectrum_comparison} (a)) are also observed here in the comparison to the HM prediction.
The \U{235} and \Pu{239} \bnue{} spectra from the present work and the EF prediction agree to within $\mathrm{\pm}$ 5\% in the energy range below $\mathrm{\sim}$7 MeV. The large differences observed at higher energies is suspected to mostly come from a different treatment of the NND component. The EF summation prediction mostly uses the gross theory of $\mathrm{\beta}$ decay, which gives smaller \bnue{} fluxes at high energy than the pool modeling proposed in the present work (see Figure~\ref{fig:ratio_impact_NND}). Last but not least, using FY evaluation from the JEFF-3.1.1 library as in the original EF prediction has been found to significantly improve the agreement between the corresponding actinide fission \bnue{} spectra, especially for the Plutonium isotopes. This last point again demonstrates the importance of a robust evaluation of the fission fragment yields for more    accurate summation calculations.
%
%

\begin{figure}[!ht]
    \centering
    \includegraphics[width=0.47\textwidth]{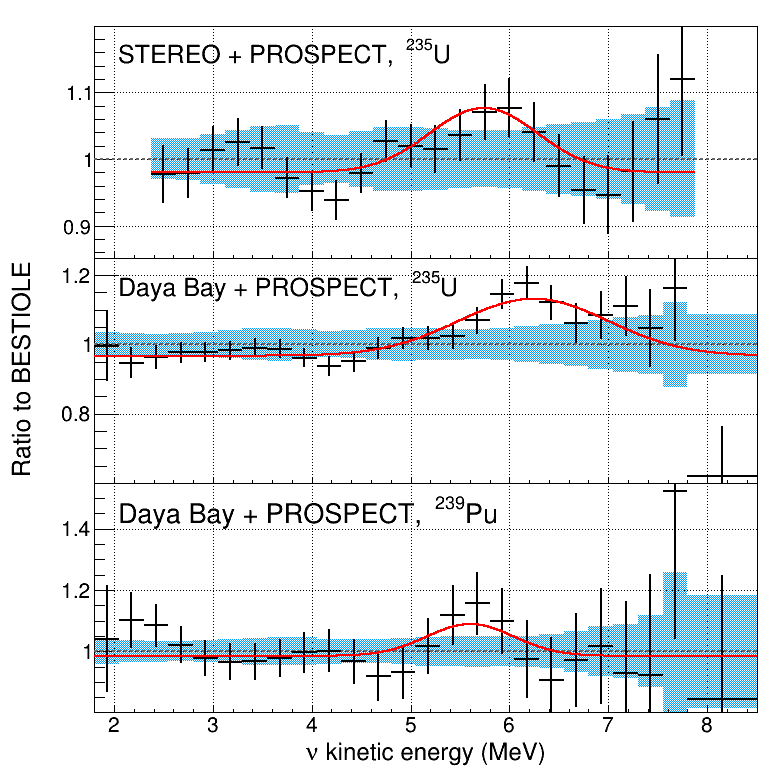}
    \caption{Shape-only comparison of the BESTIOLE summation prediction to experimentally measured \U{235} and \Pu{239} fission \bnue{} spectra. Prediction uncertainties are represented by the blue bands, and experimental uncertainties by the histogram error bars. From top to bottom are respectively displayed the ratios of the unfolded \U{235} spectrum from the STEREO and PROSPECT joint measurement \cite{Almazan2022} and of the unfolded \U{235} and \Pu{239} spectra from the Daya Bay and PROSPECT joint measurement \cite{An2022a} to the summation prediction (BESTIOLE). The red curves correspond to the best-fit Gaussian distortions (see text for more details), whose respective parameters are displayed in Table~\ref{tab:Gaussian_excess_results}.
    }
    \label{fig:IBD_spectrum_shape_comparison}
\end{figure}

The last point of comparison focuses on the shape of the predicted fission \bnue{} spectra. IBD spectrum measurements extracted from the combination of the PROSPECT data together with either the STEREO~\cite{Almazan2022} (here denoted SP) or the Daya Bay~\cite{An2022a} data (here denoted DBP) are here used as benchmarks. Figure~\ref{fig:IBD_spectrum_shape_comparison} shows how the present summation calculations compare to the unfolded \U{235} and \Pu{239} \bnue{} experimental spectra. In details, both the experimental spectra and the summation prediction are area-normalized to perform a shape-only comparison. 
To account for residual effects in the experimental data unfolding process, the summation prediction is filtered using published smearing matrices as prescribed in the supplementary materials of~\cite{Almazan2022,An2022a}. Given the uncertainties both coming from the experimental measurements and those estimated in the present work, an overall good agreement between data and prediction is observed. A $\mathrm{\chi^2}$ value was computed and respectively gave $\mathrm{\chi^2/ndf}$=19.1/21, $\mathrm{\chi^2/ndf}$=25.8/23 and $\mathrm{\chi^2/ndf}$=13.8/22 for the SP \U{235}, DBP \U{235} and \Pu{239} data, demonstrating an overall good agreement with the present summation prediction in the 1.8-7.5 MeV energy range.

In light of the spectrum deviations recently reported in the 5-7 MeV energy regime with respect to the HM model, a Gaussian distortion was searched for:
\begin{equation}
\label{eq:Gaussian_excess_eq}
    \mathrm{M(E_{\nu}) = K \cdot S_{k}(E_{\nu}) \left[ 1 + A\,e^{- \frac{(E_{\nu} - \mu)^2}{2\sigma^2}} \right],}
\end{equation}
where K is a global normalisation parameter allowing for a shape-only comparison, and $\mathrm{S_k(E_{\nu})}$ is the actinide fission spectrum as defined in eq.~\ref{eq:simplified_reactor_spectrum} and eq.~\ref{eq:reactor_spectrum}. The results are summarized in Table~\ref{tab:Gaussian_excess_results}. They are also superimposed (red solid lines) to the displayed ratios in Figure~\ref{fig:IBD_spectrum_shape_comparison}. As anticipated and shown by the $\mathrm{\chi^2/ndf}$ values reported in Table~\ref{tab:Gaussian_excess_results}, none of the experimental reactor \bnue{} datasets significantly favors the Gaussian distortion hypothesis in this energy regime. The DBP \U{235} measurement, which shows the largest deviation to the summation prediction, prefers the Gaussian distortion hypothesis at the 2.3$\mathrm{\sigma}$ level.

\begin{table*}[!htp]
\centering
\begin{tabular}{lccccc}
    \hline
    \hline
        &    A   & $\mathrm{\mu}$ [MeV]  & $\mathrm{\sigma}$ [MeV]  & $\mathrm{H_0}$ & $\mathrm{H_1}$ \\
    \hline
    SP (\U{235})   &    $\mathrm{0.098 \pm 0.050}$ & $\mathrm{5.74 \pm 0.30}$  &  $\mathrm{0.55 \pm 0.20}$  & 19.1/21 & 15.1/18 \\
    DBP (\U{235})  &    $\mathrm{0.169 \pm 0.061}$ & $\mathrm{6.24 \pm 0.30}$  &  $\mathrm{0.76 \pm 0.29}$  & 25.8/23 & 16.1/20 \\
    DBP (\Pu{239}) &    $\mathrm{0.105 \pm 0.088}$ & $\mathrm{5.60 \pm 0.21}$  &  $\mathrm{0.44 \pm 0.10}$  & 13.8/22 & 12.3/19 \\
    \hline
    \hline
\end{tabular}
    \caption{Best-fit parameters resulting from the search of a Gaussian distortion (see eq.~\ref{eq:Gaussian_excess_eq}) in the STEREO and PROSPECT (SP) \U{235} and Daya Bay and PROSPECT (DBP) \U{235} and \Pu{239} unfolded \bnue{} spectra. The last two columns, denoted $\mathrm{H_0}$ and $\mathrm{H_1}$, report the $\mathrm{\chi^2/ndf}$ values obtained for the "no Gaussian distortion" and "Gaussian distortion" hypotheses, respectively.}
\label{tab:Gaussian_excess_results}
\end{table*}
%
%
%
%
\section{Conclusions}\label{sec:Conclusions}
In this article, the summation method was deeply improved using an advanced $\mathrm{\beta}$-decay formalism together with recent evaluated nuclear data. For the first time, a complete uncertainty budget accounting for all known effects likely to impact the calculation of reactor \bnue{} spectra is proposed. The modeling of all $\mathrm{\beta}$ transitions known to contribute to a reactor \bnue{} spectrum has been greatly refined using the Behrens and Burhing formalism. A majority of these transitions are of the allowed and unique forbidden type. They are now computed accurately through the implementation of the main electro-magnetic corrections to the Fermi theory and an exact calculation of their corresponding shape factors. Furthermore, the treatment of the non-unique forbidden transitions, which were for long ago anticipated to play an important role, was realistically tackled using nuclear structure calculations. In particular, the $\xi$-approximation, which is abusively used to model these transitions, has been demonstrated to have a modest percent level impact on the prediction of isotopic IBD yields. The quality of the evaluated nuclear data was also extensively assessed. The fission yield evaluations available in the most recent \jeff{}, \ENDF{} and \JENDL{} libraries were used and compared, showing significant discrepancies in the resulting actinide fission \bnue{} spectra at $\mathrm{\gtrsim 5}$ MeV energies. These discrepancies together with a lack of a complete set of covariance matrices correctly describing the correlations between the fission fragment yields make the uncertainty budget associated to the summation calculations incomplete in that respect. Furthermore, the reliability of the fission fragment decay data was scrutinized using the most up-to-date Pandemonium free measurements available in the literature and in online databases. Although many experimental efforts are being conducted to improve the quality and the reliability of those data, the present summation model remains potentially affected by a residual Pandemonium effect still not corrected for in the present day nuclear databases. An ad hoc correction was then constructed to take it into account. This effect currently dominates the associated uncertainty budget, making it necessary to further measure or remeasure the potentially impacted fission fragments with the TAGS technique. Finally, a new approach to estimate both the contribution and uncertainty of all known fission fragments left with no decay data has been developed and validated against the usual \Qb{} effective and Gross Theory modeling. These fission fragments only impact the high energy portion of the actinide fission \bnue{} spectra, and play for now a secondary role.

Following all these improvements, this newly revised summation model was then extensively compared against a selection of state-of-the-art predictions and measurements. General good agreement is achieved with measured IBD yields at LEU and HEU reactors, especially favoring the RAA to be mostly caused by an overestimate of the \U{235} flux in the HM model. While significant discrepancies with the \U{235} and \Pu{239} aggregate $\mathrm{\beta}$ spectra measured at the ILL are observed, the present work shows closer agreement with the recently measured $\mathrm{S^5/S^9}$ aggregate spectrum ratio at the Kurchatov Institute. The summation calculations achieved in this work also demonstrate good shape agreement with a set of recently measured \bnue{} spectra from the Daya Bay, PROSPECT and STEREO experiments. All of these comparison studies show that the present summation calculations pretty well describe the most recent IBD flux and spectrum measurements. They are however unable to reconcile them with the original ILL data the HM conversion prediction relies on, further casting doubts on the reliability of these data. The present work does however not allow to favour any particular scenario for understanding the origin of the RAA. Further improvements both in the future summation modeling and measurements of reactor \bnue{} flux and spectra are necessary to firmly conclude.

Finally, the uncertainty budget constructed in the present work had to address in a realistic and conservative way the many systematic effects arising from gaps in the modeling of several fission fragments. These deficiencies either come from knowingly non-valid approximations (e.g. $\mathrm{\xi}$-approximation), unreliable information (e.g. Pandemonium effect) or even missing information. The uncertainties and/or corrections associated to these effects were all estimated following a similar strategy, which uses subsets of fission fragments having known and reliable information as proxies. Implicit in this strategy is the assumption that the uncertainties and/or corrections derived from using these proxies are fully representative of those that would apply to fragments actually having these gaps. Although the validity of this assumption could be questioned, the observed good agreement of the present summation calculations with the latest IBD flux and spectrum measurements indicates no significant bias in either the modeling of these corrections or the modeling of these uncertainties.

To conclude, the summation method made an important step toward becoming a reference tool for the prediction of reactor \bnue{} fluxes and spectra. In this respect, the calculated \bnue{} ($\beta$) fission spectrum data points for each of the four actinides are provided under the form of 25 keV binned histograms in the supplementary materials of this article.
Moreover, a total covariance matrix for the major four \bnue{} ($\beta$) actinide fission spectra, including each spectrum covariance matrix as well as cross-covariance matrices, is provided. The \bnue{} spectra and the associated covariance matrices have been extended below the IBD energy threshold to 0 MeV and at high energies up to 12.5 MeV, both allowing for a proper comparison with the forthcoming CEvNS experiments and future IBD experiments sensitive to the highest reactor \bnue{} energies.
Finally and for completeness purposes, the spectra and covariance matrix of the most relevant activation products created by the neutron irradiation of reactor fuel and structural materials, namely \iso{239}{U}, \iso{239}{Np}, \iso{28}{Al}, \iso{56}{Mn}, \iso{6}{He}, \iso{52}{V}, are also provided in the supplementary materials under the exact same format.
%
%
\section*{Acknowledgments}
The authors thank Dr.~Chikako Ishizuka for kindly providing the fission yield evaluation of the latest Japanese Evaluated Nuclear Data Library release. 
Dr.~Muriel Fallot along with the Valencia-Nantes collaboration for TAGS evaluation, and Dr.~Ãleksandra Fijałkowska along with the ORNL-Warsaw group for TAGS evaluation are also graciously thanked for providing TAGS data.
This work was done in the framework of the NENuFAR project, which was supported by the direction of the cross-disciplinary programs at Commissariat à l'énergie atomique et aux énergies alternative (CEA).

\newpage

\bibliographystyle{apsrev4-2}
\bibliography{reference}

\end{document}